\documentclass{aa}

\usepackage{gensymb}

\newcommand{\FeII}{[\ion{Fe}{ii}] } 
\newcommand{\FeIIp}{[\ion{Fe}{ii}]} 
\newcommand{\FeI}{[\ion{Fe}{i}] }
\newcommand{\FeIp}{[\ion{Fe}{i}]}

\newcommand{\SII}{[\ion{S}{ii}]}
\newcommand{\SIIp}{[\ion{S}{ii}] }
\newcommand{\OI}{[\ion{O}{i}] }
\newcommand{\OIp}{[\ion{O}{i}]}

\newcommand{\NiII}{[\ion{Ni}{ii}] }
\newcommand{\NiIIp}{[\ion{Ni}{ii}]}

\newcommand{\NeII}{[\ion{Ne}{ii}] }
\newcommand{\NeIIp}{[\ion{Ne}{ii}]}
\newcommand{\CoII}{[\ion{Co}{ii}] } 
\newcommand{\CoIIp}{[\ion{Co}{ii}]} 
\newcommand{\ClI}{[\ion{Cl}{i}] } 
\newcommand{\ClIp}{[\ion{Cl}{i}]} 
\newcommand{\ClII}{[\ion{Cl}{ii}] }
\newcommand{\ClIIp}{[\ion{Cl}{ii}]}

\newcommand{\HI}{\ion{H}{i} }
\newcommand{\HIp}{\ion{H}{i}}
\newcommand{\SI}{[\ion{S}{i}] }
\newcommand{\SiII}{[\ion{Si}{ii}]}
\newcommand{\SIp}{[\ion{S}{i}]}
\newcommand{\SIII}{[\ion{S}{iii}] }
\newcommand{\SIIIp}{[\ion{S}{iii}]}
\newcommand{\ArII}{[\ion{Ar}{ii}] }
\newcommand{\ArIIp}{[\ion{Ar}{ii}]}

\newcommand{\Av}{A$_V$ } 

\usepackage{float}
\usepackage{graphicx}
\usepackage{txfonts}
\usepackage{hyperref}
\usepackage{xcolor}

\begin{document}

    \title{JWST Observations of Young protoStars (JOYS)}
    \titlerunning{The textbook case of the HH\,211 outflow}
    \authorrunning{Caratti o Garatti et al. }

     \subtitle{HH\,211: the textbook case of a protostellar jet and outflow}


   \author{A. Caratti o Garatti \inst{1},
   T.P. Ray \inst{2}, P.J. Kavanagh \inst{3},
   M.J. McCaughrean \inst{4}, C. Gieser \inst{5}, T. Giannini \inst{6}, E.F. van Dishoeck \inst{5,7}, K. Justtanont \inst{8}, 
   M.L. van Gelder \inst{7},
   L. Francis \inst{7}, H. Beuther \inst{4}, Łukasz Tychoniec \inst{7,9}, B. Nisini  \inst{6}, M.G. Navarro \inst{6}, R. Devaraj \inst{2}, S. Reyes \inst{4}, P. Nazari  \inst{7}, P. Klaassen \inst{10}, M. G\"udel \inst{11,12}, Th. Henning \inst{4}, P.O. Lagage \inst{13}, G. \"Ostlin \inst{14}, B. Vandenbussche \inst{15}, C. Waelkens \inst{15}, G. Wright \inst{10}
\\
  }
   \institute{
   INAF-Osservatorio Astronomico di Capodimonte, Salita Moiariello 16, 80131 Napoli, Italy\\
                 \email{alessio.caratti@inaf.it}
                 \and
School of Cosmic Physics, Dublin Institute for Advanced Studies, 31 Fitzwilliam Place, D02 XF86, Dublin, Ireland
  \and
Department of Experimental Physics, Maynooth University, Maynooth, Co. Kildare, Ireland
\and
Max-Planck-Institut f\"{u}r Astronomie, K\"{o}nigstuhl 17, 69117 Heidelberg, Germany
\and
Max-Planck-Institut f{\"u}r Extraterrestrische Physik, Giessenbachstrasse 1, 85748 Garching, Germany
\and
INAF-Osservatorio Astronomico di Roma, Via di Frascati 33, 00078 Monte Porzio Catone, Italy
\and
   Leiden Observatory, Leiden University, PO Box 9513, 2300RA, Leiden, The Netherlands
\and
Department of Space, Earth and Environment, Chalmers University of Technology, Onsala Space Observatory, 439 92 Onsala, Sweden
\and
European Southern Observatory, Karl-Schwarzschild-Strasse 2, 85748 Garching bei M\"unchen, Germany
\and
 UK Astronomy Technology Centre, Royal Observatory Edinburgh, Blackford Hill, Edinburgh EH9 3HJ, UK
 \and
 Dept. of Astrophysics, University of Vienna, Türkenschanzstr. 17, 1180 Vienna, Austria
\and 
ETH Z\"urich, Institute for Particle Physics and Astrophysics, Wolfgang-Pauli-Str. 27, 8093 Z\"urich, Switzerland
\and 
Universit\'e Paris-Saclay, Universit\'e Paris Cit\'e, CEA, CNRS, AIM, 91191 Gif-sur-Yvette, France
\and
Department of Astronomy, Oskar Klein Centre; Stockholm University; SE-106 91 Stockholm, Sweden
\and
Institute of Astronomy, KU Leuven, Celestijnenlaan 200D, 3001 Leuven, Belgium
}

   \date{}


  \abstract
   {Due to the high visual extinction and lack of sensitive mid-IR telescopes, the origin and properties of outflows and jets from embedded Class\,0 protostars are still poorly constrained.}
   {We aim at characterising the physical, kinematic, and dynamical properties of the \object{HH\,211} jet and outflow, one of the youngest protostellar flows.}
   {We use the {\it James Webb Space Telescope} (JWST) and its Mid-Infrared Instrument (MIRI) in the 5--28 $\mu$m range, to study the embedded \object{HH\,211} flow. We map a 0$\farcm$95$\times$0$\farcm$22 region, covering the full extent of the blue-shifted lobe, the central protostellar region, and a small portion of the red-shifted lobe. We extract spectra along the jet and outflow and construct line and excitation maps of both atomic and molecular lines. Additional JWST NIRCam H$_2$ narrow-band images (at 2.122 and 3.235\,$\mu$m) provide a visual-extinction map of the whole flow, and are used to deredden our data.}
   {The jet driving source is not detected even at the longest mid-IR wavelengths. The overall morphology of the flow consists of a highly collimated jet, mostly molecular (H$_2$, HD) with an inner atomic (\FeIp, \FeIIp, \SIp, \NiIIp) structure. The jet shocks the ambient medium, producing several large bow-shocks, rich in forbidden atomic (\FeIIp, \SIp, \NiIIp, \ClIp, \ClIIp, \ArIIp, \CoIIp, \NeIIp, \SIIIp) and molecular lines (H$_2$, HD, CO, OH, H$_2$O, CO$_2$, HCO$^+$), and is driving an H$_2$ molecular outflow, mostly traced by low-$J$, $v=0$ transitions.
   Moreover, H$_2$ 0-0\,S(1) uncollimated emission is also detected down to 2$\arcsec$--3$\arcsec$ ($\sim$650--1000\,au) from the source, tracing a cold ($T$=200--400\,K), less dense and poorly collimated molecular wind. Two H$_2$ components (warm, $T$=300--1000\,K, and hot, $T$=1000--3500\,K) are detected along the jet and outflow.  
   The atomic jet (\FeII at 26\,$\mu$m) is detected down to $\sim$130\,au from source, whereas the lack of H$_2$ emission (at 17\,$\mu$m) close to the source is likely due to the large visual extinction (\Av$>$80\,mag). 
   Dust continuum-emission is detected at the terminal bow-shocks, and in the blue- and red-shifted jet, being likely dust lifted from the disk. }
   { The jet shows an onion-like structure, with layers of different size, velocity, temperature, and chemical composition.
   Moreover, moving from the inner jet to the outer bow-shocks, different physical, kinematic and excitation conditions for both molecular and atomic gas are observed. The jet mass-flux rate, momentum, and momentum flux of the warm H$_2$ component are up to one order of magnitude higher than those inferred from the atomic jet component.
   Our findings indicate that the warm H$_2$ component is the primary mover of the outflow, namely it is the most significant dynamical component of the jet, in contrast to jets from more evolved YSOs, where the atomic component is dominant.}


   \maketitle
%
\nolinenumbers

\section{Introduction}

Accretion and ejection are common and related processes in the formation of stars, from low- to high-mass young stellar objects (YSOs). As a matter of fact, they represent two sides of the same coin. To accrete matter onto the forming star, disks have to remove angular momentum through magneto-hydrodynamic (MHD) winds (X-winds or disk-winds). MHD winds (partly) focus into collimated jets~\citep[see, e.\,g.,][and references therein]{Ray.ea.2007,Frank.ea.2014,Bally2016}, that move at supersonic speed (100--500\,km\,s$^{-1}$).
These fast-moving protostellar jets shock the circumstellar and interstellar medium (ISM), opening large cavities in the infalling envelopes, the natural reservoir of accretion disks. As jets drive through the surrounding medium, parsec-scale, less-collimated, low-velocity outflows ($\sim$10\,\,km\,s$^{-1}$), made of swept-up gas, are formed~\citep[see, e.\,g.,][]{Reipurth.Bally2001}.

On the other hand, the slow-wind component  (1--10\,km\,s$^{-1}$), launched at large disk radii (from a few au to several tens of au), is poorly or not collimated and interacts with the protostellar envelope and outflow cavities.
MHD winds and jets are a fundamental feature in YSOs, not just because of angular momentum removal, but also because of their major role in disk dispersal, both in terms of gas and dust~\citep[see, e.\,g.,][and references therein]{Pascucci.ea.2023}.

Accretion and ejection are observed throughout all stages of low-mass star formation, namely from the protostellar phase (Class\,0 and I; 10$^4$ and 10$^5$\,yr, respectively), when protostars are heavily embedded by and highly accreting from their surrounding envelopes, to the pre-main sequence phase (Class\,II and III; 10$^6$ and 10$^7$\,yr, respectively), when envelopes disappear, accretion and ejection considerably decrease and come to an end (Class\,III), with young stars becoming optically visible. This continuous process indicates that the main physics at work is just the same at different stages of low-mass star formation and, very likely, also for high-mass young stellar objects~\citep[see, e.\,g.,][and references therein]{Caratti.ea.2015}. 

At small distances from the source (tens of au), jets are already well collimated (from a few to several au in width) and have opening angles of a few degrees~\citep[see, e.\,g.,][and references therein]{Frank.ea.2014, Pascucci.ea.2023}. 
The jet width then slowly increases with increasing distance from the driving source, reaching up to several tens of au at distances of hundreds of au~\citep[see,][]{Dougados.ea.2004,Ray.ea.2007,Podio.Tabone.ea2021} and remains collimated at parsec scales.
Measurements of the specific angular momentum in a couple of Class\,0 jets (namely B\,355 and HH\,212) suggest that the launching region is located within 0.1\,au in the inner gaseous disk~\citep[see Fig.\,7 in][]{Lee2020}, therefore within the dust sublimation radius. MHD disk-wind models and additional observations~\citep[see, e.\,g.,][and discussion therein]{deValon.ea.2022} point to a more extended launching region (up to a few au) stretching into the dusty disk.

Despite YSO jets likely originating from the same physical mechanism, namely magneto-hydrodynamic (MHD) winds, jets at protostellar and pre-main sequence stages show relevant differences. 
As jets evolve from Class\,0 to II~\citep[see, e.\,g.][]{Ray.ea.2007}, observed jet velocities increase from one hundred to several hundreds of km\,s$^{-1}$. This is possibly due to the increment in mass of the central source and the larger potential well.
Protostellar jets from Class\,0 YSOs are much brighter in molecules than those from Class\,I or Class\,II YSOs, that are mostly or fully atomic~\citep[see, e.\,g.,][and references therein]{Ray.ea.2007,Frank.ea.2014,Bally2016}.  
Indeed, several Class\,0 protostellar jets, along with strong H$_2$ emission at near-IR and mid-IR wavelengths, also show bright molecular emission at FIR wavelengths~\citep[e.\,g., H$_2$O, CO, detected with Herschel; see][]{kristensen.ea.2012,Mottram.ea.2017},
as well as in the sub-mm and mm regimes, where SiO and CO~\citep[the so-called extremely high-velocity - EHV - gas; see, e.\,g.,][]{Bachiller.ea.1990,Tafalla.ea.2010,Lee2020,Yoshida.ea.2021,Podio.Tabone.ea2021} are typically observed.

Although \FeII and \OI atomic emission is often observed from near- to far-IR wavelengths~\citep[see, e.\,g.,][]{Caratti.ea.2006,Dionatos.ea.2009,Dionatos.ea.2010,vankempen.ea.2010,Nisini.ea.2015}, it is still unclear whether or not the atomic component is a major feature in protostellar jets, since, with a couple of exceptions, the atomic mass-ejection rate is up to one order of magnitude lower than the molecular one in Class\,0 jets~\citep[see, e.\,g.,][]{Nisini.ea.2015}. This would indicate that most of the mass-flux derives from the molecular jet and the atomic jet contribution in generating the outflow is not very significant. This is at variance with Class\,II jets, where the atomic jet drives the outflow~\citep[see, e.\,g.,][]{Ray.ea.2007}.
Such findings point to an evolution in the jet properties during the different stages of star formation.

While the overall picture is now well known and accepted, several points remain unclear, especially during the first stages of star formation (i.\,e. Class\,0 YSOs).
Dust has been observed along the jets and outflows with ALMA, especially in Class\,0 YSOs~\citep[see, e.\,g.,][]{Cacciapuoti.ea.2024}. However, the origin of both dust and molecular gas along the jets remains unclear~\citep[see, e.\,g.,][]{Tabone.ea.2018,Pascucci.ea.2023}. If jets are launched within the dust sublimation radius~\citep[as in the X-wind model or in dust-free MHD disk winds; see][ respectively]{Shu.Najita.ea2000,Tabone.Cabrit.ea2020}, it is unlikely that both molecules and dust are lifted from the disk, and they must therefore form along the flow. On the other hand, their presence along the flow would be easier to explain, 
if jets originated from a disk-wind radially extending beyond the dust sublimation radius, as some observations seem to indicate~\citep[e.\,g.,][]{deValon.ea.2022}. This seems in contrast with the very narrow jet launching region within the gaseous disk seen in other Class\,0 YSOs~\citep[see][and references therein]{Lee2020}. However, \citet{Tabone.Cabrit.ea2020} show that the launching radius often inferred with ALMA is largely underestimated, and the jet outer radius in disk-wind models can extend up to $\sim$40\,au in the HH\,212 protostellar jet. These authors also stress that constraining the values for the inner and outer launching radii ($r_{in}$ and $r_{out}$, respectively) is still an open issue~\citep{Tabone.Cabrit.ea2020}.

As the picture is not completely clear, the origin of outflows and jets from embedded Class\,0 protostars needs to be investigated at IR and sub-mm wavelengths, where most of the emission from the outflowing gas arises. In particular, JWST in the mid-IR is required to peer into the flow, detect and study H$_2$ and atomic jet components, and understand whether the early stage of protostellar jets is fully molecular or, if atomic emission is detected, whether this represents the spine of the jet, namely the feature that is driving the whole jet, as in Class\,II YSOs, or it is just a minor feature of a mostly molecular flow.

Here, we investigate the Herbig-Haro 211 (hereafter \object{HH 211}) flow, driven by a Class\,0 protostar (HH\,211\, mm, AKA \object{[EES2009] Per-emb 1}), located at a distance of 321$\pm$10\,pc in the Perseus Molecular Cloud~\citep{Ortiz.ea.2018}. HH\,211\,mm is possibly a close binary~\citep[$\sim$5\,au separation, as hinted at by the jet wiggling;][]{Lee.ea.2010}, with a central mass of $\sim$0.08 M$_\sun$ and a surrounding torus of gas and dust of $\sim$0.2\,M$_\sun$~\citep[see][and references therein]{Lee.ea.2018,Lee2020}, and $L_{\rm bol}$$\sim$4.1\,L$_\sun$~\citep[rescaled to $d$=321\,pc; see][]{Froebrich2005}. The protostar drives one of the most embedded, youngest~\citep[$\sim$10$^3$\,yr in dynamical age;][]{Ray.ea.2023} and best studied protostellar outflow.

In the near-IR, the HH\,211 discovery paper~\citep{McCaughrean.ea.1994} showed large red- (tens of arcseconds NW of the source) and blue-shifted (tens of arcseconds SE of the source) bow-shocks driven by a knotty jet, mostly emitting in H$_2$ and \FeII lines~\citep[see,][]{McCaughrean.ea.1994,OConnell.ea.2005,Caratti.ea.2006}. Our latest NIRCam/JWST images have revealed a more structured H$_2$ precessing jet~\citep[already discovered and discussed in][]{Lee.ea.2010,Moraghan.ea.2016} consisting of both knots and small bow-shocks~\citep[see Figure\,1 of][]{Ray.ea.2023} with H$_2$ knots moving at high-speed ($\sim$100\,km\,s$^{-1}$) along the jet. Such velocities are similar to those measured from SiO ($J=8-7$) proper motion observations with the SMA and ALMA~\citep[100--115\,km\,s$^{-1}$; see][]{Jhan.ea.2016,Jhan&Lee2021}. These studies reveal a jet inclination angle of $\sim$11$\degr$~\citep{Jhan&Lee2021} with respect to the plane of the sky. Indeed, the flow has been detected at sub-mm and mm wavelengths in SiO, SO, and CO~\citep[e.\,g.,][]{Gueth1999,Lee.ea.2007}, as well as in H$_2$O and CO at FIR wavelengths with {\it Herschel}~\citep{Tafalla.ea.2013}. 
Mid-IR maps with {\it Spitzer}/IRS revealed cooler H$_2$ emission along the embedded jet, as well as a series of atomic lines~\citep[\FeIIp, \SIp, \SiII;][]{Dionatos.ea.2010}. Further atomic emission from [\ion{O}{I}] (at 63 and 145\,$\mu$m) was detected and studied with {\it Herschel}~\citep{Dionatos.ea.2018}. 
Finally, H$\alpha$ and \SIIp emission at optical wavelengths~\citep[see][]{Walawender.ea.2005,Walawender.ea.2006} was identified at the SE terminal bow-shocks. 

As part of our JWST Guaranteed Time Observations (GTO) programme dedicated to the study of HH\,211 (PID 1257, PI: T.P. Ray), we present new JWST MIRI-MRS spectral maps of a large portion of HH\,211. The HH\,211 programme is part of the GTO survey "JWST Observations of Young protoStars" (JOYS)~\citep[ PID 1290, PI: E. van Dishoeck; see][]{VanDishoeck.Beuther.eaGTO1290,Beuther.vanDishoeck.ea2023}, aimed at studying physical and chemical properties of a large sample of protostars and their outflows.
Thanks to JWST's sensitivity and high-spatial resolution, we are able to present an unprecedented view of the Class\,0 HH\,211 flow. 
This work is organized as follows: in Section\,\ref{sec:observations}, we introduce JWST/MIRI-MRS observations and ancillary data; Section\,\ref{sec:results} describes the results, including spectral and spatial analysis of the flow, as well as its kinematics and dynamics; Section\,\ref{sec:discussion} provides a discussion on the flow structure, origin of its different components, flow excitation conditions, and comparison with other jets from Class\,0 and more evolved YSOs; our conclusions are reported in Section\,\ref{sec:conclusions}.

\begin{figure*}[hbt!]
\begin{center}
        \includegraphics[width=0.98
        \textwidth]{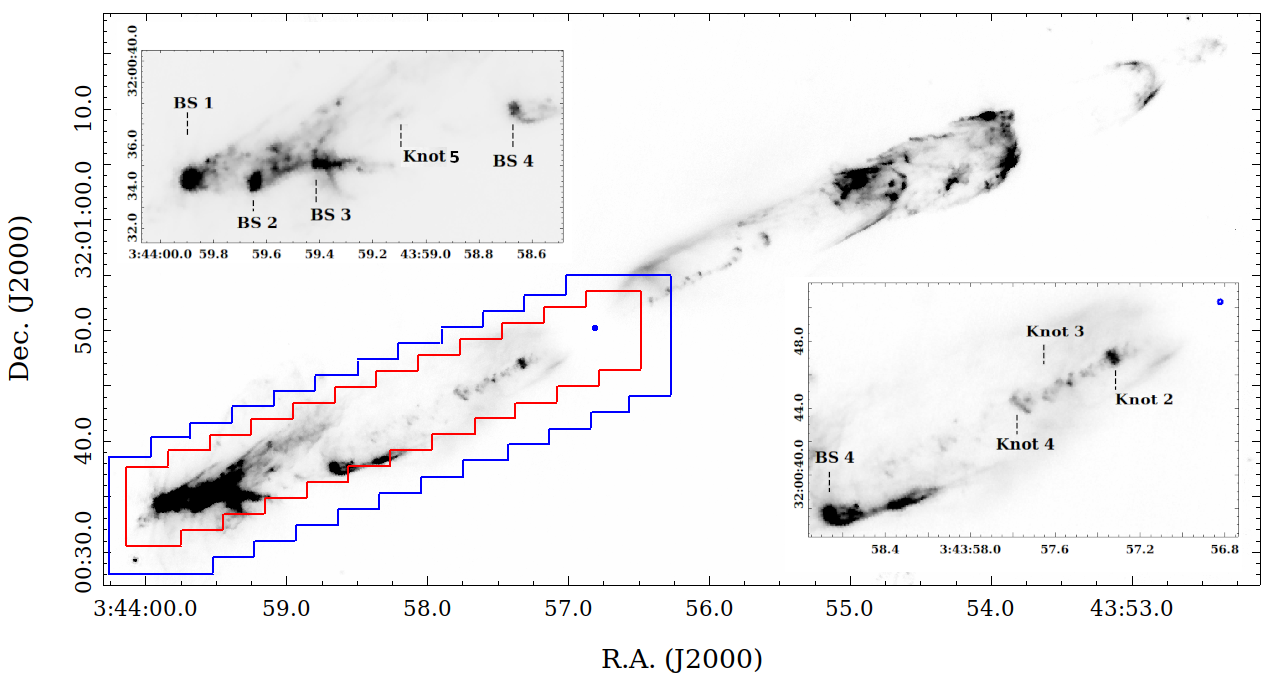}

    \caption{MIRI-MRS map coverage of the HH\,211 outflow (Channel\,1 in red and Channel 4 in blue). The F460M NIRCam image at 4.6\,$\mu$m in grey scale is from \citet{Ray.ea.2023}. The blue circle marks the HH\,211\,mm position as observed with ALMA by \citet{Lee.ea.2019}. The zoomed-in inset on the top left shows the blue-shifted terminal bow-shocks and the inset of the bottom right the main knots along the blue-shifted jet.}
    \label{fig:MIRI-footprint}
\end{center}
\end{figure*}

\section{Observations and data analysis}
\label{sec:observations}

\subsection{JWST-MIRI-MRS}
\label{sec:MIRIobs}

JWST MIRI-MRS~\citep{Wright.Rieke.ea2023} observations were obtained between the 25th and 26th of January 2023, as part of the HH\,211 Guaranteed Time Observations (GTO) PID 1257 (PI: T.P. Ray).
Our exposures were taken with 18 groups in a single integration using the FASTR1 readout pattern for all three MRS bands (SHORT, MEDIUM, LONG), which provided spectral coverage from 4.9 to 27.9 $\mu$m at a spectral resolution of $R\sim$\,4000--1500~\citep[][]{Jones.ea.2023}. The 4-point `EXTENDED SOURCE' dither pattern in the `negative' direction was used for a total exposure time of $\approx200$\,s per channel-band combination per mosaic tile. The MRS fields of view range from 3.2$\arcsec$ $\times$ 3.7$\arcsec$ in Channel 1 to 6.6$\arcsec$ $\times$ 7.7$\arcsec$ in Channel 4 \citep{Wells.ea.2015, Argyriou.ea.2023}. 

The MIRI-MRS mosaic is made of 12$\times$2 tiles with a 5\% overlap, covering an area of about 0.$\arcmin$86$\times$0.$\arcmin$15 and 0.$\arcmin$95$\times$0.$\arcmin$22 ($\sim$ 18\,300\,au$\times$4200\,au for a distance of 321\,pc) at the shortest and longest wavelengths, respectively. 
Figure\,\ref{fig:MIRI-footprint} provides the MIRI-MRS map coverage in Channel 1 (red) and 4 (blue). The underlying image shows the HH\,211 flow observed at 4.6\,$\mu$m (F460M NIRCam filter) as presented in \citet{Ray.ea.2023}. The F460M image is dominated by H$_2$ and CO emission. The blue dot shows the protostellar sub-millimetric position reported by \citet{Lee.ea.2019}. Our map covers the full extent of the blue-shifted lobe, the central source position and a small portion of the red-shifted flow. 

Dedicated background observations were taken from a field to the south of the protostar (R.A.(J2000): $03^h43^m55.^s28$; Dec.(J2000): $+32\degr00\arcmin41\farcs30$) with the same groups per integration and integrations per exposure as the science exposures, using two dithers instead of four and the `POINT SOURCE' dither pattern.


The data were reduced with the JWST Calibration Pipeline v.1.13.4 \citep{Bushouse.Eisenhamer.ea2023} using Calibration Reference Data System (CRDS) version v11.17.6 and context file \texttt{jwst\_1210.pmap}. We note that our data have been calibrated using the updated MIRI-MRS wavelength calibration reference files for channels 3C, 4A, 4B, and 4C, based on the cross-correlation analysis of observations of water in the protoplanetary disk FZ~Tau \citep{Pontoppidan.ea.2024}. The level 1b ramp files were processed through \texttt{Detector1Pipeline} with default settings. We used the dedicated background observations to build `master' detector background images for each channel/band combination and subtracted these from the science exposures. The resulting background subtracted level 2A rate files were calibrated using the \texttt{Spec2Pipeline}, with the optional detector level residual fringe correction switched on. Individual channel/band mosaics were constructed using \texttt{Spec3Pipeline} with the \texttt{mrs\_imatch} step disabled. 

To convert the observed wavelengths into radial velocities in HH\,211, we used a local standard of rest (LSR) systemic velocity of 9.2\,km\, s$^{-1}$  ~\citep{Gueth.Guilloteau1999}. As the JWST wavelength calibration is given in the barycentric reference frame, we further correct the velocity from heliocentric to v$_{LSR}$ by adding 6.7\,km\,s$^{-1}$.



\subsection{Ancillary data: JWST-NIRCam imaging}

To infer the visual extinction (A$_{\rm V}$) towards HH\,211 (both interstellar and circumstellar), we employ two NIRCam narrow-band images (using the F212N and F323N filters). 

The NIRCam images were already presented in \citet{Ray.ea.2023}. Both images were taken using the two NIRCam modules (A and B, each module has a FoV of 2.$\arcmin$2$\times$2.$\arcmin$2), with HH\,211 centred on module B. Data were taken using the BRIGHT1 readout pattern with one integration, four dithered exposures for each filter and the INTRAMODULEX pattern was used with STANDARD sub-pixel dither type for a total integration time of 664\,s per filter. A detailed description of the 1$/ \it f$ noise removal and astrometric calibration are reported in \citet{Ray.ea.2023}.

\section{Results}
\label{sec:results}

\subsection{HH\,211 visual extinction map}
\label{sec:Av}

F212N and F323N NIRCam filters cover the H$_2$ 1-0\,S(1) and  1-0\,O(5) lines, respectively. As both H$_2$ lines come from the same upper level (with $E_{\rm up}$= 6956\,K), their theoretical ratio only depends on their transition frequencies and Einstein coefficients. Therefore the observed line ratio provides us with the visual extinction, once a reddening law is applied. 
In this paper we adopt \citet{McClure2009}'s law to correct our MIRI-MRS data for extinction. For the NIRCam images, we note that \citet{McClure2009}'s law does not take into account the strong H$_2$O ice feature around 3\,$\mu$m, which affects the 3.23\,$\mu$m image. Therefore, to deredden the NIRCam images, we use the extinction curve presented in \citet{Decleir.ea.2022}, that fits the ice shape using a modified Drude profile.

\begin{figure*}
\begin{center}
        \includegraphics[width=0.95\textwidth]{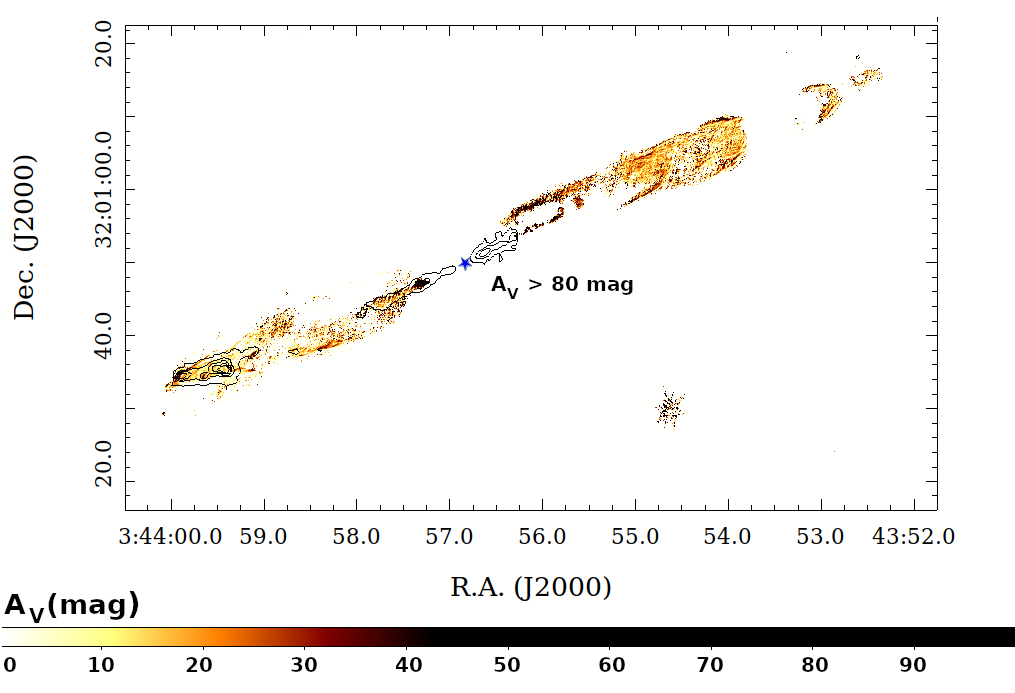}

    \caption{Visual extinction map of the HH\,211 outflow derived from the H$_2$ 1-0\,S(1) and  1-0\,O(5) lines (F212N and F323N NIRCam images). The colour bar represents the different values in mag. A value of \Av $\geq$\,80\, mag has been estimated using ro-vibrational diagrams for the source and jet inner regions where the H$_2$ 1-0\,S(1) emission is not detected. \FeII\,(26\,$\mu$m) jet contours detected in the MIRI-MRS map are shown in black (see Sect.\,\ref{sec:maps}). The position of the protostar from ALMA continuum data~\citep[see][]{Lee.ea.2019} is also marked.}
    \label{fig:Avmap}
\end{center}
\end{figure*}

To obtain the \Av\,map, the following steps were adopted.
The F212N image was resampled to the F323N one, given their different pixel-scale (31 vs 63 milliarcseconds/pixel). 
To avoid dividing by noise, pixels with a density flux below 3$\sigma$ (1.5 and 0.5 MJy\,sr$^{-1}$ for F212N and F323N, respectively) were not taken into account in the maps.
Astrometric images were then matched and divided by each other and divided by their theoretical line ratio ($\sim$0.394). The logarithm of the resulting image, rescaled for the reddening law, provides the final \Av\,map.
The resulting pixel-by-pixel \Av\,map is shown in Figure\,\ref{fig:Avmap}.

\Av increases from 5--15\,mag at the terminal bow-shocks to 20--50\,mag moving along the jet towards the central source.
As the innermost jet regions (within a $\sim$6$\arcsec$ radius from the source) are not measured in the extinction map (see Fig.\,\ref{fig:Avmap}), we use an average value of 80\,mag through the paper to correct for visual extinction in those regions. Such a value is inferred from the H$_2$ ro-vibrational diagrams (see Sect.\,\ref{sec:rovibrational}).

\subsection{MIRI-MRS maps: flow morphology}
\label{sec:maps}

As mentioned in Sect.\,\ref{sec:MIRIobs}, the MIRI-MRS mosaic covers the whole blue-shifted lobe, the central region around the protostar and a small portion of the red-shifted lobe (see Fig.\,1). NIRCam images show that the collimated blue-shifted jet (precessing with a 3.5$\degr$ opening angle, measured from the NIRCam images) is made of several knots and small bow-shocks~\citep[see Fig.\,\ref{fig:MIRI-footprint} and NIRCam image in Fig.\,1 of][]{Ray.ea.2023}. The blue-shifted jet first drives an extended bow-shock (labelled BS\,4 in the insets of Fig.\,1) and subsequently produces a large terminal bow-shock, which is actually made of three distinct bow-shocks~\citep[labelled BS\,1, 2 and 3 in the upper-left inset of Fig.\,1; BS\,1 and BS\,3 are also known as knot j - \object{[MRZ94] j} - and knot i - \object{[MRZ94] i}, respectively, in][]{McCaughrean.ea.1994} with larger precession opening angles ($\sim$6$\degr$, measured from the NIRCam images).

\begin{figure*}
\begin{center}
        \includegraphics[width=0.95\textwidth]{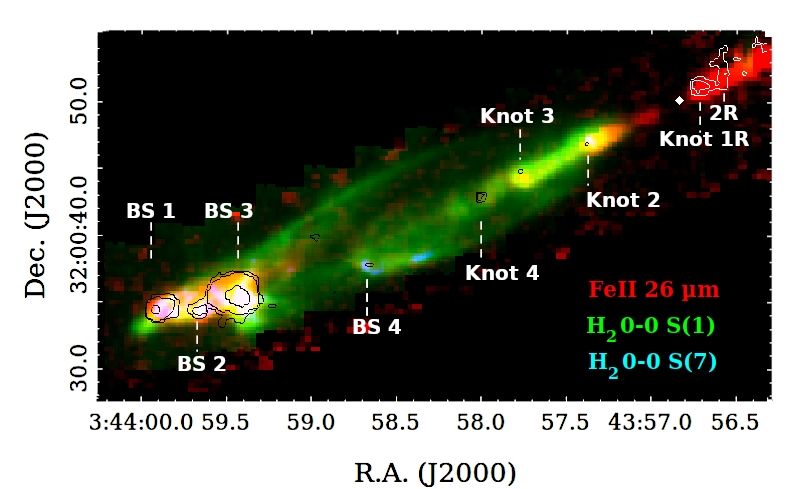}

    \caption{Tricolour MIRI-MRS map of H$_2$\,0-0\,S(7) (at 5.5\,$\mu$m, in blue), H$_2$\,0-0\,S(1) (at 17\,$\mu$m, in green), and \FeII (at 26\,$\mu$m, in red) emission lines. The white circle marks the position of the ALMA mm continuum source. Black and white contours indicate the position (on the blue- and red-shifted lobe side, respectively) of continuum emission integrated between 25.3 and 25.9\,$\mu$m (displayed contours are at 3, 5, and 50\,$\sigma$; 1\,$\sigma$=4\,MJy\,sr$^{-1}$). Knots and bow-shocks (BS) showing continuum emission are indicated.}
    \label{fig:MIRI-tricolor}
\end{center}
\end{figure*}

Figure\,\ref{fig:MIRI-tricolor} displays a tricolour image of the H$_2$\,0-0\,S(7) (5.5\,$\mu$m), H$_2$\,0-0\,S(1) (17\,$\mu$m), and \FeII (26\,$\mu$m) emission lines in blue, green, and red, respectively.
Our MIRI maps show that the jet is both atomic and molecular and it is driving a large molecular outflow (see Figure\,\ref{fig:MIRI-tricolor}). 

The inner jet, within $\sim$2.5$\arcsec$ from the source position (marked by the white circle in Fig.\,\ref{fig:MIRI-tricolor}), is mostly traced by atomic emission (in red), and the lack of H$_2$ emission is likely due to the large visual extinction (\Av$>$80\,mag) close to source. The \FeII emission at 26\,$\mu$m is detected down to $\sim$130\,au and 300\,au from the source on the jet red- and blue-shifted sides, respectively. Such a difference might be due to different visual extinction or excitation conditions in the two lobes.

The outer jet and bow-shocks show both atomic and molecular emission, whereas the outflow, located at rear and wings of the bow-shocks and likely made of entrained ambient gas, is fully molecular (H$_2$ emission only) and well traced by H$_2$ pure-rotational transitions (in green) at low-energy excitation. Unfortunately, due to the poor MIRI-MRS sensitivity beyond 27\,$\mu$m, the 0-0\,S(0) line is not detected in our maps.
The atomic jet is more compact in diameter, whereas the cold H$_2$ molecular component is more extended (see Fig.\,\ref{fig:MIRI-tricolor} and Sect.\,\ref{sec:jet_bs_outflow}), indicating a jet onion-like structure with different layers, where the atomic component is at the jet core, nested in a more extended molecular jet~\citep[see e.\,g.][and references therein]{Shang.ea.2006,Machida2014,Shang.ea.2023}. This structure is readily visible in Fig.\,\ref{fig:Tri-jet} (see Appendix\,\ref{sec:add_maps}), which shows a tricolour map of the H$_2$\,0-0\,S(7) (at 5.5\,$\mu$m, in blue), \SI (at 25.2\,$\mu$m, in green), and \FeII (at 26\,$\mu$m, in red) emission lines. The combination of atomic and hot molecular emission provides a better view of the jet emission. In contrast, the
H$_2$ emission (0-0\,S(1) line at 17.0\,$\mu$m) is overplotted with magenta contours. At high $\mbox{SNR}$ ($>$20), the H$_2$ emission overlaps and cocoons the atomic jet, whereas at lower $\mbox{SNR}$, it traces a less-collimated wind, as well as bow-shock wings and outflow.

\begin{figure}

        \includegraphics[width=0.5\textwidth]{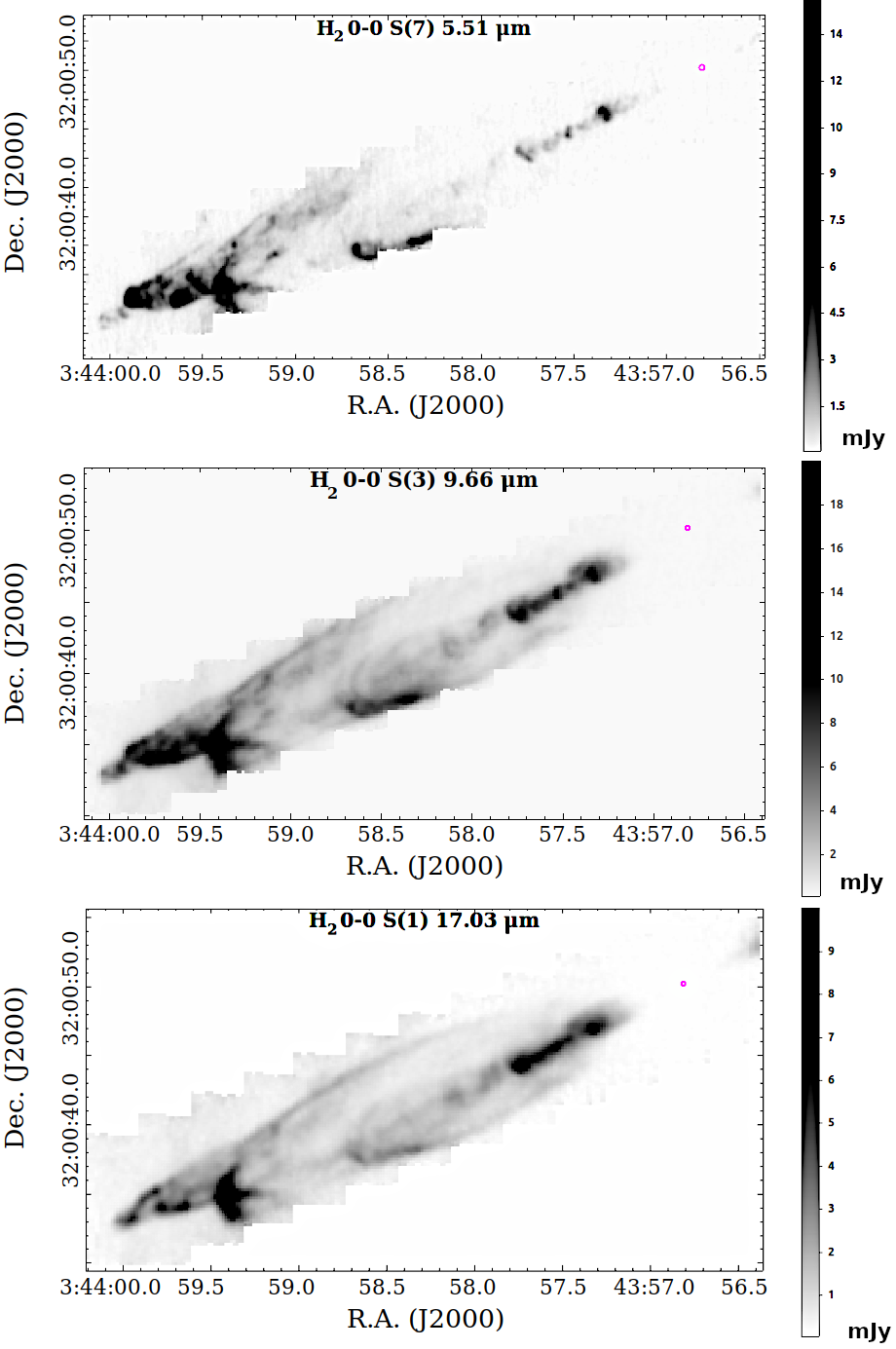}

    \caption{H$_2$ line maps of the brightest transitions detected along the flow.
    From top to bottom: 0-0\,S(7) (5.5\,$\mu$m), 0-0\,S(3) (9.7\,$\mu$m), and 0-0S(1) (17.0\,$\mu$m) lines. The magenta circle shows the position of the ALMA mm continuum source. Integrated flux is in mJy\,pixel$^{-1}$.}
    \label{fig:H2-maps}
\end{figure}

Indeed, the cold H$_2$ component at 17\,$\mu$m (i.\,e. the 0-0\,S(1) transition - $E_{\rm up}$=1015\,K, Figure\,\ref{fig:MIRI-tricolor} in green and Fig.\,\ref{fig:H2-maps}, bottom panel, and magenta contours of Fig.\,\ref{fig:Tri-jet}) is also well detected in the outflow cavity (down to 2$\arcsec$--3$\arcsec$ from the source), likely originating from a less-collimated wind. From the extent of the H$_2$ emission, the half-opening angle of the wind is $\leq$20$\degr$. 
Curved narrow emission delineates the boundaries of the outflow cavity and the interaction between outflow and ISM. These features become less visible in the H$_2$ transitions at higher excitation energy (e.\,g., 0-0\,S(3) - $E_{\rm up}$=2504\,K; see middle panel of Fig.\,\ref{fig:H2-maps}) and disappear in those at the highest energy (e.\,g., 0-0\,S(7) - $E_{\rm up}$=7197\,K; see upper panel of Fig.\,\ref{fig:H2-maps}).

Notably, no continuum emission is detected on source at the longest JWST wavelengths (26\,$\mu$m; below 36 MJy\,sr$^{-1}$; i.\,e. $F_{protostar}\,(26\,\mu m)\leq$0.85\,mJy), nor towards the outflow cavities at 5\,$\mu$m (below 20 MJy\,sr$^{-1}$). The latter value is about one order of magnitude larger than that of scattered emission ($\sim$2.1\,MJy\,sr$^{-1}$) detected towards the outflow cavities with the NIRCam F460M filter (see Fig.\,\ref{fig:MIRI-footprint}). This explains the non-detection in our data. 
Nevertheless, strong continuum emission longward of 10\,$\mu$m is observed at BS\,1, BS\,2, and BS\,3, and much fainter emission (\mbox{SNR}$\sim$5\,$\sigma$, longward of 25\,$\mu$m) along the red-shifted jet (knots 1 and 2 red) and at Knot\,4 and, marginally (\mbox{SNR}$\sim$3\,$\sigma$), at Knot\,3, Knot\,2 and BS\,4 in the blue-shifted jet (see black and white contours in Fig.\,\ref{fig:MIRI-tricolor}). Towards BS\,3, the continuum emission is extended, elongated towards the west (see black contours in Fig.\,\ref{fig:MIRI-tricolor}), and has a full-width-half-maximum (\mbox{FWHM}) of $\sim$1$\farcs$6, which is much larger than that of the nominal point spread function (${\rm PSF}\sim$1$\arcsec$) at 25\,$\mu$m.

\subsubsection{The jet, the bow-shocks and the outflow}
\label{sec:jet_bs_outflow}

\begin{figure*}
\centering
        \includegraphics[width=0.96\textwidth]{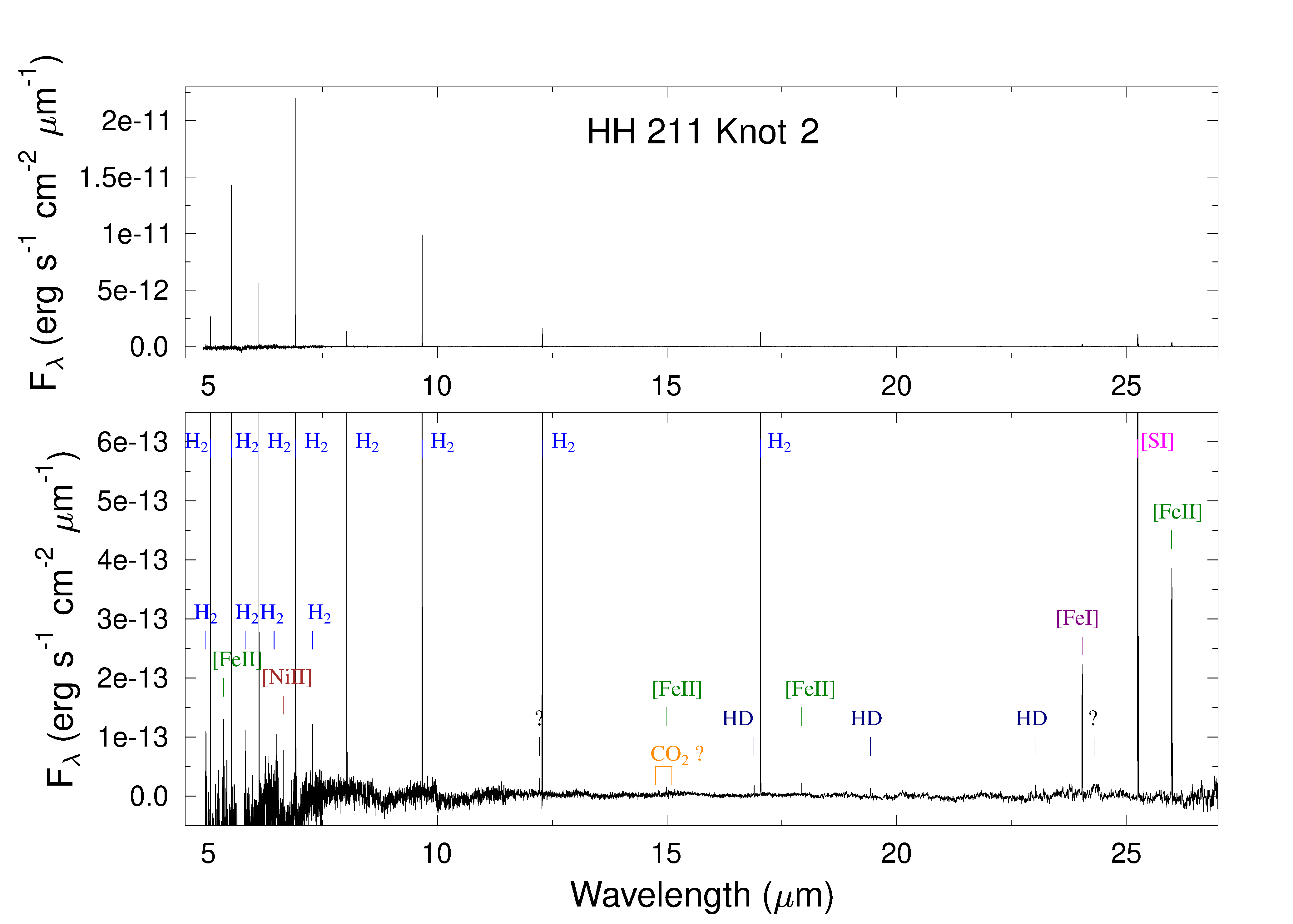}

    \caption{Spectrum of the HH\,211 blue-shifted jet extracted at Knot\,2 (R.A.(J2000): $03^h43^m57.^s331$, Dec.(J2000): $+32\degr00\arcmin47.\arcsec04$). The top panel shows the full flux-density range of the spectrum (up to 2.5$\times$10$^{-11}$\,erg\,s$^{-1}$\,cm$^{-2}$\,$\mu$m$^{-1}$), and the bottom panel shows a close-up (up to 6.5$\times$10$^{-13}$\,erg\,s$^{-1}$\,cm$^{-2}$\,$\mu$m$^{-1}$). Detected lines are labelled. Different colours indicate different species.}
    \label{fig:knot2_spec}
\end{figure*}

Line maps and spectra indicate that H$_2$ is the brightest and most abundant species along the flow, even along the jet (see top panel of Fig.\,\ref{fig:knot2_spec}). A large number of H$_2$ transitions ($v=0-0$ and $v=1-1$, $J$=1--9, with upper energy levels from $\sim$1000 to 16\,000\,K) are observed and listed in Table\,\ref{tab:h2_lines}, along with their theoretical wavelength (in $\mu$m), energy of the upper level (in K), and corresponding MIRI Channel/Grating.

Another molecule detected along the jet in the MIRI-MRS data is HD (see spectrum in Fig.\,\ref{fig:knot2_spec} and top panel of Fig.\,\ref{fig:all_maps}). Several faint 0-0\,R  transitions (see Table\,\ref{tab:molecule_lines}) are detected both along the blue-shifted jet and in the outer bow-shocks. HD emission is analysed in detail in a forthcoming paper (Francis et al. in prep.). CO$_2$ at 15\,$\mu$m is also faintly detected (\mbox{SNR}$\sim$3$\sigma$) in Knot\,2~(see Fig.\,\ref{fig:knot2_spec} and Tab.\,\ref{tab:line_fluxes_Knot2}).

\begin{figure}
        \includegraphics[width=0.47\textwidth]{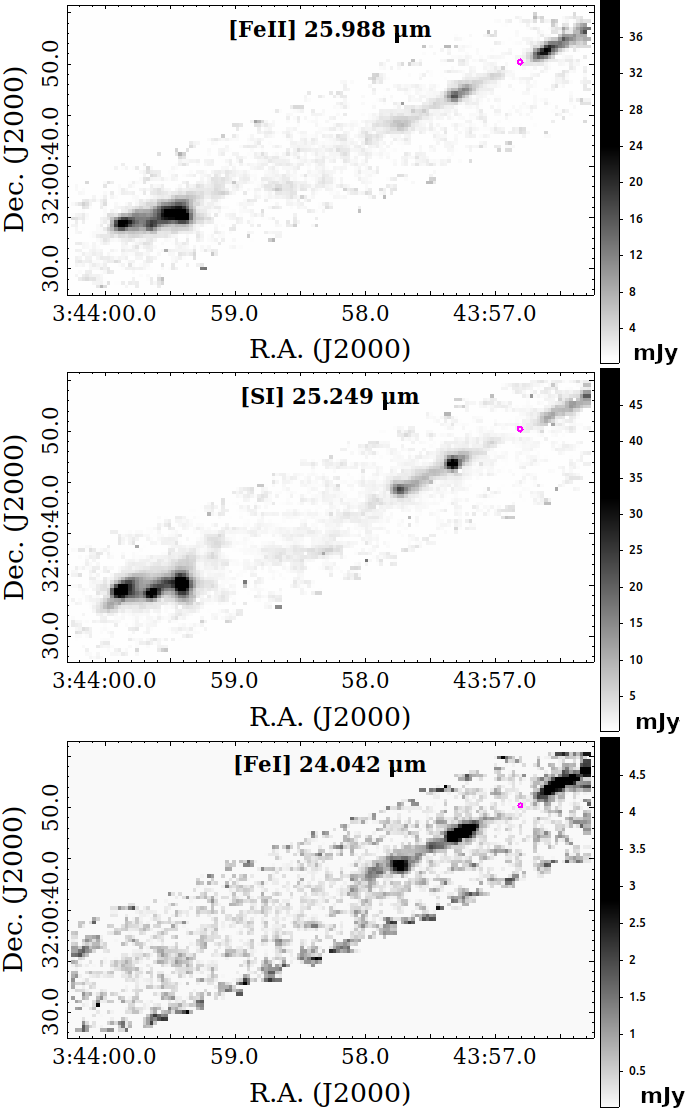}

    \caption{From top to bottom: \FeII (26\,$\mu$m), \SI (25\,$\mu$m), and \FeI (24\,$\mu$m) continuum-subtracted emission lines along the jet. The magenta circle shows the position of the ALMA mm continuum source. Integrated flux is in mJy\,pixel$^{-1}$.}
    \label{fig:atomic-maps}
\end{figure}

In addition to the molecular emission, several atomic species are detected along the jet. Besides the many transitions of \FeII with the excitation energy of the upper level ($E_{\rm up}$) ranging from $\sim$500 to $\sim$4000\,K (see Fig.\,\ref{fig:knot2_spec} and Tab.\,\ref{table:atomic_lines}), strong \SI and \FeI emission (see middle and bottom panels of Figure\,\ref{fig:atomic-maps}) at 25.25 and 24.04\,$\mu$m, respectively, are detected. 
Note that these transitions have $E_{\rm up}$ similar to the \FeII line at 26\,$\mu$m ($\sim$550--600\,K).
These features match the jet very well, delineated by the \FeII emission (see top panel of Figure\,\ref{fig:atomic-maps}), although the intensity of such lines largely vary along the flow, and likely follow the different excitation conditions along the jet. Indeed, the continuum-subtracted map in Fig.\,\ref{fig:atomic-maps} show that \FeI emission, seen for the first time in a protostellar jet, is mostly detected along the jet, whereas faint (signal-to-noise ratio - \mbox{SNR}$\leq$5) or no emission is seen at the outer bow-shocks, where \FeII and \SI emission lines are strongest. This almost certainly reflects an increase in the ionisation fraction along the flow.
The other atomic species detected along the jet is \NiII at 6.6\,$\mu$m (see Fig.\,\ref{fig:all_maps} in Appendix\,\ref{sec:add_maps}). Two more transitions from \NiII are also observed at BS\,3 (see Tab.\,\ref{table:atomic_lines}).

\begin{figure*}
\begin{center}
        \includegraphics[width=0.96\textwidth]{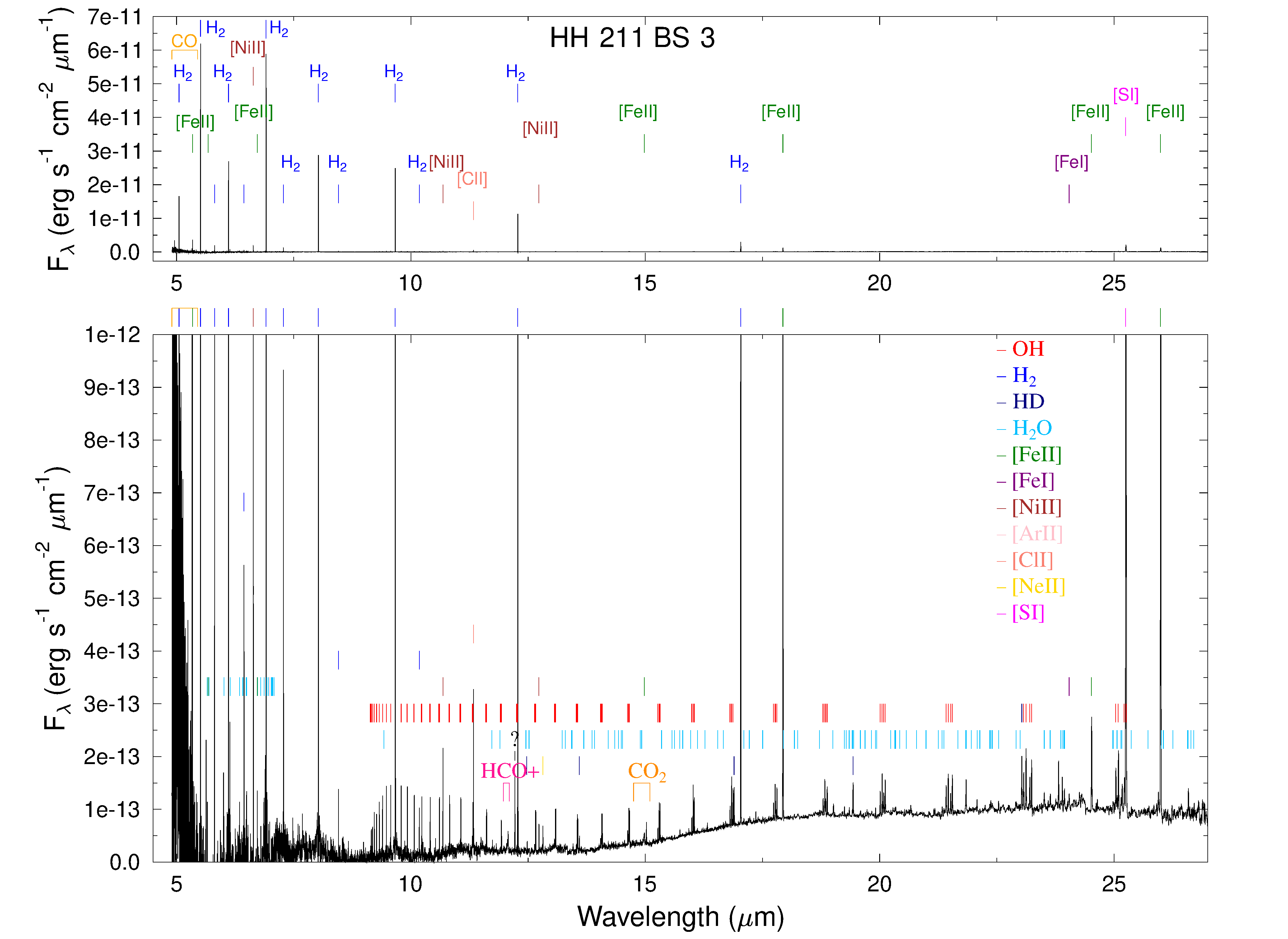}

    \caption{Spectrum of HH\,211 BS\,3 (see Fig.\,\ref{fig:MIRI-footprint}) extracted at R.A.(J2000): $03^h43^m59.^s413$, Dec.(J2000): $+32\degr00\arcmin35.\arcsec27$. The top panel shows the full flux-density range of the spectrum (up to 7$\times$10$^{-11}$\,erg\,s$^{-1}$\,cm$^{-2}$\,$\mu$m$^{-1}$), and the bottom panel shows a close-up (up to 10$^{-12}$\,erg\,s$^{-1}$\,cm$^{-2}$\,$\mu$m$^{-1}$). Detected lines are labelled. Different colours indicate different species.}
    \label{fig:BS3_spec}
\end{center}
\end{figure*}

The terminal bow-shocks (BS\,1--BS\,3) are clearly richer in terms of chemistry, especially BS\,3, which is the brightest (see Fig.\,\ref{fig:BS3_spec}). In addition to the species visible along the jet, many other molecular and atomic forbidden lines are detected (see Tab.\,\ref{table:atomic_lines} and Tab.\,\ref{tab:molecule_lines}). In particular, the tail (i.\,e. $J$\,$\geq$25 up to $J$=59) of the P-branch CO fundamental (i.\,e. $v=1-0$, up to $\sim$5.4\,$\mu$m) is the brightest molecular emission after H$_2$ (see Fig.\,\ref{fig:BS3_spec}, and middle upper panel of Fig.\,\ref{fig:all_maps}), although P- and R-branch low-$J$ lines at shorter wavelengths (i.\,e. between $\sim$4.4 and $\sim$5\,$\mu$m) are brighter~\citep[their total flux is $\sim$4--5 times larger than that from the tail; see HH\,211 BS\,1 NIRSpec spectrum in Fig.\,2 of][]{Ray.ea.2023}. Our MIRI-MRS spectra only show CO on the four bow-shocks (BS\,1--BS\,4, see Fig.\,\ref{fig:all_maps}, middle panel), but not along the jet or the outflow~\citep[see NIRCam image in Fig.\,4 of][]{Ray.ea.2023}, since the integrated CO emission in the NIRCam image of \citet{Ray.ea.2023} is about one order of magnitude fainter than our map 3$\sigma$ threshold sensitivity ($\sim$0.4\,mJy\,arcsec$^{-2}$ or $\sim$17\,MJy\,sr$^{-1}$).

Plenty of OH lines (between 9.1 and 25\,$\mu$m) are detected in the spectra of the terminal bow-shocks (see Fig.\,\ref{fig:BS3_spec}), coming from pure rotational states ($v=0$, $J^\prime \rightarrow J^\prime$-1) arising in the $^2\Pi_{3/2}$ and $^2\Pi_{1/2}$ ladders and cross-ladder. These OH mid-IR lines~\citep[suprathermal OH rotational emissions; see][]{Neufeld.ea.2024} originate from water photodissociation by 114--143\,nm UV radiation, produced in this case by strong jet shocks ($\rm{v}\geq$\,40\,km\,s$^{-1}$)~\citep[see, e\,g.][]{Tabone.vanHemert.ea2021,Zannese.ea.2024}. This emission was already observed in low-resolution {\it Spitzer}/InfraRed Spectrograph (IRS) spectra of HH\,211~\citep[see][]{Tappe.ea.2008} and also predicted and modelled by \citet{Tabone.vanHemert.ea2021}. In addition, these mid-IR lines were also observed with {\it Spitzer} in DG\,Tau~\citep[at $\lambda>$13\,$\mu$m][]{Carr.ea.2014} and recently detected with MIRI also in the HOPS\,370 jet~\citep[see][]{Neufeld.ea.2024}.
Additional H$_2$O transitions ($v=0-0$ and $v_2=1-0$), as well as faint HCO$^+$ ($v_2=1-0$) at 12\,$\mu$m and CO$_2$ at 15\,$\mu$m are also detected in the spectra of the three outer bow-shocks (see Fig.\,\ref{fig:BS3_spec}). 

Other atomic forbidden lines in emission detected in BS\,3 include bright \ClI at 11.3\,$\mu$m (also detected in BS\,1 and 2, see Fig.\,\ref{fig:all_maps} in Appendix\,\ref{sec:add_maps}), \NeII at 12.8\,$\mu$m, barely visible in BS\,1, as well as faint (\mbox{SNR}$\leq$3$\sigma$) emission of \ArIIp, \ClIIp, \CoIIp, and \SIII (see Fig.\,\ref{fig:BS3_spec} and Tab.\,\ref{table:atomic_lines}). Notably, all these atomic lines and their intensities were predicted in \citet[][hereafter, HM89]{Hollenbach.McKee1989} $J$-shock models.

\subsection{Jet radius}
\label{sec:radius}

A noteworthy result from our line maps is that the size of the inner jet is resolved, or marginally resolved, in both atomic species (\FeII at 26 and 17.9\,$\mu$m, \FeI, and \SI) and H$_2$ lines (0-0\,S(7) and 0-0\,S(1)).
We compute the diameter of the jet for the different lines for various knots (knot id. and coordinates are listed in Table\,\ref{tab:neTe}), measuring the \mbox{FWHM} orthogonal to the jet axis at each knot position, after collapsing the image over the knot size along the jet axis. The spatial line-profile is then fitted with a 1D Gaussian and the resulting deconvolved diameter (or jet size) is $d=\sqrt{\mbox{FWHM}^2 -\mbox{PSF}^2}$, where $\mbox{PSF}$ is the point-spread function value at wavelengths close to those of the emission line.
For the atomic lines and the H$_2$ 0-0\,S(1) line, we measure the $\mbox{PSF}$ of the continuum emission towards BS\,1, that is not spatially resolved.
As no continuum is detected at 5.5\,$\mu$m, to infer a reference $\mbox{PSF}$ for the H$_2$ 0-0\,S(7) line we use a set of faint H$_2$O lines towards BS\,1, which do not seem to be spatially resolved. The obtained value is 0$\farcs$3, which is similar to the nominal one reported in \citet{Law.Morrison.ea2023} (0$\farcs$28). It is also worth noting that, as the H$_2$ jet is fully resolved (0$\farcs$5--1$\arcsec$) at this wavelength, a difference of 0$\farcs$02 would not significantly affect its inferred size.

\begin{figure}
        \includegraphics[width=0.495\textwidth]{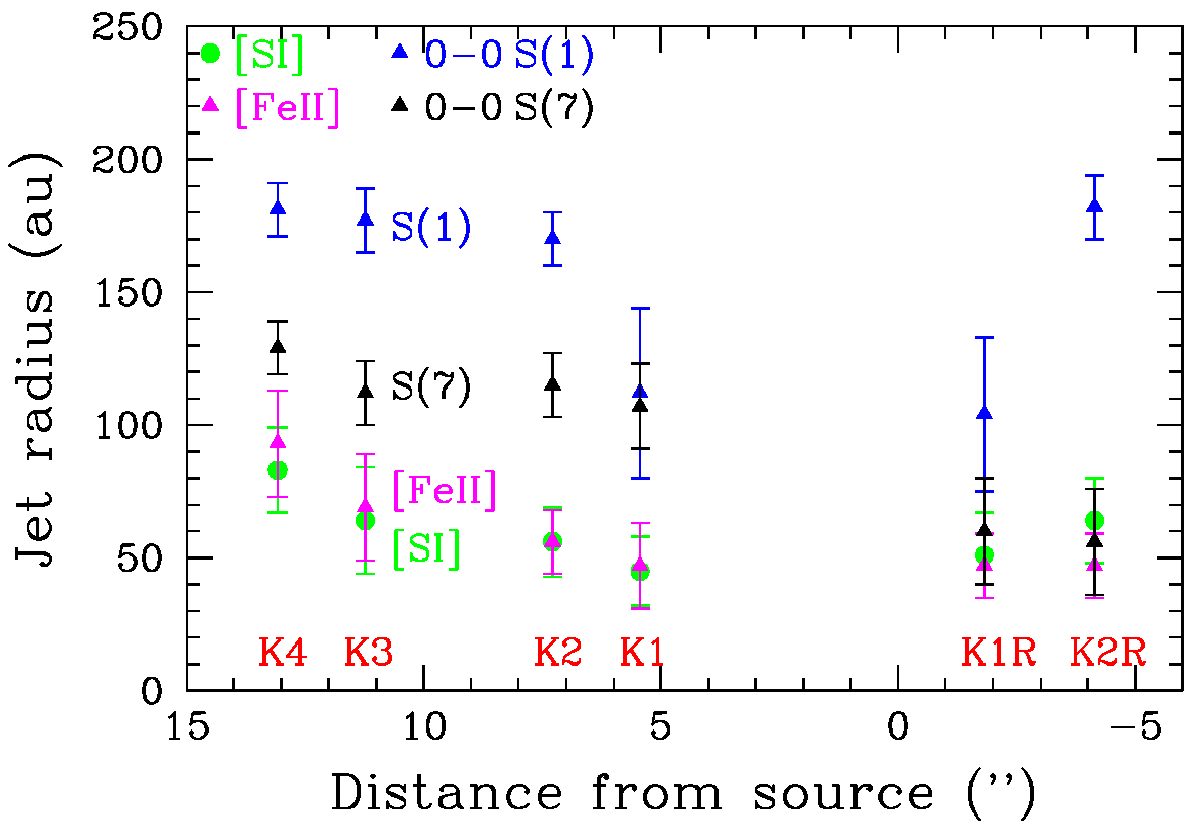}

    \caption{Inferred jet radius (in au) vs. distance (in arcseconds) from the source for different lines. Green dots, magenta triangles, blue and black triangles show \SIp, \FeII (at 26\,$\mu$m), H$_2$ 0-0\,S(1) and S(7) lines, respectively. Knot identification is displayed in red.}
    \label{fig:size knots}
\end{figure}

Measured \mbox{FWHM}s (in $\arcsec$), deconvolved sizes (in $\arcsec$) and radii (in au) of the jet for the different species and at different positions are reported in Table\,\ref{tab:size}. Figure\,\ref{fig:size knots} shows that the jet radius varies for the different species (i.\,e., \SIp, \FeII 26\,$\mu$m, H$_2$ 0-0\,S(1) and S(7) lines, depicted as green dots, magenta triangles, blue, and black triangles, respectively) at different distances from the source. The position and name of each knot are labelled in red at the bottom of the figure. 

Overall, Figure\,\ref{fig:size knots} confirms the onion-like structure of the jet, with the atomic jet displaying smaller radii and the molecular component larger radii. The different atomic lines show similar values in radius (ranging from $\sim$45 to $\sim$100\,au), possibly because the angular resolution is not sufficient to separate neutral and ionised gas. On the other hand, the H$_2$ lines have radii similar or larger than the atomic jet. The radius of the H$_2$ 0-0\,S(7) line ranges from $\sim$60 to $\sim$130\,au, whereas the 0-0\,S(1) line is positioned on the outer layers of the jet ($\sim$100--180\,au).

In most cases, the jet size is just marginally resolved (see Tab.\,\ref{tab:size}). Therefore these trends can be hardly seen in our maps, with the exception of the H$_2$ 0-0\,S(1) emission line, that overlaps and encloses both atomic and hot H$_2$ molecular emission (see Fig.\,\ref{fig:MIRI-tricolor} and Fig.\,\ref{fig:Tri-jet}).
 
\subsection{H$_2$ ro-vibrational diagrams}
\label{sec:rovibrational}

The large number of H$_2$ rotational transitions from $v=0$ and $v=1$ levels and their ample range of excitation energies (1000\,K$\lesssim E_{\rm up} \lesssim$16\,000\,K) allow us to infer both gas temperature ($T({\rm H_2})$) and column density ($N({\rm H_2})$) along the flow~\citep[see, e.g.][]{Giannini.ea.2004,Caratti.ea.2006} by means of ro-vibrational diagrams. 
Extinction-corrected line column densities, divided by their statistical weights, are plotted against their excitation energies using a semi-logarithmic scale.
For a gas in local thermal equilibrium (LTE), the gas excitation follows a Boltzmann distribution ($N_{\rm v,J}$/$g_{\rm v,J} \propto \exp(-E_{\rm j}/kT_{\rm ex})$) and points in the diagram align in a straight line, whose slope is the reciprocal of the gas excitation temperature (if $T_{\rm ex}$=$T_{\rm gas}$). The y-axis intercept provides the gas column density.
Often, the H$_2$ gas shows stratification in temperature. Generally, transitions at low excitation ($E_{\rm up}\lesssim$4000--5000\,K) trace a cold component ($T({\rm H_2})\lesssim$1000\,K), those at higher energy ($E_{\rm up}\lesssim$10\,000--12\,000\,K) a warm component ($T({\rm H_2})\lesssim$2000--2500\,K), and those at the highest energy a hot component ($T({\rm H_2})\sim$ 3000--4000\,K). 

Lines tracing the cold, warm, and hot components can be detected at MIR wavelengths (see Tab.\,\ref{tab:h2_lines}).
Therefore, the MIRI-MRS regime can trace up to three H$_2$ components, and we might expect to measure up to three different temperatures and column densities~\citep[see e.\,g.][]{Neufeld.Nisini.ea2009, Dionatos.ea.2010}, with the cold component tracing the highest column densities and the hot component the lowest.

Unfortunately, it is not possible to directly measure the visual extinction with our MIRI-MRS data, as we do not detect any pair of H$_2$ transitions arising from the same upper level in the MIRI wavelength range. However, a rough estimate (usually within a 5-10\,mag uncertainty) can be also inferred by varying \Av in the ro-vibrational diagrams and maximising the correlation coefficient in the fits of the Boltzmann plots. We use this technique to infer the visual extinction of the inner jet region of HH\,211 (namely Knot\,1, 1R, and 2R), too embedded to be detected in the 1-0\,S(1) NIRCam filter (see Fig.\,\ref{fig:Avmap}). The ro-vibrational diagram shows that the blue jet closest to source has an \Av value of 80\,mag (see Fig.\,\ref{fig:Av_jet} in Appendix\,\ref{sec:Av_jet}) and we thus adopt this visual extinction value within an $\sim$6$\arcsec$ ($\sim$1930\,au) radius from the source, namely where no meaningful extinction value is measured in our \Av map (see Fig\,\ref{fig:Avmap}). It is worth noting, however, that close to the source (i.\,e. within 2$\arcsec$--3$\arcsec$ from the source, where no H$_2$ emission is detected) and on-source the visual extinction is very likely much higher that 80\,mag (probably \Av $>$100\,mag; as hinted by the lack of continuum mid-IR emission on source at $\lambda \leq$27\,$\mu$m, see Sect.\,\ref{sec:maps}, as well as the disappearing of the \FeII atomic jet close to the source).

To measure the H$_2$ excitation conditions, we employ a pixel-by-pixel ro-vibrational diagram analysis as described in \citet{Gieser.ea.2023}. Briefly, the python routine first provides sub-cubes around the H$_2$ lines of interest (see Tab.\,\ref{tab:h2_lines}) and resamples each sub-cube to a common (and worst) spatial resolution of 0.$\arcsec$7 (0.$\arcsec$2 pixel-scale size) of MIRI-MRS Channel\,3. Each sub-cube is then dereddened using the \Av map presented in Sec.~\ref{sec:Av_jet}.

Line fluxes are derived for each line by extracting the spectrum from each spaxel of the corresponding sub-cube and by fitting a 1D-Gaussian profile, as the H$_2$ lines are barely spectrally resolved with MIRI-MRS. To avoid spurious detection, a flux threshold of 40\,MJy\,sr$^{-1}$ per spaxel is set in the original cube. As a further constraint, fits to the ro-vibrational diagrams are done only if five or more transitions are detected in a spaxel. Ro-vibrational analysis is performed using the \texttt{pdrtpy}\footnote{\texttt{pdrtpy} is developed by Marc Pound and Mark Wolfire. This project is supported by NASA Astrophysics Data Analysis Program grant 80NSSC19K0573; from JWST-ERS program ID 1288 provided through grants from the STScI under NASA contract NAS5-03127; and from the SOFIA C+ Legacy Project through a grant from NASA through award \#208378 issued by USRA} Python package.

In our ro-vibrational plots we detect a mixture of temperatures. The
\texttt{prdtpy} package was set to fit two components in our diagrams. 
For simplicity, we call them warm and hot components, following the nomenclature of \citet{Gieser.ea.2023}. This is to distinguish the warm H$_2$ gas from the colder outflow traced by other species at sub-mm wavelengths.
However, it is worth noting that both temperature and column of each component largely vary along the flow, depending on the gas excitation conditions, that is on the number of detected lines and their excitation energies (see examples in Fig.\,\ref{fig:RV_components} of Appendix\,\ref{sec:rovibrational}). Therefore our simplification is highly reductive.

\begin{figure*}
        \includegraphics[width=1\textwidth]{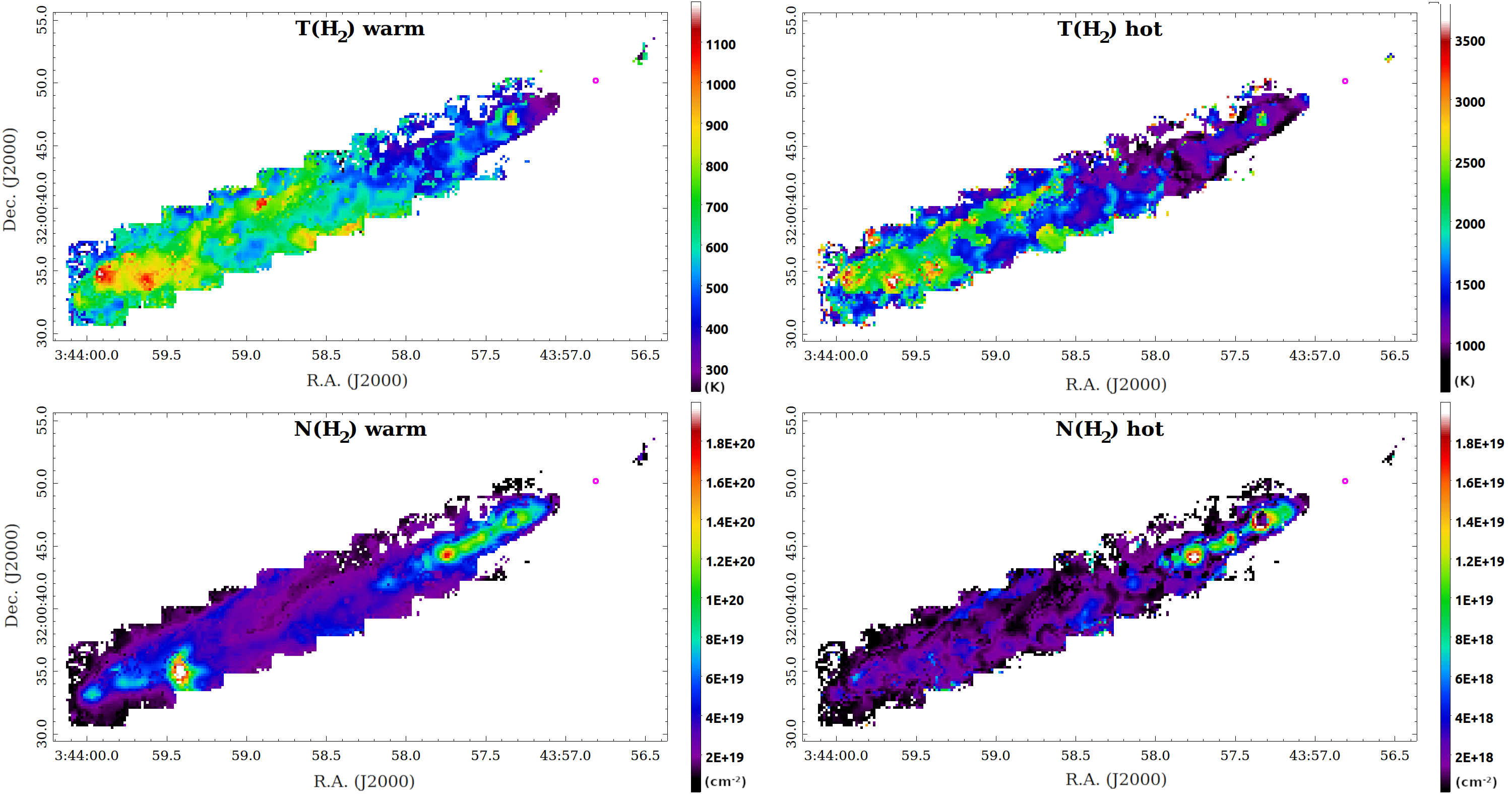}
    \caption{Temperature (top panels) and column density (bottom panels) maps of the warm (left panels) and hot (right panels) H$_2$ components. The magenta circle shows the position of the ALMA mm continuum source. }
    \label{fig:Boltzmann_maps}
\end{figure*}

Figure\,\ref{fig:Boltzmann_maps} shows temperature (top panels) and column density (bottom panels) maps for the warm (left panels) and hot components (right panels) of the gas. Overall, the gas is colder and denser in the inner jet, whereas it becomes warmer and less dense in the outer bow-shocks. Moreover, the warm component has column densities one order of magnitude larger than the hot component (see bottom panels of Fig.\,\ref{fig:Boltzmann_maps}).

The temperature of the warm component varies from $\sim$300\,K, in the inner jet close to source, to 500--700\,K along the jet, and up to 900--1000\,K in the terminal bow-shocks, and its column density changes from 10$^{19}$ to 10$^{20}$ \,cm$^{-2}$ along the jet, while it is just some 10$^{19}$\,cm$^{-2}$ along the bow-shocks, with the exception of BS\,3, which has the highest column densities ($\sim$3$\times$10$^{20}$\,cm$^{-2}$).

The hot component varies from 1000 to 2000\,K along the jet, whereas it is much higher (2000--3500\,K) at the bow-shocks. On the other hand, its column density is higher along the jet (1--2$\times$10$^{19}$\,cm$^{-2}$) and drops in the outer jet and bow-shocks (10$^{18}$--10$^{19}$\,cm$^{-2}$).

Less collimated, colder (200--400\,K) and less dense (10$^{18}$--10$^{19}$\,cm$^{-2}$) gas (showing a U or V shape at the rear of the jet) is detected in the inner regions (bottom panel of Fig.\,\ref{fig:H2-maps} and left panels of Fig.\,\ref{fig:Boltzmann_maps}), likely tracing a poorly collimated wind. In addition, the outflow gas (i.\,e. the entrained gas) appears less dense and colder than that in the jet and bow-shocks, with the exception of the shocked outflow-ISM interface (blue coded in the $N({\rm H_2})$ maps), where column density and temperature appear to be higher than those of the entrained gas.

\subsection{Physical properties of the flow from the atomic species}
\label{sec:atomic_res}

We can use the different atomic species detected along the flow to infer the main physical parameters of the atomic gas. 

\subsubsection{Electron density and temperature from \FeII }

Electron density ($n_e$) and temperature ($T_e$) of the atomic gas can be derived from the many \FeII transitions detected along the flow in the MIRI-MRS data. 
For our analysis, we use a non-local thermal equilibrium (nLTE) excitation model presented in \cite{Giannini.ea.2013}, here updated to include the MIRI-MRS transitions at low $E_{\rm up}$. The model assumes electronic collisional excitation/de-excitation and spontaneous radiative decay. It employs the atomic database of the XSTAR
tool~\citep{Bautista2001}, which provides energy levels, Einstein coefficients, and collision rates (for temperatures between 2000\,K and 20\,000\,K) for the first 159 fine-structure levels of Fe$^+$. 
Our nLTE model provides a line intensity grid for all transitions from the 159 levels for 100\,$\leq n_e \leq$\,10$^7$\,cm$^{-3}$ (in steps of log$_{10}$ ($\delta n_e$/$cm^{-3}$) = 0.06) and 400\,$\leq T_e \leq$\,10$^5$\,K (in steps of $\delta T_e$=200\,K). 
 
The observed line fluxes are de-reddened using the values reported in our \Av map (see Sec.\,\ref{sec:Av}), and their line ratios are used to find the best fit to our model, leaving $T_e$ and $n_e$ as free parameters. Fits with the lowest chi-square ($\chi^2$) value provide the best $T_e$ and $n_e$ solutions.

Along the jet we just detect the \FeII lines at 5.3, 17.9, and 26\,$\mu$m. On the other hand, for the external bow-shocks more lines have been used in our fits (see Column\,6 of Tab.\,\ref{table:atomic_lines}).

Spectra with 1$\arcsec$ radii were extracted from eleven regions, the four outer bow-shocks (BS\,1--4), five knots along the blue-shifted (Knot 1--5), and red-shifted jet (Knot 1 red and 2 red) (see Fig.\,1 and Column\,2 of Tab.\,\ref{tab:neTe} for feature identification and coordinates, respectively). 
Only in seven of these spectra (i.\,e. the four bow-shocks and Knot 2, 3, and 5) the \FeII 5.3\,$\mu$m line is bright enough for our analysis ($\geq$5$\sigma$).

Columns\,3 and 4 of Table\,\ref{tab:neTe} report $n_e$ and $T_e$ values of the fits for each feature, while Column\,5 lists the $\chi^2$ of the best fit along with its degrees of freedom (i.\,e. number of line ratios used minus the two variables, $n_e$ and $T_e$). Temperature increases moving away from the source, notably from $T_e\sim$1000\,K in the inner jet to $T_e\sim$1400\,K in BS\,4, $T_e\sim$2800\,K moving further out to Knot\,5, and it reaches its peak at BS\,3 ($T_e\sim$3800\,K). $T_e$ finally drops in the two most external bow-shocks, BS\,2 and BS\,1, at 1800 and 2400\,K, respectively. The trend of $n_e$ is similar, low values (100--230\,cm$^{-3}$) along the jet and at BS\,4 and higher values (350--800\,cm$^{-3}$) at the three terminal bow-shocks (BS\,1--BS\,3). 

\begin{figure}

        \includegraphics[width=0.48\textwidth]{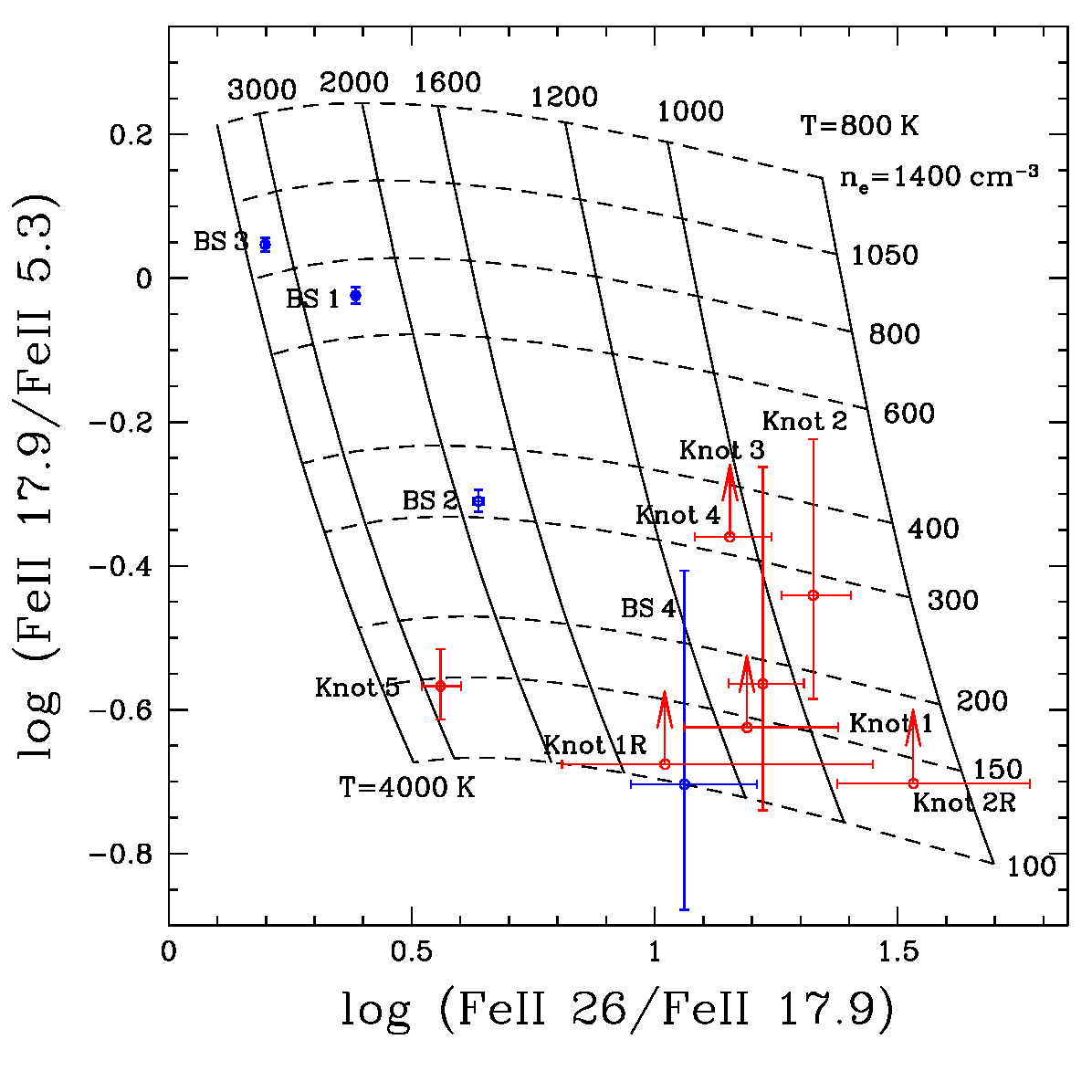}

    \caption{Logarithmic grid of the \FeII 26\,$\mu$m/17.9\,$\mu$m line ratio (x-axis), sensitive to $T_e$ and \FeII 17.9\,$\mu$m/5.3\,$\mu$m line ratio (y-axis), sensitive to $n_e$. Line ratios and error bars for the studied features (bow-shocks in blue, knots in red) are shown in the plot.} 
    \label{fig:Te_ne_plot}
\end{figure}

It is also worth noting that the three brightest \FeII lines (i.\,e. at 5.3, 17.9, and 26\,$\mu$m) can be combined to infer both parameters, as the 17.9/5.3\,$\mu$m ratio is sensitive to $n_e$ and the 26/17.9\,$\mu$m ratio to $T_e$. Figure\,\ref{fig:Te_ne_plot} shows a plot of the two line ratios (26\,$\mu$m/17.9\,$\mu$m line on the x-axis and 17.9\,$\mu$m/5.3\,$\mu$m line ratio on the y-axis) in logarithmic scale and the grid of $T_e$ and $n_e$ values derived from our model. Line ratios and uncertainties of each analysed feature are displayed (bow-shocks in blue, knots in red). Derived values and errors for $n_e$ and $T_e$ are listed in Column\,6 and 7 of Tab.\,\ref{tab:neTe}, respectively. Within the error bars, these results are the same as for the fits, however these numbers provide a better constraint on the uncertainties. The larger uncertainties are those on $n_e$ along the inner jet and this is due to the fact that the 5.3\,$\mu$m line has a relatively low \mbox{SNR} ($\leq$5) there.
For four more knots (Knot\,1 and Knot\,4 blue-shifted, and Knot\,1 and Knot\,2 red-shifted, the latter labelled with an additional R for distinguishing purposes in Fig.\,\ref{fig:Te_ne_plot} and other figures and tables of the paper), we report lower limits on $n_e$ (as the 5.3\,$\mu$m line is not detected or is below 3$\sigma$ in our spectra) and the corresponding temperature upper limits. The upper limit to the 5.3\,$\mu$m line flux is inferred by multiplying the 3$\sigma$ noise of the spectrum at the line wavelength by the nominal \mbox{FWHM} of the line.

Finally, we employ continuum-subtracted line images of the three bright \FeII lines to construct both log\,(26\,$\mu$m/17.9\,$\mu$m) and log\,(17.9\,$\mu$m/5.3\,$\mu$m) maps, to visualise how $T_e$ and $n_e$ vary along the flow. Images were sampled at the lowest pixel scale of the 26\,$\mu$m image, and, to avoid spurious detection, pixels with fluxes below 3$\sigma$ were masked.
Furthermore, in the resulting maps, only pixels within the 3$\sigma$ line contours are displayed.

\begin{figure}

        \includegraphics[width=0.48\textwidth]{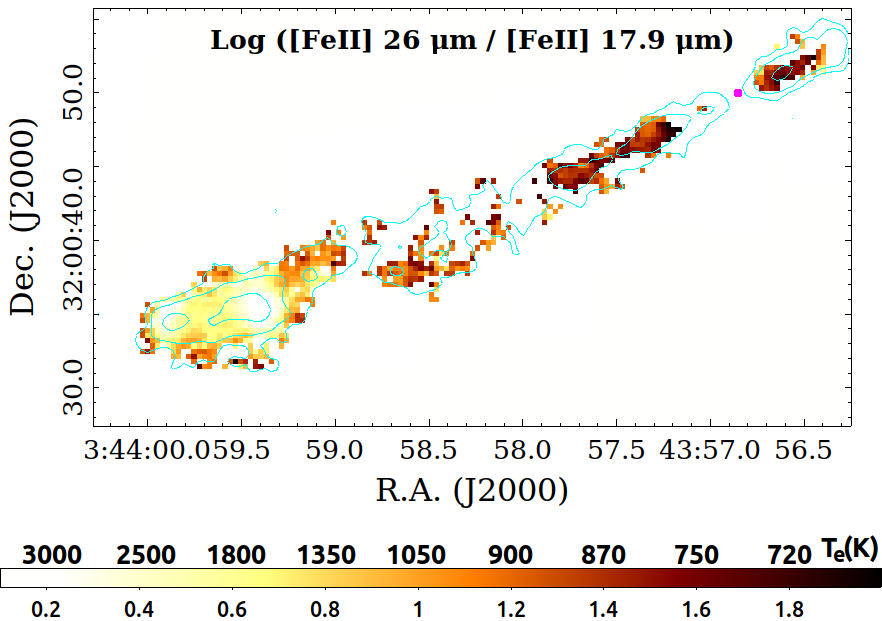}

    \caption{Logarithmic map of the ratio of \FeII lines at 26\,$\mu$m and 17.9\,$\mu$m. Cyan contours show the \FeII (26\,$\mu$m) line flux at 3, 10, 50\,$\sigma$ (0.48, 1.6, 8\,mJy\,pixel$^{-1}$).
    Colour code displays the logarithmic value of the ratio (bottom) and the corresponding $T_e$ (in Kelvin) for an average $n_e$ of 500\,cm$^{-3}$. Only pixels within 3$\sigma$ line contours are displayed. The magenta dot shows the position of the ALMA mm continuum source.} 
    \label{fig:Te_map}
\end{figure}

\begin{figure}
        \includegraphics[width=0.485\textwidth]{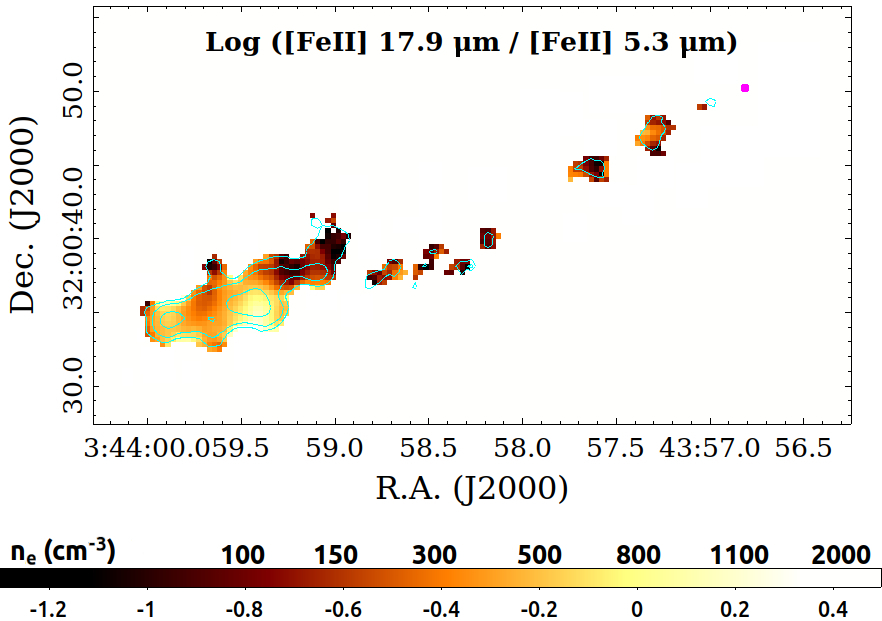}

    \caption{Logarithmic map of the ratio of \FeII lines at 17.9\,$\mu$m and 5.3\,$\mu$m. Cyan contours show the \FeII (5.3\,$\mu$m) line flux at 3, 10, 50\,$\sigma$ (0.1, 0.4, 2.1\,mJy\,pixel$^{-1}$). Colour code reports the logarithmic value of the ratio (bottom) and the corresponding $n_e$ (in cm$^{-3}$) for an average $T_e$ of 1000\,K. Only pixels within 3$\sigma$ line contours are displayed. The magenta dot shows the position of the ALMA mm continuum source.}
    \label{fig:ne_map}
\end{figure}

Figure\,\ref{fig:Te_map} shows the logarithmic map of the 26\,$\mu$m/17.9\,$\mu$m line ratio.
The colour-coded bar reports both logarithmic values and the corresponding $T_e$ for a gas with $n_e$=500\,cm$^{-3}$ (i.\,e. an average of the measured range of values along the flow). Therefore, the corresponding temperature is slightly underestimated along the jet and overestimated along the terminal bow-shocks. Notably, the counter-jet has $T_e$ similar to the jet and the increasing $T_e$ trend towards the terminal shocks is also visible.
Fig.\,\ref{fig:ne_map} shows the logarithmic map of the 17.9\,$\mu$m/5.3\,$\mu$m. 
The colour-coded bar indicates the corresponding $n_e$ of a gas at $T_e$=1000\,K. Electron densities are lower along the jet and in BS\,4 whereas they are higher in the outer bow-shocks.


\begin{table*}
\centering
\caption{$n_e$ and $T_e$ values from \FeII analysis along the HH\,211 flow }
\label{tab:neTe}
\begin{tabular}{lcccccc}
\hline \hline
Feature & Coordinates (J2000)         & $n_e$ (fit)    & $T_e$ (fit) & $\chi^2$ (dof$^a$) & $n_e$ (17.9/5.3 ratio) & $T_e$ (26/17.9 ratio)\\
Name    & RA(\degr) ; DEC.(\degr)  &  (cm$^{-3}$)  & (K) &   &  (cm$^{-3}$)  & (K)\\
\hline
Knot\,2 red (2R)  &   55.9855237 ; +32.0144652 & $\cdots$   & $\cdots$ 	&  $\cdots$   &  $>$150	&  $<$980\\ \\[-9pt]
Knot\,1 red (1R)  &   55.9861638 ; +32.0141036 & $\cdots$   & $\cdots$ 	&  $\cdots$   &  $>$160	&  $<$1900\\ \\[-9pt]
Knot\,1  &   55.9883338 ; +32.0132814 & $\cdots$   & $\cdots$ 	&  $\cdots$   &  $>$200		&  $<$1300 \\ \\[-9pt]
Knot\,2  &   55.9888571 ; +32.0130183 & 230  & 1000 & 0.7 (1) & 280$^{+270}_{-100}$ & 930$^{+100}_{-80}$ \\ \\[-9pt]
Knot\,3  &   55.9899822 ; +32.0124774 & 100  & 1000 & 0.5 (1) & 180$^{+270}_{-180}$ & 1060$^{+100}_{-80}$\\ \\[-9pt]
Knot\,4  &   55.9905371 ; +32.0122722 & $\cdots$   & $\cdots$ 	&  $\cdots$   &  $>$350	&  $<$1100\\ \\[-9pt]
Knot\,5  &   55.9962485 ; +32.0104424 & 150  & 2800 & 0.8 (2) & 150$^{+25}_{-20}$ & 2800$\pm$400 \\ \\[-9pt]
BS\,4    &   55.9944886 ; +32.0105781 & 100  & 1400 & 0.3 (1) & 100$^{+170}_{-100}$ & 1400$^{+200}_{-300}$\\ \\[-9pt]
BS\,3    &   55.9975519 ; +32.0097905 & 800  & 3800 & 169 (3) & 860$^{+70}_{-60}$ & 3600$^{+200}_{-100}$\\ \\[-9pt]
BS\,2    &   55.9985417 ; +32.0095831 & 350  & 1800 & 20 (3) & 350$^{+70}_{-60}$ & 1800$^{+200}_{-100}$\\ \\[-9pt]
BS\,1    &   55.9995251 ; +32.0095191 & 690  & 2400 & 1.6 (3) & 690$^{+70}_{-60}$ & 2400$^{+200}_{-100}$\\ \\[-9pt]
 \hline
\end{tabular}
\tablefoot{$^a$ dof= degrees of freedom}\\
\end{table*}

\subsubsection{Gas-phase iron abundance}
\label{sec: depletion}

Gas-phase iron abundance, and thus the observed \FeII and \FeI line intensities, are regulated by the shock efficiency in eroding the dust grains, through processes like sputtering and grain-grain collision, which release iron into the gas-phase~\citep[see, e.\,g.][]{Seab1987,Jones2000,Colangeli.ea.2003}. Studies of nearby protostellar jets at near- and mid-IR wavelengths have shown that the gas-phase abundance of Fe is much lower than the typical solar abundance~\citep[i.\,e. (Fe/H)$_{\sun}$=2.88$\times$10$^{-5}$; see,][]{Asplund.ea.2021}, indicating that, if solar abundance is assumed, metals are still partially locked onto grains~\citep[see, e.\,g.,][]{Nisini.ea.2002,Podio.ea.2009,Dionatos.ea.2009,Dionatos.ea.2010}. Typically, this type of analysis is conducted by comparing \FeII line intensities with those from a non-refractory species (e.\,g., Ne, S, O, P, Cl) emitted under the same physical conditions and assuming solar abundances.

An analysis of Fe and Si gas-phase depletion in HH\,211 was accomplished by \citet{Dionatos.ea.2010} with {\it Spitzer}/IRS, using sulphur as a non-refractory species. They found a gas-phase Fe abundance between 3\%--10\% and 2\%--7\% with respect to the solar one for the blue- and red-shifted jet, respectively (see their Table\,5). However, as no \FeI emission was detected in their spectra, they assumed that the iron was fully ionised, therefore those values should be considered as an upper limit of the iron abundance (or a lower limit value of the Fe gas-phase depletion) along the flow.

Constraints on the gas-phase iron abundance can be also inferred by comparing observed dereddened line ratios with those predicted in dissociative shock models~\citep[see, e.\,g.,][]{Nisini.ea.2002}. 
To trace the iron abundance, we use the observed \FeII 26\,$\mu$m/\SI and \FeI/\SI line ratios, as both \FeI and \FeII lines are equally depleted along the jet. We assume solar abundance for the different species and, most importantly, that S is a non-refractory species, and thus it is all in the gas phase.
It is worth noting, however, that the latest studies in nearby molecular clouds have shown that sulphur is a semi-refractory species, and its depletion depends on the environment and star formation activity~\citep[see, e\,g.,][]{Fuente.ea.2023}.
In particular, \citet{Fuente.ea.2023} show that S is, on average, depleted by a factor of 20 in the Perseus molecular clouds. 

For different knots along the jet, we compare the observed line ratios against those predicted by HM89 $J$-shock models for different values of the gas-phase iron abundance and a range of shock velocities (${\rm v_s}$=30--100\,km\,s$^{-1}$). 

\begin{figure}
        \includegraphics[width=0.495\textwidth]{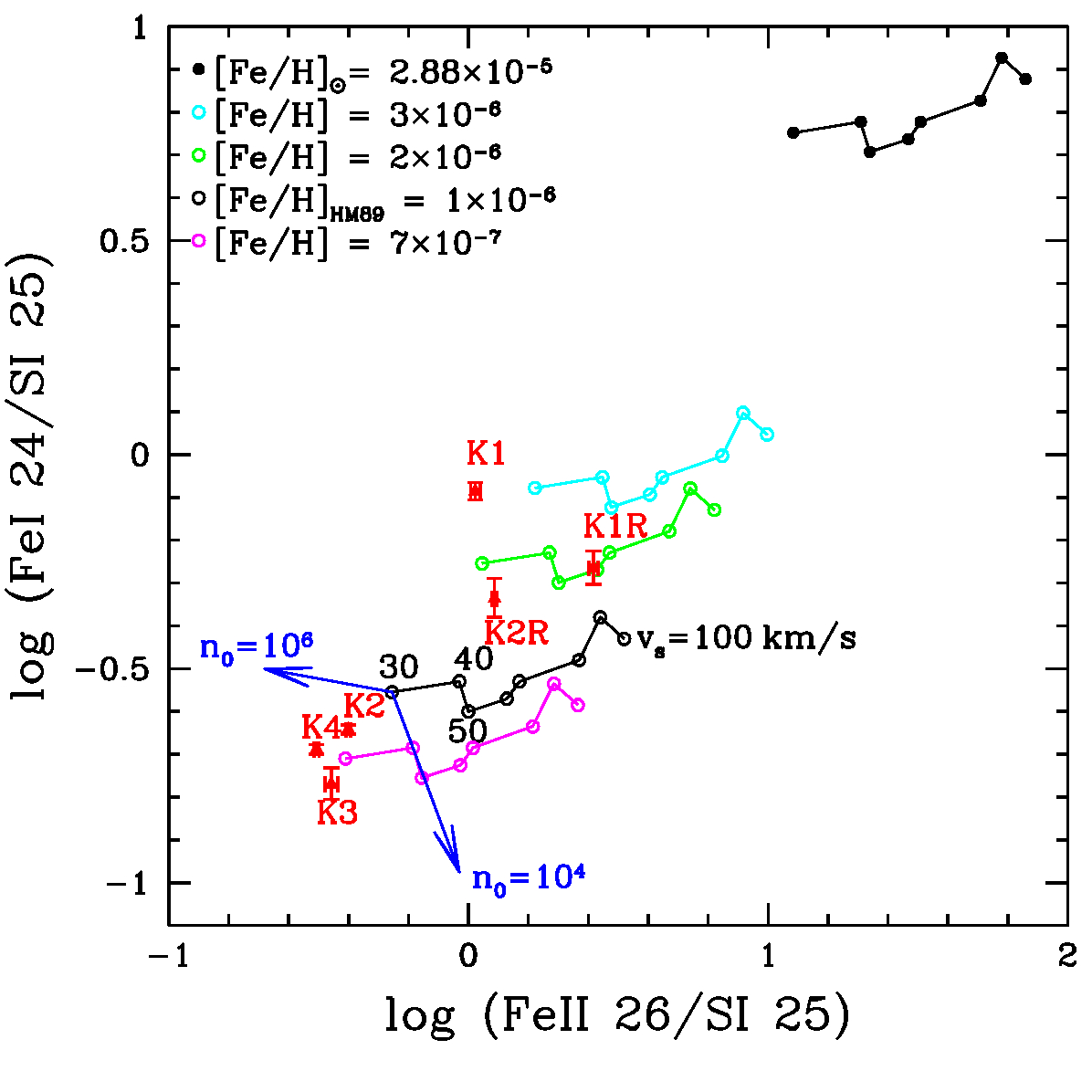}

    \caption{Observed vs. predicted \FeII 26\,$\mu$m/\SI and \FeI/\SI line ratios for different knots (red triangles) along the jet. Each curve reproduces the \citet{Hollenbach.McKee1989} dissociative models for different values of the gas-phase iron abundance (coloured curves, from 2.88$\times$10$^{-5}$ - black dots - to 7$\times$10$^{-7}$ - magenta open circles), pre-shock density $n_0$=10$^5$\,cm$^{-3}$, and a range of shock velocities (${\rm v_s}$=30--100\,km\,s$^{-1}$), as reported in the labels. The two blue arrows show how the plots move by varying $n_0$ from 10$^4$ to 10$^6$\,cm$^{-3}$}
    \label{fig:abundance_Fe}
\end{figure}

Figure\,\ref{fig:abundance_Fe}
shows the observed line ratios for different knots (red triangles) plotted over the HM89 original values (black open circles), calculated for \HI pre-shock density of $n_0$=10$^5$\,cm$^{-3}$ and an Fe abundance of 10$^{-6}$. Predicted \SI line fluxes have been modified to take into account the slightly different S abundance in the original HM89 model (10$^{-5}$) with respect its most recent solar abundance~\citep[1.318$\times$10$^{-5}$;][]{Asplund.ea.2021}. The same curve is plotted for a range of gas-phase iron abundances (plotted in different colours, as labelled in the figure): from solar (2.88$\times$10$^{-5}$, black dots curve, upper right) to 7$\times$10$^{-7}$ (magenta open circles curve, lower left). Changes in pre-shock density moves the curves as shown by the blue arrows in the plot. 

Figure\,\ref{fig:abundance_Fe} shows that Fe is largely depleted (i.\,e. knots fall far away from the curve with solar abundance - black dots curve) and just a small amount is in gas-phase (between $\sim$2\% and $\sim$10\%), as abundances range from $\sim$7$\times$10$^{-7}$ to $\sim$3$\times$10$^{-6}$. Knots close to the source (namely Knot\,1, Knot\,1 red and 2 red) seem to be less depleted than those positioned further away from the source (i.\,e. Knot\,2, 3 and 4). Although we cannot be certain about the 
absolute values of Fe depletion, which rely on the assumption that S is not depleted, we can be confident regarding the Fe differential depletion along the jet.

Nevertheless, one would expect the opposite of what we find, namely that the Fe gas-phase increases along the flow moving away from the source~\citep[see, e.\,g.,][]{Nisini.ea.2005,Podio.ea.2006}, as Fe is being released from grains via shocks along the flow. Our measure of a decrement in the Fe gas-phase along the jet (see Fig.\,\ref{fig:abundance_Fe}) suggests that this difference is local and related to the different strength of the shocks along the flow, rather than a continuous destruction of grains along the flow.

\subsubsection{Density, shock-velocity and ionisation fraction of the atomic jet}
\label{sec: atomic params}

Figure\,\ref{fig:abundance_Fe} already provides us with some indications about pre-shock densities ($n_0$) and shock velocities (${\rm v_s}$) along the flow. However, as both iron abundance and ionisation fraction are not well constrained by such analysis, it is not possible to properly infer ${\rm v_s}$ and $n_0$ values. The latter parameter is particularly important to define the dynamical properties of the atomic jet component and thus its relevance with respect to the molecular component.

We can use other atomic lines predicted by the HM89 models, to constrain these two parameters. In particular, we note that the \NeII line intensity strongly depends on ${\rm v_s}$, but it is less sensitive to $n_0$ variations (see Figure\,7 in HM89). In contrast, \SI and \ClI (at 11.4\,$\mu$m) line intensities are strongly dependent on $n_0$ variations but not on ${\rm v_s}$ (at least for shock velocities $\lesssim$60\,km\,s$^{-1}$).

Therefore, we first constrain the shock velocity along the flow using the dereddened \NeII line intensity as observed in different bow-shocks and knots of HH\,211.
As the line is detected below 3$\sigma$ (or not detected) in the knots along the jet, here we use a 3$\sigma$ upper limit to estimate its line intensity. Shock velocities of 40$\pm$5\,km\,s$^{-1}$ are measured for the bow-shocks (35$\pm$5\,km\,s$^{-1}$ for BS\,2), whereas ${\rm v_s}<$50--60\,km\,s$^{-1}$ is inferred along the jet. Results are listed in Column\,2 of Table\,\ref{tab:jet_atomic_model}.

We then employ \SI and \ClI line fluxes to infer the gas pre-shock density. Values derived from \SI are more reliable, as \ClI is only weakly detected along the jet (\mbox{SNR}$\leq$3$\sigma$). Despite this, $n_0$ values derived from \ClI are consistent with those derived from \SI (see Column\,3 of Table\,\ref{tab:jet_atomic_model}).

Pre-shock density along the jet ranges from 7$\times$10$^4$ to 2$\times$10$^5$\,cm$^{-3}$, being denser (1--2$\times$10$^5$\,cm$^{-3}$) in knots close to the source (i.\,e. Knot\,1, Knot\,1 red and 2 red). The outer jet and bow-shocks show a lower density (4--9$\times$10$^4$\,cm$^{-3}$). 

By combining $n_e$ and $n_0$, it is then possible to infer the \HI\, ionisation fraction ($x_e$=$n_e$/$n_0$). Values along the jet are a few 10$^{-3}$ (see Column\,4 of Table\,\ref{tab:jet_atomic_model}), whereas at the terminal bow-shocks it is slightly higher (from several 10$^{-3}$ to 10$^{-2}$). We note that these values are similar to what was inferred in HH\,211 with {\it Spitzer} by \citet{Dionatos.ea.2010}. Such small $x_e$ values are typical of Class\,0 protostellar jets~\citep[see, e.\,g.][]{Dionatos.ea.2009,Dionatos.ea.2010}, in contrast to Class\,I and II jets, where $x_e$ ranges from 0.03 to 0.9  ~\citep[see, e.\,g.,][and references therein]{Ray.ea.2007,Frank.ea.2014}.
\begin{table*}
\caption{Physical and dynamical parameters along the HH\,211 atomic flow}
\label{tab:jet_atomic_model}
\begin{tabular}{lccccccc}
\hline \hline \\
Feature & ${\rm v_s}$           & $n_0$ \SI - \ClI  & $x_e$ & $\dot{M}_{\rm knot}$ &  $P_{\rm knot}$ & $\dot{P}_{\rm knot}$ & $\tau_{\rm atomic}$ \\
        &   (km\,s$^{-1}$)  & (10$^5$\,cm$^{-3}$)    & (10$^{-3}$)& (10$^{-7}$ M$_{\odot}$\,yr$^{-1}$) & (10$^{-4}$\,M$_{\odot}$\,km\,s$^{-1})$ & (10$^{-5}$\,M$_{\odot}$\,yr$^{-1}$\,km\,s$^{-1}$) & (yr)\\
\hline
Knot\,2R &  $<$50 &  2$\pm$0.5 - $<$3 & $>$0.8  & 3$\pm$1  & 8$\pm$2 & 3$\pm$2 & 50$\pm$10 \\
Knot\,1R &  $<$60 &  1$\pm$0.1 - 0.8--3 & $>$2  & 0.9$\pm$0.4  & 2.4$\pm$0.5 & 1.1$\pm$0.6 & 22$\pm$4 \\
Knot\,1 &   $<$50 &  2$\pm$0.5 - $<$5 & $>$1 &  1.3$\pm$0.9  & 2.8$\pm$0.9 & 2$\pm$1 & 65$\pm$12 \\
Knot\,2 &   $<$45 &  0.8$\pm$0.1 - 0.3--0.9 & 4$^{+4}_{-2}$  & 0.8$\pm$0.4  & 1.2$\pm$0.2 & 1.1$\pm$0.6 & 87$\pm$17 \\
Knot\,3 &   $<$50 &  0.7$\pm$0.1 - $<$0.5 & 3$^{+4}_{-3}$ & 0.9$\pm$0.5  & 1.1$\pm$0.3 & 1.2$\pm$0.7 & 134$\pm$26 \\
Knot\,4 &   $<$50 &  0.8$\pm$0.1 - $<$0.5 & $>$1 &  1.8$\pm$0.7  & 3.1$\pm$0.7 & 2.3$\pm$0.9 & 156$\pm$30 \\
BS\,4 &   40$\pm$10 &  0.4$\pm$0.1 - 0.3$\pm$0.2 & 3$^{+5}_{-3}$ & $\cdots$ & $\cdots$  & $\cdots$ & $\cdots$ \\
BS\,3 &   45$\pm$5 &  0.8$\pm$0.1 - 0.7$\pm$0.2 & 11.3$\pm$0.5 & $\cdots$ & $\cdots$  & $\cdots$ & $\cdots$ \\
BS\,2 &   35$\pm$5 &  0.7$\pm$0.1 - 0.7$\pm$0.3 & 5$\pm$2 & $\cdots$ & $\cdots$  & $\cdots$ & $\cdots$  \\
BS\,1 &   40$\pm$5 &  0.9$\pm$0.1 - 0.9$\pm$0.3 & 7.7$\pm$0.1 & $\cdots$ & $\cdots$  & $\cdots$ & $\cdots$  \\
\hline
\end{tabular}
\end{table*}


\subsection{Flow kinematics}
\label{sec: kinematics}
\begin{figure}
        \includegraphics[width=0.49
        \textwidth]{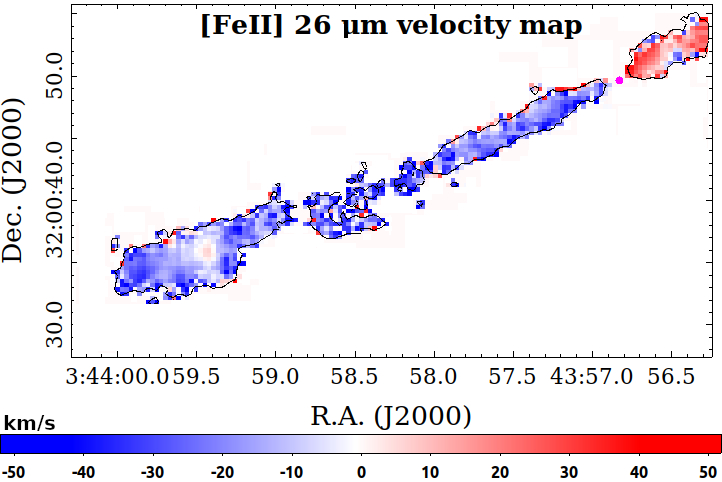}

    \caption{A \FeII\,26\,$\mu$m radial-velocity map. Black contours show the integrated continuum-subtracted line intensity at 5$\sigma$ (0.8\,mJy\,pixel$^{-1}$). The magenta dot shows the position of the ALMA mm continuum source.}
    \label{fig:FeII_vmap}
\end{figure}

MIRI-MRS datacubes also allow us to explore the velocity structure along the flow for the brightest atomic (\FeIIp, \SIp, \FeIp) and molecular (H$_2$) lines.
To derive radial velocity maps of these lines, we employ the python routine \texttt{bettermoments}  ~\citep{Teague.ea.2018} and fit a Gaussian line profile to each pixel with a \mbox{SNR}$\geq$5$\sigma$ in each line image.
Depending on the \mbox{SNR} of the spectral line, this method typically provides a radial velocity precision ($\sim$$\Delta {\rm v}_{\rm r}/\sqrt{\mbox{SNR}_{line}}$, where $\Delta {\rm v}_{\rm r} = c/R$) much higher than the nominal spectral resolution ($R$) of the instrument.

Radial velocity maps for \FeII at 26\,$\mu$m (see Figure\,\ref{fig:FeII_vmap}) and 17.9\,$\mu$m, \SIp, \FeI (see Figure\,\ref{fig:atomic_vmap} in Appendix\,\ref{sec:add_vmaps}), and H$_2$ 0-0\,S(1) and S(7) (see Figure\,\ref{fig:H2_vmap} in Appendix\,\ref{sec:add_vmaps}) lines were constructed.

Figure\,\ref{fig:FeII_vmap} shows the radial-velocity (${\rm v}_{\rm r}$) map of the \FeII at 26\,$\mu$m. The red and blue lobes of the jet are detected straddling the protostar's position as derived by ALMA~\citep[magenta circle; see][]{Lee.ea.2019}. Radial velocities have an average value of ${\rm v}_{\rm r}$= -25$\pm$5 and +25$\pm$5\,km\,s$^{-1}$ in the blue and red-shifted inner jet respectively (see Column\,2 in Table\,\ref{tab:FeII_kinematics}).
These values translate to a total velocity (${\rm v}_{\rm tot}$) of about 130$\pm$25\,km\,s$^{-1}$, assuming a jet inclination angle ($i$) of 11$\degr$ with respect the plane of the sky, as derived from the SiO analysis  ~\citep[see][]{Jhan&Lee2021}. At $\sim$20$\arcsec$ from the source radial velocity values increase up to $\sim$-30$\pm$5\,km\,s$^{-1}$ (i.\,e. ${\rm v}_{\rm tot}$=160$\pm$25\,km\,s$^{-1}$, assuming that $i$ is constant along the jet. As the jet shocks the ambient medium forming BS\,4, ${\rm v}_{\rm r}$ drops to $\sim$-20\,km\,s$^{-1}$. 

Indeed, the terminal bow-shocks have lower velocities, which might be real or just caused by slightly different jet inclination angles; ${\rm v}_{\rm r}(BS1)$=-20$\pm$5\,km\,s$^{-1}$, ${\rm v}_{\rm r}(BS2)$=-15$\pm$5\,km\,s$^{-1}$, and, at BS\,3, the radial velocity becomes slightly red-shifted (${\rm v}_{\rm r}$=5$\pm$5\,km\,s$^{-1}$\,km\,s$^{-1}$), possibly indicating that the direction of the flow has changed and the inclination with respect to the observer is larger than 90$\degr$ (see Column\,2 in Table\,\ref{tab:FeII_kinematics}).

We note that the position of such red-shifted emission is slightly offset with respect of BS\,3, namely $\sim$1$\arcsec$ north of the BS\,3 peak.
We also note a similar red-shifted radial velocity in the \FeII 17.9\,$\mu$m velocity map at the same position (see bottom panel of Fig.\,\ref{fig:atomic_vmap} in Appendix\,\ref{sec:add_vmaps}). However, both \SI (upper panel of Fig.\,\ref{fig:atomic_vmap} in Appendix\,\ref{sec:add_vmaps}) and H$_2$ (Fig.\,\ref{fig:H2_vmap}) velocity maps do not show any red-shifted emission at this location (\FeI emission at BS\,3 is below 5$\sigma$), indicating a different geometry for those species or a different origin for the \FeII red-shifted emission. This might otherwise suggest the presence of a second, independent flow, traced by the \FeII in the BS\,3 region, as a N-S crossing flow was possibly detected in the NIRCam images of \citet{Ray.ea.2023}.

Overall, both \SI and \FeI velocity maps show, within the uncertainties, radial velocities similar to those in the \FeII maps (see Fig\,\ref{fig:atomic_vmap} and Fig.\,\ref{fig:FeII_vmap}).
The H$_2$ lines (see Fig.\,\ref{fig:H2_vmap}) have a similar behavior, but show radial velocities $\sim$10\,km\,s$^{-1}$ lower than the atomic species (see Column\,2 and 3 of Table\,\ref{tab:H2_kinematics}). Overall, gas traced by the H$_2$ 0-0\,S(1) line moves at slightly lower speed than that traced by the 0-0\,S(7) line (see Columns 2 and 3 of Tab.\,\ref{tab:H2_kinematics}).
Jet radial velocities of the 0-0\,S(7) line range between  $\sim$14 and $\sim$20\,km\,s$^{-1}$ (but the same velocity within the error bar) and then ${\rm v}_{\rm r}$ drops to $\sim$8--12\,km\,s$^{-1}$ at the terminal bow-shocks. On the other hand, the molecular outflow shows radial velocities lower than the molecular jet (see Fig.\,\ref{fig:H2_vmap}), ranging from -10 to -3$\pm$5\,km\,s$^{-1}$, and the poorly collimated (wind) emission close to source has radial velocities $\leq$-5\,km\,s$^{-1}$ (see Fig.\,\ref{fig:H2_vmap}). 

Assuming that the H$_2$ 0-0\,S(7) line ($E_{\rm up}$=7197\,K) is tracing the same gas and velocities as the 1-0\,S(1) line ($E_{\rm up}$=6956\,K), we can derive the flow inclination at different positions by combining the tangential velocities (${\rm v}_{\rm tg}$), measured in \citet{Ray.ea.2023} (reported in Column\,4 of Table\,\ref{tab:H2_kinematics}), and the radial velocities measured here. The inclination angle with respect to the plane of the sky ranges from 9$\fdg$8$\pm$1$\fdg$5 to 12$\degr$$\pm$1$\degr$ along the jet (Column\,7 of Table\,\ref{tab:H2_kinematics}), and the weighted mean is 11.$\degr$6$\pm$0.$\degr$6, which perfectly matches the inclination value from SiO~\citep{Jhan&Lee2021}, confirming that our previous assumption on $i$ was correct. On the other hand, the terminal bow-shocks have a much larger spread in $i$, ranging from 5$\fdg$5$\pm$4$\fdg$4 to 19$\fdg$4$\pm$2$\fdg$1 (see BS\,1 to 4 in Column\,7 of Table\,\ref{tab:H2_kinematics}). These differences are not unexpected, as they reflect the larger precession angle measured in the outer bow-shocks (see Sect.\,\ref{sec:maps}).

We can also infer the total velocity of each feature for the 0-0\,S(7) line from both inclination and 1-0\,S(1) tangential velocities (Column\,5 of Table\,\ref{tab:H2_kinematics}). For those knots where no ${\rm v}_{\rm tg}$ nor $i$ are available, ${\rm v}_{\rm r}$ and an average value of $i$=11$\degr$ are assumed. The corresponding uncertainties are therefore larger. Similarly, we compute total velocities for the 0-0\,S(1) line (Column\,4 of Table\,\ref{tab:H2_kinematics}). Given the small inclination of the flow, within the uncertainties, ${\rm v}_{\rm tot}$ of the H$_2$ 0-0\,S(7) line is the same as ${\rm v}_{\rm tg}$ inferred in \citet{Ray.ea.2023} (see Columns 5 and 4 of Tab.\,\ref{tab:H2_kinematics}). Total velocities inferred from the 0-0\,S(1) line are similar or slightly smaller than those inferred for the H$_2$ 0-0\,S(7) line (see Columns 6 and 5 of Tab.\,\ref{tab:H2_kinematics}).

An interesting feature, detected in the atomic species (\FeII at 26\,$\mu$m, \SIp, and \FeI) along the inner jet (within $\sim$10$\arcsec$ from the source position) is a a mirror symmetry in radial velocities between the two sides of both blue- and red-shifted jet (see Fig.\,\ref{fig:FeII_vmap} and Fig\,\ref{fig:atomic_vmap}). The bottom side (towards SE) of the blue-shifted jet has an average ${\rm v}_{\rm r}$ of -40$\pm$10\,km\,s$^{-1}$, whereas the top side (towards NW) has ${\rm v}_{\rm r}$ of -15$\pm$10\,km\,s$^{-1}$. Conversely, the top side (NW) of the red-shifted jet has ${\rm v}_{\rm r}$$\sim$40$\pm$10\,km\,s$^{-1}$ and bottom side (SE) of 15$\pm$10\,km\,s$^{-1}$. Although the differences are per-se small ($\Delta {\rm v}_{\rm r}$=25$\pm$15\,km\,s$^{-1}$) and almost within the uncertainties, they may well be significant because they are detected in all the three maps and the shifts in the red- and blue-shifted lobes are reversed. One possible explanation is that we are detecting jet rotation (counterclockwise, i.e. the bottom side is approaching and the top is receding) from a few hundred to a several thousand au from the source. Alternatively, jet precession  ~\citep{Cerqueira.ea.2006}, asymmetrical jet shocks  ~\citep{DeColle.ea.2016}, or the, less likely, presence of a twin jet  ~\citep[e.\,g.][]{Soker.ea.2022}
could also explain the observed shifts.
We note a similar velocity gradient (1.5$\pm$0.8\,km\,s$^{-1}$ at 30$\pm$15\,au from the jet axis) was detected by \citet{Lee.ea.2007} in SiO with the SMA. However, ALMA observations at similar resolution,  but with higher sensitivity, in \citet{Lee.ea.2018} could not find any clear rotation signal and instead found only an upper limit of $\sim$27\,au\,km\,s$^{-1}$ for the inferred jet speciﬁc angular-momentum. This upper limit is more than one order of magnitude smaller than what was found here, thus making it very unlikely we are detecting jet rotation.

\subsection{Molecular and atomic mass-flux rates along the jet}
\label{sec:mass-flux}

Using both physical and kinematic parameters derived 
from the H$_2$ and atomic lines, mass ejection rates along the jet can be inferred.

For the warm and hot H$_2$ components we assume that the jet has laminar flow across the observed pixels in each considered knot along the blue-shifted jet  ~\citep[see, e.\,g.,][]{Dionatos.ea.2010}:

\begin{equation}
\dot{M}_{\rm jet}(H_2)= 2\mu m_H\times(\overline{N}_{H_2} A)\times ({\rm v}_{\rm tg}/l_{tg}) 
\label{eq:mjetH2}
\end{equation}

where $\mu$ (=1.35) is the mean atomic weight, $m_H$ is the proton mass, $\overline{N}_{H2}$
is the H$_2$ column density (warm and hot, see Column\,3 and 4 in Tab.\ref{tab:H2_dynamics}, respectively) averaged over the knot emitting area ($A$; assumed circular and derived from the radii reported in Tab.\,\ref{tab:size}), ${\rm v}_{\rm tg}$ is the knot tangential velocity (see Tab.\,\ref{tab:H2_kinematics}) and $l_{tg}$ (see Column\,2 of Tab.\ref{tab:H2_dynamics}) is the measured knot cross section. Only values for knots along the blue-shifted jet are reported, as the H$_2$ column densities are poorly constrained in the red-shifted jet (see Fig.\,\ref{fig:Boltzmann_maps}).

\begin{table*}
\caption{H$_2$ column densities and mass flow along the HH\,211 flow for the warm (W) and hot (H) components}
\label{tab:H2_dynamics}
\begin{tabular}{lcccccccc}
\hline \hline 
Feature &  $l_{\rm tg}$ & $\overline{N}_{\rm H_2}(\rm W)$ & $\overline{N}_{\rm H_2}(\rm H)$ & $\dot{M}_{\rm H_2}(\rm W)$ & $\dot{M}_{\rm H_2}(\rm H)$ & $P(\rm W)$ & $\dot{P}(\rm W)$ & $\tau(\rm H_2)$ \\
 &  ($\arcsec$) & \multicolumn{2}{c}{(10$^{19}$ cm$^{-2}$)}  & \multicolumn{2}{c}{(10$^{-7}$ M$_{\odot}$\,yr$^{-1}$)} & (10$^{-4}$ M$_{\odot}$\,km\,s$^{-1})$ & (10$^{-5}$ M$_{\odot}$\,yr$^{-1}$\,km\,s$^{-1}$) & (yr)\\
\hline
Knot\,1 &   0.45  &  12$\pm$1 & 1.2$\pm$0.2 & 7$\pm$1 & 0.4$\pm$0.1  & 5$\pm$1 & 6$\pm$1 & 100$\pm$10 \\
Knot\,2 &   0.45  &  11$\pm$1 & 1.1$\pm$0.2 & 5.3$\pm$0.8 & 0.3$\pm$0.1  & 3.8$\pm$0.6 & 4$\pm$1 & 140$\pm$20 \\
Knot\,3 &    0.4  &  12$\pm$2 & 1.2$\pm$0.4 & 7$\pm$1  & 0.4$\pm$0.1  & 5$\pm$1 & 6$\pm$1 & 200$\pm$30 \\
Knot\,4 &    0.45  &  15$\pm$2 & 1.8$\pm$0.3 & 8$\pm$2 & 0.5$\pm$0.2  & 6$\pm$1 & 6$\pm$2 & 270$\pm$40 \\
\hline
\end{tabular}
\end{table*}

Mass-flux rates for the warm and hot H$_2$ components are reported in Column 5 and 6 of Table\,\ref{tab:H2_dynamics}, respectively. Derived $\dot{M}_{H_2}(W)$ range from 5 to 8$\times$10$^{-7}$\,M$_\odot$\,yr\,$^{-1}$, whereas $\dot{M}_{H_2}(H)$ are about one order of magnitude smaller, given the lower column density and (typically) smaller emitting size of the hot component. Overall, our mass flux rates are typically two or three times smaller than those found by \cite{Dionatos.ea.2010} with {\it Spitzer}, likely due to the poor spatial resolution of the {\it Spitzer}/IRS modules, whereas both column densities and velocities are similar.
No mass-loss rates are derived for the external bow-shocks, because they would likely be overestimated, as part of the observed material is made of entrained gas from the ISM.

Assuming that both blue- and red-shifted jet are symmetric (i.\,e. red-shifted knots have same mass-flux), we infer a mass-ejection rate along the jet of $\dot{M}_{\rm jet}(H_2)$=1--1.6$\times$10$^{-6}$\,M$_\odot$\,yr\,$^{-1}$. These values perfectly match those derived from the SiO and CO jet by \citet{Jhan&Lee2021} (1.1$\times$10$^{-6}$\,M$_\odot$\,yr\,$^{-1}$ with $v$=100\,km\,s$^{-1}$) and \citet{Lee.ea.2010} (1.8$\times$10$^{-6}$\,M$_\odot$\,yr\,$^{-1}$ with $v$=170\,km\,s$^{-1}$).
Using the mass accretion rate inferred by \citet{Lee.ea.2010} from $L_{\rm bol}$ ($\dot{M}_{\rm acc}$=8.5$\times$10$^{-6}$), $\dot{M}_{\rm jet}$/$\dot{M}_{\rm acc}$ varies from 12\% to 19\%, consistently with what was found with ALMA. Such a high $\dot{M}_{\rm jet}$/$\dot{M}_{\rm acc}$ efficiency is predicted by MHD disk-winds at the protostellar stage  ~\citep[see, e.\,g.,][]{Ferreira.ea.2006}.

As column density, mass flux, and velocities are known, other important dynamical properties of the flow can be retrieved.
In particular, momentum ($P= M \times {\rm v}_{\rm tot}$) and momentum flux ($\dot{P}=\dot{M}\times {\rm v}_{\rm tot}$) of the molecular component can be inferred and compared with those of the atomic component, to examine whether the molecular jet is made of entrained material or it belongs to the jet, being launched from the disk. Column\,7, 8, and 9 of Table\,\ref{tab:H2_dynamics} report momentum, momentum flux, and dynamical time ($\tau$) for the warm H$_2$ component of the analysed knots.

To compare molecular and atomic dynamics, the mass-loss rates of the various knots ($\dot{M}_{\rm knot}$) were derived from the physical and kinematic parameters inferred in Sects\,\ref{sec: atomic params} and \ref{sec: kinematics}. 
$\dot{M}_{\rm knot}$ can be expressed in terms of the pre-shock density ($n_0$), speed of the gas entering the shock, and jet radius ($r_j$, as inferred in Table\,\ref{tab:size}):

\begin{equation}
\dot{M}_{\rm knot}(\rm atomic)= \mu m_H \times n_0 \times \pi r_j^2 \times {\rm v}_{\rm tot}
\label{eq:mjetatomic}
\end{equation}

As $n_0$ was obtained from \SIp, ${\rm r}_{\rm j}$ values are also taken from \SI in Table\,\ref{tab:size}. In any event, this assumption should not change our results as \SI and \FeII have the same ${\rm r}_{\rm j}$ values within the error bars (see also Fig.\,\ref{fig:size knots}).

The average jet velocity is assumed to be ${\rm v}_{\rm tot}$=130$\pm$25\,km\,s$^{-1}$ (see also Sec.\,\ref{sec: kinematics})\footnote{Note that, at variance with Eq.~\ref{eq:mjetH2}, radius and velocity in Eq.~\ref{eq:mjetatomic} are de-projected.}. Namely, as \SI and \FeII radial velocities are similar, knot radial velocities from Table\,\ref{tab:FeII_kinematics} are converted to velocities assuming a jet average inclination angle of 11$\degr$.

\begin{figure}
        \includegraphics[width=0.49\textwidth]{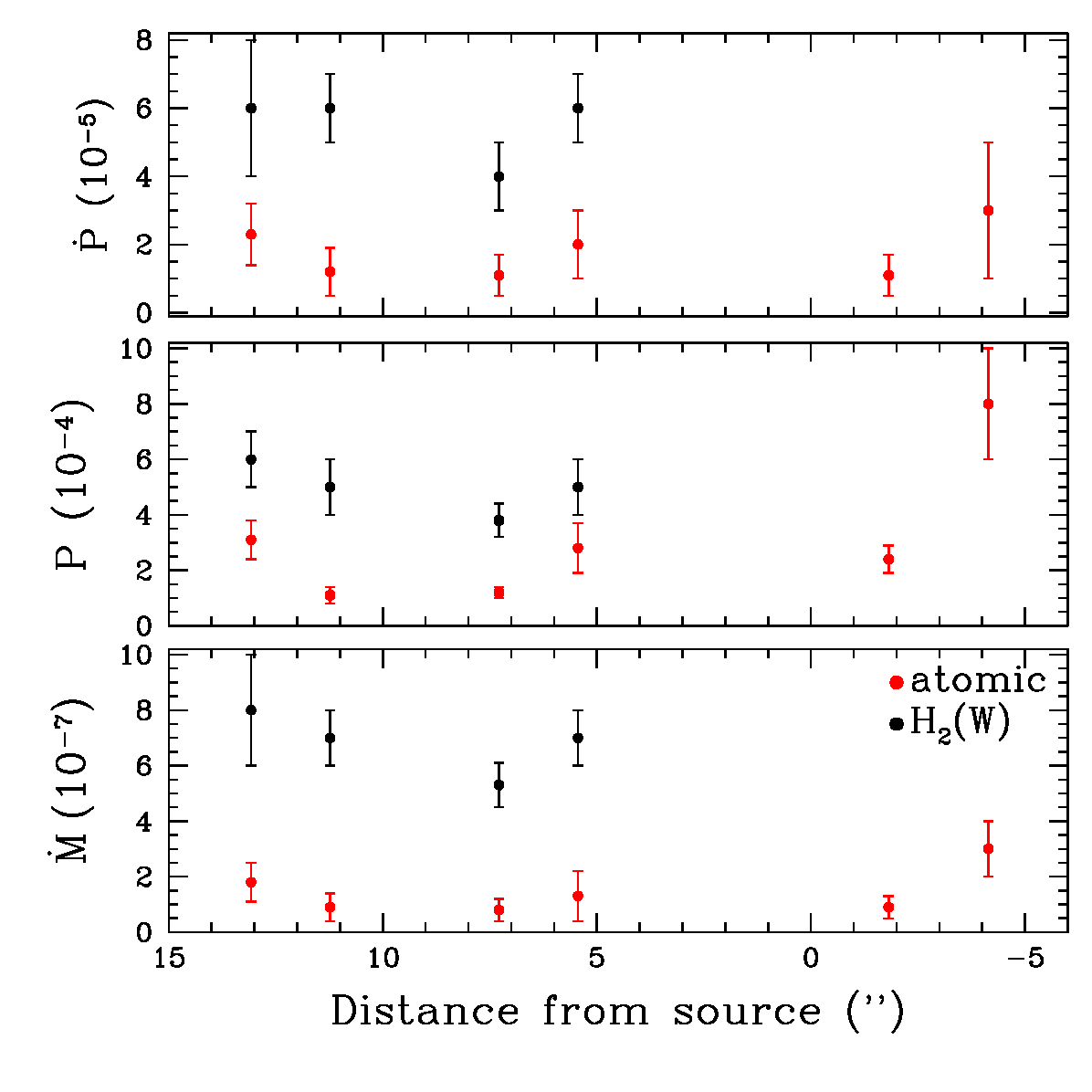}
    \caption{Comparison between atomic (red dots) and warm H$_2$ molecular (H$_2$(W), black dots) dynamical parameters for different knots of the HH\,211 jet (see Tab.\,\ref{tab:jet_atomic_model} and Tab.\,\ref{tab:H2_dynamics}). Bottom, middle, and top panels display mass-flux rates (in 10$^{-7}$\,M$_{\odot}$\,yr$^{-1}$), momenta (in 10$^{-4}$ M$_{\odot}$\,km\,s$^{-1}$), and momentum fluxes (in 10$^{-5}$ M$_{\odot}$\,yr$^{-1}$\,km\,s$^{-1}$), respectively.}
    \label{fig:jet_dynamics}
\end{figure}

As in the case of H$_2$, $\dot{M}$, $P$, $\dot{P}$, and $\tau$ (Columns 5, 6, 7, and 8 of Tab.\,\ref{tab:H2_dynamics}, respectively) are computed for the atomic component. $\dot{M}$, $P$, and $\dot{P}$ values of the atomic (red dots) and warm H$_2$ component (black dots) components of each knot are plotted in Figure\,\ref{fig:jet_dynamics} (bottom, middle, and top panel, respectively) as a function of the distance to the source. It is clear that the dynamical parameters of the atomic jet are always smaller than those of the warm H$_2$ component (H$_2$(W)), indicating that most of the thrust of jet derives from its molecular component. Most importantly, this confirms that the molecular jet is not entrained but originates from the disk.

\subsection{HH\,211: a dusty flow}
\label{sec:dusty}

Beyond 10\,$\mu$m, MIRI-MRS maps spatially resolve continuum emission at the three terminal bow-shocks (BS\,1--BS\,3). Their position matches the brightest \FeII emission (at 26\,$\mu$m), with the bulk of emission at BS\,3 (\mbox{SNR}$\geq$50\,$\sigma$ or $\geq$200\,MJ\,sr$^{-1}$ at $\lambda \geq$25\,$\mu$m, see contours in Fig.\,\ref{fig:MIRI-tricolor}). More faint continuum emission (detected beyond 24--25\,$\mu$m with  \mbox{SNR}$\geq$5\,$\sigma$ or $\geq$20\,MJ\,sr$^{-1}$) is also observed along the jet at Knot\,4 and along the counter-jet close to the protostar (Knot\,1R and Knot\,2R; see contours in Fig.\,\ref{fig:MIRI-tricolor}). Much fainter continuum emission (\mbox{SNR}$\geq$3\,$\sigma$ or $\geq$12\,MJ\,sr$^{-1}$) is also marginally detected at Knot\,2, 3 and at BS\,4 (see contours in Fig.\,\ref{fig:MIRI-tricolor}).

The shape of such continuum emission is prominent in the BS\,3 spectrum of Fig.\,\ref{fig:BS3_spec}, where a rising continuum, roughly peaking around 25--26\,$\mu$m, is detected under the strong emission lines. This shape can be fitted with a modified black-body emission, which provides a temperature of $\sim$90\,K. By fitting the continuum spectral energy distribution (SED), we derive a mid-IR luminosity for BS\,3 ($L_{\rm MIR}$(BS3)) of 0.0035\,L$_\sun$ and a total bolometric luminosity ($L_{\rm bol}$(BS3)) of 0.009\,L$_\sun$. Assuming optically thin dust, a dust emissivity spectral index ($\beta)$ equal to 1.8~\citep[typical of star forming regions; see, e.\,g.][]{Schnee.ea.2010}, and adopting a dust mass opacity coefficient k($\lambda$)=(850\,$\mu$m/$\lambda$)$^\beta$$\times$ k(850\,$\mu$m)~\citep{Millard.ea.2020}, where k(850\,$\mu$m) = 0.077\,m$^2$\,kg$^{-1}$  ~\citep{Dunne.ea.2000}, we get a very rough estimate of the dust mass in BS\,3 of $M_{\rm dust}\sim$0.044\,M$_\oplus$. For BS\,1 dust emission, the second brightest spot, we infer $L_{\rm bol}$(BS1)=0.0023\,L$_\sun$ ($L_{\rm MIR}$=0.0009\,L$_\sun$) and $M_{\rm dust}\sim$0.011\,M$_\oplus$. As the continuum flux in BS\,2 is almost one order of magnitude fainter than in BS\,1, we infer $L_{\rm bol}$(BS2)=0.0003\,L$_\sun$ ($L_{\rm MIR}$=0.00012\,L$_\sun$) and $M_{\rm dust}\sim$0.005\,M$_\oplus$, and in Knot R\,1 $M_{\rm dust}\sim$0.001\,M$_\oplus$. For BS\,1 and Knot R\,1 we assume that temperatures and SED shapes are the same as in BS\,3 and BS\,1, as the shape of the continuum emission is to faint to be properly fitted.
It is worth stressing that the reported $M_{\rm dust}$ values are probably lower limits, as dust is unlikely to be optically thin in the MIR and $k(\lambda)$ largely depends on the size of the dust particles (here unknown).

We note that such continuum emission was already observed towards the BS\,1--BS\,3 region with {\it Spitzer}/IRS by \citet{Tappe.ea.2008} (see their Fig.\,4), but it was not spatially resolved, due to the lower resolution of {\it Spitzer}. \citet{Tappe.ea.2008} fitted the continuum by thermal dust emission at a temperature of $\sim$85\,K.
The detection of continuum emission in the inner jet indicates that a large quantity of dust grains are present along the flow and thermally heated, but not fully destroyed, by the shocks. This dust is likely lifted from the protostellar disk, and, apparently, can survive the transport along outflows and jets as also seen at sub-millimetric wavelengths in a sample of Class\,0 YSOs~\citep[see][]{Cacciapuoti.ea.2024}.

The proposed scenario is also supported by the low gas-phase iron abundance along the HH\,211 flow, the detection of other atomic species with low ionisation-potential (namely \FeIp, \ClI and \SIp, see bottom and middle panels of Figure\,\ref{fig:atomic-maps} and bottom panel of Fig.\,\ref{fig:all_maps}), the very low ionisation fraction along the flow, as well as the low $T_e$ and $n_e$ values. 

While we cannot quantify how much of the dust observed in the external bow-shocks originates from entrained dust coming from the circumstellar or interstellar environment, the continuum emission detected in the inner jet must come from dust lifted from the disk, likely by a disk-wind. Assuming that the inferred dust mass is correct, our conclusions are also supported by the dust-to-gas mass ratio derived for Knot R\,1, which is less than 10$^{-3}$, similar to what expected from MHD disk-wind model predictions~\citep[see, e.\,g.,][]{Giaccalone.ea.2019,Franz.ea.2020,Rodenkirch.ea.2022}. On the other hand, this ratio is much higher ($\sim$10$^{-2}$) towards BS\,3 and BS\,1, similarly to the expected dust-to-gas mass ratio in the ISM ($\sim$0.01), confirming that most of the dust at the external bow-shocks has an ISM origin.

\section{Discussion}
\label{sec:discussion}

MIRI-MRS observations have revealed the fine structure of a low-mass protostellar flow, its chemistry, physics and dynamics. As HH\,211 is one of the best studied protostellar outflows, from optical to millimetric wavelengths, we can use ours and previous observations to draw the most up-to-date and complete picture of a low-mass protostellar flow. 

\subsection{A textbook case of a protostellar outflow}
\label{sec:textbook}

\begin{figure}
        \includegraphics[width=0.49\textwidth]{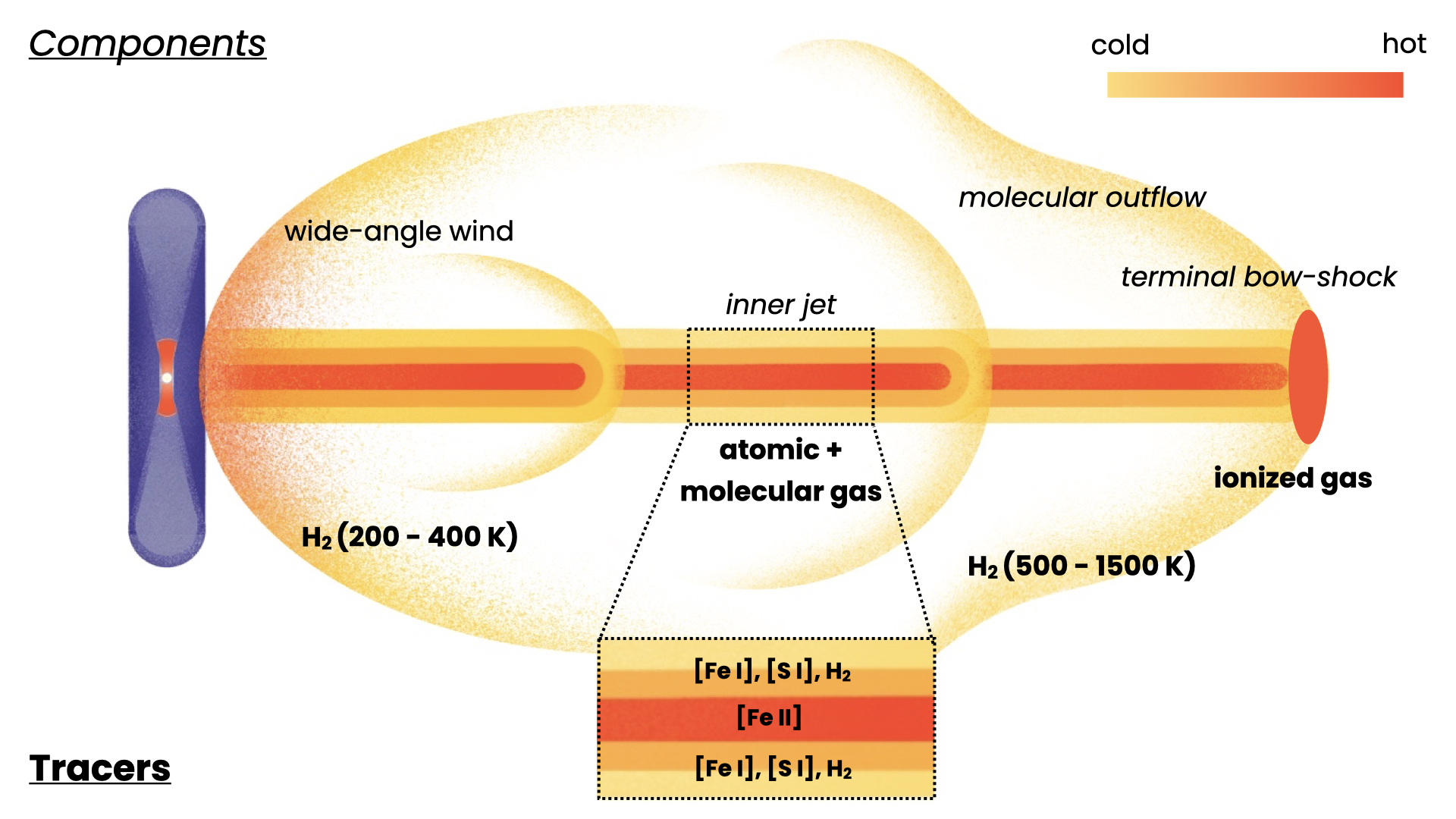}
    \caption{Not-to-scale schematic cartoon displaying our findings on the HH\,211 protostellar jet and outflow. Different outflow components, line tracers, and temperature gradients (not-to-scale) are labelled.}
    \label{fig:cartoon}
\end{figure}

HH\,211 can be considered a textbook case of a Class\,0 protostellar flow.
Indeed, our observations reveal all the typical jet/outflow structures: a poorly collimated molecular wind, traced by cold H$_2$ at 200--400\,K; a stratified (i.\,e. onion-like structured) jet, traced by atomic and molecular gas; large terminal bow-shocks, that sweep up the circumstellar and interstellar medium, forming a large (less collimated) molecular outflow, that is traced by the warm/cold H$_2$ (500--1500\,K).
A not-to-scale schematic sketch of our results is shown in Figure\,\ref{fig:cartoon}, which depicts the main outflow components, line tracers and temperature gradients observed.

The jet onion-like radial structure is probably one of the most striking features observed in HH\,211. We observe jet stratification in size, velocity, temperature, and chemistry, as predicted for jets originating from MHD disk-winds~\citep[see, e.\,g.,][]{Panoglou.Cabrit.ea2012,Pascucci.ea.2023}. 
In extended MHD disk-wind models, the wind has an “onion-like” kinematic and thermo-chemical structure, with streamlines launched from larger disk radii having lower velocity, temperature, and ionization, as well as a higher H$_2$ abundance ~~\citep[see, e.\,g.,][]{Panoglou.Cabrit.ea2012,Wang.ea.2019, Pascucci.ea.2023}.
According to MHD disk-wind theory, such thermo-chemical gradients arise from a radially extended ($\sim$0.1--20\,au) disk-wind. As Keplerian velocities, chemical composition, density, and physical properties change across the disk radius, this would naturally produce an onion-like structured jet with different layers. Therefore, the different features observed are related to each other. 

The jet's innermost atomic component of HH\,211 is produced by the fast flow (${\rm v}_{\rm tot}\sim$130\,km\,s$^{-1}$), likely ejected from the inner gaseous disk. The inner atomic jet has electron temperatures of $\sim$800--2000\,K, very low electron densities $\sim$100--400\,cm$^{-3}$, high pre-shock densities up to a few 10$^5$\,cm$^{-3}$, and very low ionisation fraction ($\lesssim$10$^{-3}$).
The most ionised region, traced by \FeII emission at higher excitation, is the core of the atomic jet, possibly surrounded by \SI and \FeI at lower excitation energies. Unfortunately, MIRI does not have enough spectral and/or spatial resolution to differentiate these two possible components, which, in our data, have similar size and velocity.

The molecular jet layer is made of H$_2$ that also shows a stratification in size, temperature and column density: the warm/hot component at higher temperature (1000--1500\,K) and lower column density (10$^{18}$--10$^{19}$\,cm$^{-2}$) has a varying radius of 60--130\,au and thus positioned in between the atomic and the cold/warm component at lower temperature (400--1000\,K) and higher column density (10$^{19}$--10$^{20}$\,cm$^{-2}$) and radii of 100--180\,au.

One could argue that SiO, and the fast CO jet components, detected at sub-millimetric and millimetric wavelengths~\citep[see, e\,g.,][]{Gueth.Guilloteau1999,Lee.ea.2007,Lee.ea.2018}, possibly arise from the H$_2$ warm/hot region, on the basis of their observed velocities~\citep[$\sim$100\,km\,s$^{-1}$; see][]{Jhan.ea.2016,Jhan&Lee2021}, and temperatures~\citep[$T_{CO}=$250--950\,K and $T_{SiO}>$250\,K see][, respectively]{Giannini.ea.2001,Nisini.ea.2002} similarly to the warm/hot H$_2$ component (see Tab.\,\ref{tab:H2_kinematics} and Fig.\,\ref{fig:Boltzmann_maps}).

Another important feature predicted by disk-wind models is dust removal from the disk and the presence of dust along the flows~\citep[see][]{Panoglou.Cabrit.ea2012}. As disk winds extend well beyond the dust sublimation radius, they can lift dusty particles up to $\sim$1\,$\mu$m in size~\citep{Booth.ea.2021}, that could be seen along the flow. Indeed, this is likely what we are observing in HH\,211. We note that dust has also been recently detected in the environment of other Class\,0 protostars with ALMA~\citep[see][]{Cacciapuoti.ea.2024},
but here it is also detected in a protostellar jet. Future MIRI/MRS observations from this and other programmes will tell whether this is a common feature in protostellar jets, previously unseen with {\it Spitzer}, or rather it can be ascribed to the youth of this particular flow.


Our observations indicate significant differences in the dynamical properties of the jet components as seen from their onion-like structure. 
We note that the mass-ejection rates measured from the warm H$_2$ component of the jet ($\dot{M}_{\rm jet}(\rm H_2)$=1--1.6$\times$10$^{-6}$\,M$_\odot$\,yr\,$^{-1}$), match very well those from SiO and fast CO (1.1--1.8$\times$10$^{-6}$\,M$_\odot$\,yr\,$^{-1}$; see Sect.\,\ref{sec:mass-flux}), showing that the warm molecular component is the primary mover of the outflow. In contrast, the mass-flux rate, momentum and momentum flux derived from the atomic component are up to one order of magnitude smaller than those inferred from the warm H$_2$ component (see Tab.\,\ref{tab:jet_atomic_model} and \ref{tab:H2_dynamics}). This confirms that the molecular jet is the dynamically most important component of the outflow contrary to the more evolved Class\,II jets, where the atomic component of the jet is the most important and drives the outflow~\citep[see, e.\,g.][]{Ray.ea.2007}. A similar result for young outflows was reported by \citet{Nisini.ea.2015,Sperling.ea.2021} who investigated [OI]\,63\,$\mu$m emission using SOFIA.




\subsection{An onion-like flow launched by a disk-wind?}
\label{sec:disk-wind}

An interesting feature seen along the jet and depicted in Figure\,\ref{fig:size knots} is the increase in the jet radius with distance to the source, evident for the atomic component, as expected in MHD wind models~\citep[e.\,g.][and references therein]{Pascucci.ea.2023}.
Despite the large uncertainties in the radius estimate of the different components (see Fig.\,\ref{fig:size knots}), the atomic jet seems to show a larger opening angle with respect to the H$_2$ components, which appear to be almost flat and surely collimated beyond 1000--1500\,au in Figure\,\ref{fig:size knots}. In MHD wind models it should be the other way round, namely the opening angle of the atomic component should be narrower than that of the molecular one. These results might indicate that the H$_2$ emission is tracing extended structures (likely small bow-shocks rather than the jet) that are not fully spatially resolved (see, e.\,g. the shape of Knot\,4 and Knot\,3 in Fig.\,\ref{fig:MIRI-footprint} and upper panel of Fig.\,\ref{fig:H2-maps}),
or that the molecular jet recollimates at smaller distances from the source.

In any case, fitting the jet radius as a function of the distance to the source just provides a very crude approximation of the jet launching region (in X-winds; or the launching inner radius - $r_{in}$ - in extended disk-winds).
To get a better estimate of $r_{in}$, we can use the observed jet velocity, assuming that it is equal to the asymptotic speed of the jet poloidal velocity~\citep[see, e.g.,][]{Ferreira.ea.2006,Panoglou.ea.2012}.
In this case, we can express the launching foot point ($r_{0}$) as a function of the magnetic lever arm ($\lambda$), stellar mass ($M_*$) and jet velocity (${\rm v_{tot}}$):

\begin{equation}
\label{eq:r_0}
    r_{0}=(2\lambda-3) GM_*/{\rm v_{tot}}^2
\end{equation}

If the jet is ejected from a very small region (e.\,g., in X-winds and very compact disk-winds) $\lambda \simeq \dot{M}_{\rm acc}$/$\dot{M}_{\rm jet}$~\citep[see, e.\,g.,][]{Pelletier.ea.1992,Pascucci.ea.2023}.
From Sect.~\ref{sec:mass-flux}, we estimated $\dot{M}_{\rm jet}$/$\dot{M}_{\rm out}\sim$0.12--0.19, thus we infer $\lambda\sim$5--8, and we get $r_{0}\sim$0.03-0.05\,au from Eq~\ref{eq:r_0}.
This region is well within the dust sublimation radius ($\gtrsim$0.1\,au) of a low-mass Class\,0 protostar~\citep[e.\,g.,][]{Lee2020}. 
This is in contrast with the detection of both dust and H$_2$ emission along the jet of HH\,211.
It is then more likely that we are dealing with an extended disk-wind and thus $r_{0}\sim r_{in}$ (i.\,e. $r_{0}$ matches the innermost radius of the jet launching region).

It is not possible to properly derive $r_{out}$ (i\,e. the most external launching radius) from the data at our disposal, unless a disk-wind model is used~\citep[see][]{Tabone.Cabrit.ea2020,Lee.ea.2021}.
As the radius of the toroid/disk in HH\,211 is $\sim$20\,au~\citep[][]{Lee.ea.2019}, we can assume this as an upper limit for $r_{out}$.  

In the past few years, ALMA observations at high-angular resolution have shown small-scale poorly-collimated rotating molecular winds, launched from the disks of YSOs~\citep[e\,g., HH\,212, DG\, Tau\,B, L1448-mm; see][]{Lee.ea.2021,deValon.Dougados.ea2020,Nazari.ea.2024}. When both jet and molecular disk-wind ejection rates can be estimated, $\dot{M}_{\rm wind}$ is about one order of magnitude higher than $\dot{M}_{\rm jet}$, suggesting that it is the slow-velocity disk-wind that removes much of the material from the disk rather than the jet itself~\citep[i.\,e. the re-collimated part of the disk-wind; see][and discussion therein]{Pascucci.ea.2023}.

As we detect slow-moving, low density, poorly-collimated warm H$_2$ emission down to $\sim$2$\arcsec$ ($\sim$650\,au) from the source, we can assume that this H$_2$ is tracing a poorly collimated disk-wind, rather than being circumstellar material swept-up by the jet.
From the 0-0\,S(1) velocity map, we infer radial velocities in the range of 4--5($\pm$5)\,km\,s$^{-1}$. We cannot adopt an inclination angle of 11$\degr$ as the wind is not collimated. However, we can assume a typical wind angle of $\sim$30$\degr$ with respect of the disk plane, therefore $\sim$\,40$\degr$ of inclination, and the wind velocity will be 6--8($\pm$8)\,km\,s$^{-1}$. The emitting radius is about 4 times larger than the jet radius, and the column density is 3--6$\times$10$^{19}$\,cm$^{-2}$, namely 2--4 times lower than that of the jet. Therefore, the inferred disk-wind mass ejection rate ($\dot{M}_{\rm DW}$) is of the same order of $\dot{M}_{\rm H_2}(\rm W)$, but can not be much larger as measured in other sources\,\citep[see][and references therein]{Pascucci.ea.2023}. However, it is worth noting that we are measuring the wind at 650-1000\,au from the driving source, as the visual extinction is too high to detect any emission closer to it. Therefore our $\dot{M}_{\rm DW}$ estimate is likely a lower limit of its real value.

\subsection{Excitation conditions along the flow}
\label{sec:excitation}

Besides the onion-like structure discussed in Sect.~\ref{sec:textbook} and \ref{sec:disk-wind}, we also detect different physical conditions along the flow, moving from the inner jet to the outer bow-shocks, in both molecular and atomic emission. 

One of the most striking features in the H$_2$ maps of Fig.\,\ref{fig:Boltzmann_maps} is the increase (by a factor of 2--3) in temperature, moving from the inner jet to the outer bow-shocks as well as the symmetric decrease (by a factor of 2--3) in column density in both components. The atomic component shows a similar trend in $T_e$, increasing by a factor of 2--3 from the jet to the bow-shocks. On the other hand, the pre-shock density slightly decreases and $n_e$ slightly increases, and, overall, $x_e$ increases from $\sim$10$^{-3}$ to $\sim$10$^{-2}$ moving along the flow (see Tab.\,\ref{tab:neTe} and \ref{tab:jet_atomic_model}). 
As the fast jet shocks the ambient medium at the four external bow-shocks, the flow velocity drops, and radial velocity gradients are detected in both molecular and atomic lines (see Figures\,\ref{fig:FeII_vmap}, \ref{fig:atomic_vmap}, \ref{fig:H2_vmap}, and Tables~\ref{tab:FeII_kinematics} and \ref{tab:H2_kinematics}). 

This change in the physical excitation conditions along the flow is also seen in its chemistry, namely in the diversity of the spectra observed in the terminal bow-shocks in contrast to the spectra seen along the jet. 

For comparison, the jet emission is mostly characterised by atomic forbidden lines with low excitation energies ($E_{\rm up}\lesssim$2700\,K) and low ionisation potentials ($I.P\leq$7.9\,eV; see Table\,\ref{table:atomic_lines}). In particular, it is the first time that \FeI is detected along a protostellar jet, and its coexistence with [FeII] is very indicative of the low-ionisation of the flow. Besides relatively faint HD emission, the only molecular emission along the jet is H$_2$ (the main cooler along the whole flow) with transitions at relatively low excitation energies ($E_{\rm up}\lesssim$7200\,K; see Tab.\,\ref{tab:h2_lines}).

In contrast to the jet, the emission at the terminal bow-shocks (BS\,1--BS\,3) shows atomic forbidden lines at much higher excitation energies ($E_{\rm up}\lesssim$14\,600\,K), and ionisation potentials ($I.P\leq$23.81\,eV). Notably, \FeI emission mostly or completely disappear and \SIII is just about detected, indicating that the atomic species are much more ionised here. 
It is also worth noting that \SI at 25\,$\mu$m is still extremely bright, having similar $I.P.$ and $E_{\rm up}$ of \FeI line at 24\,$\mu$m. This is due to the fact that, as predicted in $J$-shock models~\citep[e.\,g.][]{Hollenbach.McKee1989}, the \SI line intensity is 1--2 orders of magnitude higher than that of \FeI for pre-shock densities of $n_0\sim$10$^4$\,cm$^{-3}$~\citep[see Fig.\,7 in][]{Hollenbach.McKee1989}. Nevertheless, \SIIp optical emission has been observed at the bow-shocks~\citep[see][]{Walawender.ea.2005,Walawender.ea.2006}, which provides further evidence that the atomic component is partially ionised.  

Moreover, the terminal bow-shocks (BS\,1--BS\,3) display an extremely rich molecular emission.
Along with HD and H$_2$ transitions at high excitation energies~\citep[$E_{\rm up}$ up to 16\,000\,K; however NIRSpec spectra reveal $v=5$ transitions with $E_{\rm up}$ up to $\sim$30\,000\,K;][]{Ray.ea.2023}, our MIRI spectra show bright CO (1-0) emission from the high-J P-branch at 2000--3000\,K~\citep[see also NIRSpec spectra in][]{Ray.ea.2023}, OH, H$_2$O, CO$_2$, and HCO$^+$.
In particular, the detection of suprathermal OH rotational emission, as well as \HI emission in BS\,3, namely H$\alpha$~\citep[][]{Walawender.ea.2005,Walawender.ea.2006} and Pa$\alpha$ in the NIRSpec spectra (but no Br$\alpha$; Ray priv. comm.), indicates that a UV radiation field, produced by a relatively fast $J$-shock, must be present to photodissociate H$_2$O by Ly$\alpha$ radiation. Notably, the inferred shock velocity at the brightest bow-shock (BS\,3, ${\rm v_s}$=45$\pm$5\,km\,s$^{-1}$; see Tab.\,\ref{tab:jet_atomic_model}) matches well the shock velocity predicted by \citet{Tabone.Cabrit.ea2020} 
(50\,km\,s$^{-1}$) to reproduce the suprathermal OH rotational emissions in HH\,211 terminal bow-shocks. 
Intermediate velocity (25--60\,km\,s$^{-1}$), stationary, weakly magnetised, $J$-type, molecular shocks are also able to reproduce many other molecular (e.\,g., H$_2$, HCO$^+$, CO, CO$_2$, H$_2$O) as well as atomic features observed in HH\,211~\citep[e.\,g., \HIp, \SIp, \SII, \OIp, \FeIIp; see][]{Lehmann.ea.2020,Lehmann.ea.2022}. 
However, the \citet{Lehmann.ea.2020} fiducial shock-model (${\rm v_s}$=40\,km\,s$^{-1}$, $n_{\rm H}$=10$^4$\,cm$^{-3}$)
provides electron temperatures of about one order of magnitude higher than what we measured in BS\,3 (10$^4$ vs 10$^3$\,K), suggesting that we are not exactly probing the same regions of the shock with the \FeIIp.

The overall picture of the HH\,211 blue-lobe indicates that a relatively fast-moving, mostly molecular and weakly-shocked jet is impinging the ambient medium, generating moderate $J$-shocks at the terminal bow-shocks.
Unfortunately, only upper limits of the jet shock-velocity (${\rm v_s}<$60\,km\,s$^{-1}$) are available from our analysis. Nevertheless, assuming that the shock velocity broadens the line profile, we can compare the \mbox{FWHM} of the \FeII 5.3\,$\mu$m line along the jet and at the terminal bow-shocks and check whether the line is narrower along the jet. At the shortest wavelengths of MIRI, the line width is spectroscopically marginally resolved at the bow-shocks, whereas the H$_2$ lines are not resolved. We measure \mbox{FWHM}(\FeIIp)=19$\pm$0.9\,\AA, and \mbox{FWHM}(H$_2$)=17$\pm$0.1\,\AA, which provides a deconvolved \mbox{FWHM} of 8.5$\pm$0.9\,\AA~(i.\,e. $\Delta$v=48$\pm$5\,km\,s$^{-1}$). On the other hand, the \FeII line is not spectrally resolved along the jet, namely at Knot\,5 and Knot\,2, where it is detected at relatively high \mbox{SNR}, as its \mbox{FWHM} matches that of the H$_2$ 0-0\,S(7), indicating that the shock velocity along the jet is smaller than at BS\,3.




\subsection{Comparing HH\,211 flow properties with other Class\,0 and more evolved flows}

Spatially resolved flows in Class\,0, I, and II sources share many features in common, including collimated jets, wider winds, and outflows, implying that the accretion/ejection mechanisms are similar over a wide range of stellar masses and over the whole star formation process.
However, a closer look shows that physical, chemical, and dynamical properties of the flow change as the central source evolves from the protostellar to the pre-main sequence phase. 
In particular, it is worth noting that the jet's atomic component becomes dominant moving from the protostellar to the Class\,II stage, when the molecular component of the jet mostly or totally disappears, being fully dissociated, and the molecular emission is mostly or fully relegated to winds and entrained outflows. Jets become gradually faster (from 100\,km\,s$^{-1}$ in Class\,0 to 300-400\,km\,s$^{-1}$ in Class\,II) and less dense (from 10$^5$-10$^6$ to 10$^3$-10$^4$\,cm$^{-3}$), the gas excitation conditions along the flow increase, showing a higher ionisation fraction (from 10$^{-3}$-10$^{-2}$ to 0.2-0.5) and electron temperature (from 10$^2$-10$^3$ to 10$^4$\,K)~\citep[see, e.\,g.][]{Ray.ea.2007,Nisini.ea.2015,Podio.Tabone.ea2021}. At the same time, mass-flux rates decrease (from 10$^{-5}$-10$^{-6}$ to 10$^{-8}$-10$^{-10}$\,M$_\odot$\,yr$^{-1}$), as less matter is available in the disk to be ejected or accreted onto the forming star.

While the overall picture becomes more and more clear, as the studied sample of YSOs increases in number, still, flows driven by the same class of objects show significant differences in terms of chemistry, excitation conditions, and morphology.
Although some differences might be related to the different mass of the accreting source~\citep[see, e.\,g.,][]{Federman.ea.2024}, different ages and binarity of the central protostar~\citep{Tychoniec.ea.2024}, or differences in the accretion and ejection rates, similar objects at the same evolutionary stage still present variations.
Indeed, as different JWST programmes are providing us with new and exciting mid-IR data of spatially and spectrally resolved protostellar jets~\citep[see, e\.g.][]{Yang.ea.2022,Narang.ea.2024,Federman.ea.2024,Tychoniec.ea.2024,Nisini.ea.2024,Rubinstein.ea.2023}, we have now the chance to study and compare sources with similar mass and age, using similar tracers in the mid-IR regime.

For example, we can now compare the flow properties of HH\,211 with those of IRAS\,16253-2429~\citep{Narang.ea.2024}, a Class\,0 low-luminosity protostar ($L_{\rm bol}$=0.2\,L$_\sun$, $M_*$=0.12--0.17\,M$_\sun$), observed with both MIRI and NIRSpec. Despite the similar mass and evolutionary stage of these protostars, there are several striking differences between the two flows.
The IRAS\,16253-2429 jet is more atomic than molecular and more ionised than HH\,211, as no \FeI emission is detected, \SI emission is confined to close to the source, and \NeII and Br$\alpha$ emission are well detected along the flow~\citep[see Fig.\,1, 2, and 3 in][]{Narang.ea.2024}.
The pre-shock density is two orders of magnitude lower~\citep[2$\times$10$^3$\,cm$^{-3}$;][]{Narang.ea.2024}, shock velocities are similar or slightly higher ($\sim$54\,km\,s$^{-1}$) than in HH\,211, jet velocities of the \FeII component are similar or higher (${\rm v_{tot}}$=169$\pm$15\,km\,s$^{-1}$), mass-flux rates are extremely low~\citep[$\sim$0.4--1.1$\times$10$^{-10}$\,M$_\odot$\,yr$^{-1}$;][]{Narang.ea.2024} with respect to what measured in HH\,211. This is very likely connected to IRAS 16253-2429 having a very low mass-accretion rate~($\dot{M}_{\rm acc}$=2.4$\pm$0.8$\times$10$^{-9}$\,M$_\odot$\,yr$^{-1}$; Watson et al. in prep.), and it is reflected in the one order of magnitude difference in bolometric luminosity between the two Class\,0 sources. 

On the other hand the size of the two atomic jets seem to be roughly comparable. The diameter of the IRAS\,16253-2429 jet at 400\,au from the source (i.\,e. at the largest distance measured there) is $\sim$60\,au~\citet{Narang.ea.2024}, whereas the HH\,211 jet width at Knot\,1R ($\sim$580\,au), the closest marginally-resolved knot, has a size of 90$\pm$30\,au. Unfortunately, further comparison is not possible, as the HH\,211 innermost jet emission (in \FeIIp) is not spatially resolved.

Overall, the disparity in accretion/ejection activity, reflected in the YSO bolometric luminosity, might be key to explain such observational differences. However, comparing the flow morphology and chemistry of HH\,211 with the outflows associated with B335 and HOPS\,153~\citep[Class\,0 YSOs, slightly more massive than HH\,211 but with similar $L_{\rm bol}$; see Table\,1 in][]{Federman.ea.2024} presented in \citet{Federman.ea.2024}, we still note significant differences in terms of their morphology and jet excitation. In particular, in B335 and HOPS\,153, \HI Br$\alpha$ emission from the jet is detected, H$_2$ emission is more confined towards the winds and outflow cavities and is faint along the jet~\citep[see Fig.\,3 and 4 in][]{Federman.ea.2024}. \citet{Hodapp.ea.2024} JWST/NIRCAM images of B335 reveal CO and H$_2$ emission in several knots along the jet.

While some morphological differences among HH\,211, B335 and HOPS\,153 can be ascribed to the different wavelength regimes (NIRSpec 2.8--5\,$\mu$m $vs.$ MIRI 5--28\,$\mu$m),  tracing somewhat different excitation conditions of the gas, nevertheless the excitation conditions appear to be dissimilar.

In contrast, by comparing the HH\,211 jet properties with those of Class\,0 sources with similar $L_{\rm bol}$, presented in \citet{Podio.ea.2021} (i.\,e. the CALYPSO IRAM-PdBI survey), we note that, in several cases (e.\,g. IRAS\,4A and 4B sources), jet width and dynamical properties (i.\,e. $\dot{M}_{\rm jet}$ and $\dot{P}_{\rm jet}$) of the HH\,211 warm  H$_2$ component are similar to those derived for SiO and CO high-velocity components~\citep[see Table\,3 and Fig.\,5 in][]{Podio.ea.2021}, despite observing different wavelength regimes. Nevertheless, although $\dot{M}_{\rm jet}$ is correlated to $L_{\rm bol}$, 
the CALYPSO survey shows significant dispersion (at least one order of magnitude) in the mass ejection rates of Class\,0 sources with similar $L_{\rm bol}$.

The cause of such a scattering is not yet clear.
Episodic accretion could be one of the possible explanations, and the presence of several knots along the flows surely hints to some variability.
Nevertheless, as $L_{\rm bol}$ of Class\,0 sources derives from their accretion luminosity ($L_{\rm acc}\sim G\dot{M}_{\rm acc}M_*/R_*$), the latter should be strictly correlated with their mass accretion 
$and ejection$
rates, and it is thus difficult to explain such large spreads. Of course one possible cause of such discrepancies could be that the $\dot{M}_{\rm jet}$/$\dot{M}_{\rm acc}$ efficiency varies from source to source depending, for example, on magnetic properties.  Alternatively our $M_*$/$R_*$ estimates are wrong, or the selected samples of Class\,0 YSOs include protostars with different ages and accretion rates.
As our sample of Class\,0 protostars observed with JWST becomes larger, we might have the chance to answer these questions.

\section{Conclusions}
\label{sec:conclusions}

We have presented JWST MIRI-MRS spectral maps (5--28\,$\mu$m) of the HH\,211 flow, covering the blue-shifted lobe, the central protostar, and a small portion of the red-shifted lobe.
The analysis of these maps provides an unprecedented view of the flow physics, chemistry, and excitation conditions, its kinematics and dynamical properties. The central protostar is not detected. Ancillary JWST NIRCam H$_2$ narrow-band images at 2.12 and 3.25\,$\mu$m (1-0\,S(1) and 1-0\,O(5) lines, respectively) provide a visual-extinction map of the whole flow and are used to deredden our images and spectra in the analysis. The visual extinction ranges from from 5--15\,mag at the terminal bow-shocks up to 20--60\,mag along the bipolar jet towards the central source. As the inner jet regions (within $\sim$5$\arcsec$ from the source) are not detected at 2.12\,$\mu$m, we use a value of 80\,mag, inferred from H$_2$ ro-vibration diagrams, to correct for visual extinction in these regions.
Our main conclusions are reported below.

\begin{itemize}
    \item The overall morphology of the flow consists of a highly collimated jet, both atomic (\FeIp, \FeIIp, \SIp, \NiIIp) and molecular (H$_2$, HD), that shocks the ambient medium producing several large bow-shocks, as well as driving a large molecular (H$_2$) outflow, mostly traced by H$_2$ 0-0 transitions at low-$J$. Further H$_2$ 0-0\,S(1) uncollimated emission is also detected down to 2$\arcsec$--3$\arcsec$ from the source, tracing a less collimated wind. 

    \item The inner jet, within $\sim$2.5$\arcsec$ from the source, is mostly traced by atomic emission, and the lack of H$_2$ emission is likely due to the large visual extinction (\Av$>$80\,mag) close to source. The \FeII emission at 26\,$\mu$m is detected down to $\sim$130\,au and 300\,au from the source on the jet red- and blue-shifted side, respectively.

    \item In contrast to the relatively few atomic and molecular species detected in the jet, the terminal bow-shocks (especially BS\,3) are very rich in chemistry with many forbidden atomic (\ClIp, \ClIIp, \ArIIp, \CoIIp, \NeIIp, \SIIIp) and molecular (CO, OH, H$_2$O, CO$_2$, HCO$^+$) species detected.
    In particular, suprathermal OH rotational emission (between 9.1 and 25\,$\mu$m), originating from water photodissociation (114--143\,nm UV radiation produced by strong - $\mathrm{v_s}\geq$40\,km\,s$^{-1}$ - shocks) is observed.
    
    \item Dust continuum-emission (at $T\sim$90\,K), roughly matching the \FeII spatial emission at 26\,$\mu$m, is detected at the terminal bow-shocks (BS\,1--3), and more faint emission in the blue- and red-shifted jet. Although some or most of the dust in the external bow-shocks likely originate from the circumstellar or interstellar environment, the continuum emission detected in the inner jet must arise from dust lifted from the disk. This is also supported by the dust-to-gas mass ratio ($\leq$10$^{-3}$) inferred along the jet.

    \item Line maps show that the HH\,211 jet is marginally resolved in diameter. The analysis shows an onion-like structure of the jet, with the atomic jet displaying a smaller size and the molecular component extends to larger radii. The different atomic lines show similar values in radius (ranging from $\sim$45 to $\sim$100\,au), and the H$_2$ lines have radii similar or larger than the atomic jet. The radius of the 0-0\,S(7) line ranges from $\sim$60 to $\sim$130\,au, whereas the 0-0\,S(1) line is positioned on the outer layers of the jet ($\sim$100--180\,au).
        
    \item Maps of the H$_2$ excitation conditions ($N_{\rm H_2}$, $T_{\rm H_2}$) along the flow are built. The H$_2$ emission shows a stratification in temperature, and two components, referred to as the warm (W) and hot (H) component, are fitted. $T_{\rm H_2}(\rm W)$ along the jet varies from $\sim$300\,K, close to source, to 500--700\,K reaching 900--1000\,K in the outer bow-shocks. Its column density changes from 10$^{19}$ to 10$^{20}$ \,cm$^{-2}$ along the jet, while it is only 10$^{19}$\,cm$^{-2}$ along the bow-shocks, except in BS\,3 ($\sim$3$\times$10$^{20}$\,cm$^{-2}$). $T_{\rm H_2}(\rm H)$ varies from 1000 to 2000\,K along the jet, whereas it is much higher (2000--3500\,K) at the terminal bow-shocks. Its column density is higher along the jet (1--2$\times$10$^{19}$\,cm$^{-2}$) and drops in the outer jet and bow-shocks (10$^{18}$--10$^{19}$\,cm$^{-2}$).

    \item The less collimated wind close to the source is colder (200--400\,K) and less dense (10$^{18}$--10$^{19}$\,cm$^{-2}$) than the jet. The outflow gas (i.\,e. the entrained gas) is less dense and colder than the jet and bow-shocks. 
    
    \item $T_e$ and $n_e$ are derived from \FeII line fluxes using an nLTE code. Moving away from the source, $T_e$ increases from $\sim$1000\,K to $\sim$1400\,K at BS\,4. The terminal bow-shocks show the highest electron temperatures (1800--3800\,K). A similar trend is seen for $n_e$, with low electron densities (150--400\,cm$^{-3}$) along the jet and at BS\,4, and higher values (350--800\,cm$^{-3}$) at the three terminal bow-shocks.
    
    \item By employing \citet{Hollenbach.McKee1989} $J$-shock models, \FeII gas-phase abundance along the jet is inferred. Iron is largely depleted along the flow and just a small amount is in gas-phase (between $\sim$2\% and $\sim$10\%), with abundances ranging from $\sim$7$\times$10$^{-7}$ to $\sim$3$\times$10$^{-6}$. Surprisingly, jet knots close to the source seem to be less depleted than those positioned further away from the source. We argue that this might be the result of different shock conditions along the jet.
    
    \item The \citet{Hollenbach.McKee1989} $J$-shock models also allow us to infer pre-shock density, shock velocity, and ionisation fraction along the flow. 
    $n_0$ varies from $\sim$1--2$\times$10$^5$\,cm$^{-3}$ in the inner jet to $\sim$0.4--0.9$\times$10$^5$\,cm$^{-3}$. Shock velocities at the terminal bow-shocks range from 35$\pm$5 to 45$\pm$5\,km\,s$^{-1}$, whereas only upper limits (${\rm v_s}<$45--60\,km\,s$^{-1}$) can be derived for the jet. 
    Ionisation fraction varies from 10$^{-3}$ along the jet to 10$^{-2}$ at the terminal bow-shocks.

    \item Radial velocities maps for both molecular (H$_2$) and atomic lines (\SIp, \FeIp, \FeIIp) are also constructed. Overall, the atomic lines show similar velocities. Average values of ${\rm v}_{\rm r}$= -25$\pm$5 and 25$\pm$5\,km\,s$^{-1}$ in the blue and red-shifted inner jet, which correspond to a total velocity of $v$= 130$\pm$25\,km\,s$^{-1}$, for a jet inclination of 11$\degr$. The radial velocity of the atomic component peaks at -30$\pm$5\,km\,s$^{-1}$ (${\rm v_{tot}}$=160$\pm$25\,km\,s$^{-1}$) in the outer jet and drops at the terminal bow-shocks. The kinematics of the H$_2$ shows a behaviour similar to the atomic lines, but has radial velocities $\sim$10\,km\,s$^{-1}$ lower than the atomic component. An average jet inclination angle (with respect to the plane of the sky) of 11$\fdg$6$\pm$0$\fdg$6  is inferred from H$_2$ radial and tangential velocities~\citep{Ray.ea.2023}. On the other hand, the terminal bow-shocks have a much larger spread in $i$, ranging from 5$\fdg$5$\pm$4$\fdg$4 to 19$\fdg$4$\pm$2$\fdg$1.

    \item Mass flux rate, momentum, and momentum flux of different knots of the jet are inferred for the H$_2$ (warm and hot) and atomic components. $\dot{M}$, $P$, and $\dot{P}$ values of the warm H$_2$ component are up to one order of magnitude higher than those inferred from the atomic component, whereas the hot  H$_2$ component shows jet dynamical values similar to the atomic component. Our findings indicate that the warm H$_2$ component is the primary mover of the outflow, namely it is the most significant dynamical component of the jet, in contrast to jets from more evolved YSOs.

\end{itemize}

Overall, our JWST MIRI/MRS maps of HH\,211 show a textbook case of a protostellar flow, revealing a collimated jet, with an onion-like structure, comprised of layers of different size, velocity, temperature, and chemical composition. The jet is dusty, possibly launched by a disk-wind, mostly molecular, and its warm H$_2$ molecular component is the primary mover of the  molecular outflow. The jet close to the source is surrounded by a U- or V-shape, less-dense and colder molecular wind. Excitation conditions also vary along the flow.
Further JWST observations of other young Class\,0 and Class\,I jets and outflows will show whether they share the same picture presented here for HH\,211 or alternatively that the flow properties markedly evolve with time, as early indications suggest.

\begin{acknowledgements}

We would like to thank the referee for their helpful suggestions as well as Marta Tychoniec for drawing the HH\,211 cartoon shown in Fig.\,16.

A.C.G. would like to thank David Hollenbach for helpful discussions and for sharing the HM89 code, as well as Juan Alcal\'a for his useful comments.

This work is based on observations made with the NASA/ESA/CSA James Webb Space Telescope. The data were obtained from the Mikulski Archive for Space Telescopes at the Space Telescope Science Institute, which is operated by the Association of Universities for Research in Astronomy, Inc., under NASA contract NAS 5-03127 for JWST. These observations are associated with program ID 1257.

The NIRCam observations presented here were made from the Guaranteed Time Allocation to MJM upon selection as an Interdisciplinary Scientists on the JWST Science Working Group in response to NASA AO-01-OSS-05 issued in 2001.

The following National and International Funding Agencies
funded and supported the MIRI development: NASA; ESA; Belgian Science
Policy Office (BELSPO); Centre Nationale d’Etudes Spatiales (CNES); Danish
National Space Centre; Deutsches Zentrum fur Luft- und Raumfahrt (DLR);
Enterprise Ireland; Ministerio De Economiá y Competividad; Netherlands Research
School for Astronomy (NOVA); Netherlands Organisation for Scientific
Research (NWO); Science and Technology Facilities Council; Swiss Space Office;
Swedish National Space Agency; and UK Space Agency.

A.C.G. acknowledges support from PRIN-MUR 2022 20228JPA3A “The path to star and planet formation in the JWST era (PATH)” funded by NextGeneration EU and by INAF-GoG 2022 “NIR-dark Accretion Outbursts in Massive Young stellar objects (NAOMY)” and Large Grant INAF 2022 “YSOs Outflows, Disks and Accretion: towards a global framework for the evolution of planet forming systems (YODA)”. T.P.R. acknowledges support from ERC grant 743029 EASY. 
H.B. acknowledges support from the Deutsche Forschungsgemeinschaft in the Collaborative Research Center (SFB 881) “The Milky Way System” (subproject B1). EvD, MvG, LF, KS, WR and HL acknowledge support from ERC Advanced grant 101019751 MOLDISK, TOP-1 grant 614.001.751 from the Dutch Research Council (NWO), the Netherlands Research School for Astronomy (NOVA), the Danish National Research Foundation through the Center of Excellence “InterCat” (DNRF150), and DFGgrant 325594231, FOR 2634/2. P.J.K. acknowledges financial support from the Science Foundation Ireland/Irish Research Council Pathway programme under
Grant Number 21/PATH-S/9360. K.J. acknowledges the support from the Swedish National Space Agency (SNSA).  G.P. gratefully acknowledges support from the Max Planck Society. 
This research has made use of NASA's Astrophysics Data System Bibliographic Services. This research made use of NumPy \citep{Harris.Millman.ea2020}; Astropy, a community-developed core Python package for Astronomy \citep{ AstropyCollaboration.Robitaille.ea2013,AstropyCollaboration.PriceWhelan.ea2018}; Matplotlib \citep{Hunter2007}; \texttt{pdrtpy} \citep{Kaufman.Wolfire.ea2006,Pound.Wolfire2008,Pound.Wolfire2011,Pound.Wolfire2023}

\end{acknowledgements}

\bibliographystyle{aa}
\bibliography{bibjoys.bib}

\appendix

\section{Additional MIRI-MRS maps}
\label{sec:add_maps}
\begin{figure}
    \centering
    \includegraphics[width=0.49\textwidth]{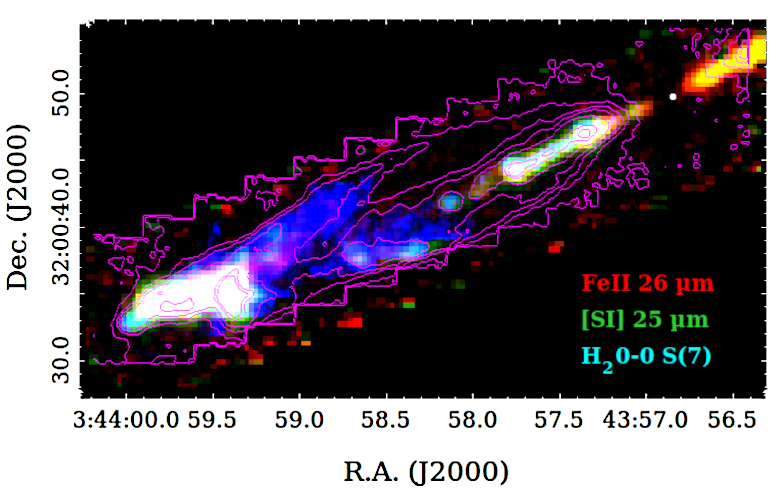}\\  
    \caption{Tricolour MIRI-MRS map of H$_2$\,0-0\,S(7) (at 5.5\,$\mu$m, in blue), \SI (at 25.2\,$\mu$m, in green), and \FeII (at 26\,$\mu$m, in red) emission lines. H$_2$\,0-0\,S(7) (at 17.0\,$\mu$m) magenta contours (at 3, 10, 20, 50, 100, 200, and 500\,$\sigma$; 1\,$\sigma$=40\,MJy\,sr$^{-1}$) are also overplotted.
    The white circle marks the position of the ALMA mm continuum source. }
    \label{fig:Tri-jet}
\end{figure}

Figure\,\ref{fig:Tri-jet} shows a tricolour MIRI-MRS map of H$_2$
of H$_2$\,0-0\,S(7) (at 5.5\,$\mu$m,in blue), \SI (at 25.2\,$\mu$m,in green), and \FeII (at 26\,$\mu$m, in red) emission lines. H$_2$ emission (0-0\,S(1) line at 17.0\,$\mu$m) is overplotted with magenta contours (at 3, 10, 20, 50, 100, 200, and 500\,$\sigma$; 1\,$\sigma$=40\,MJy\,sr$^{-1}$).

\begin{figure}
    \centering
    \includegraphics[width=0.47\textwidth]{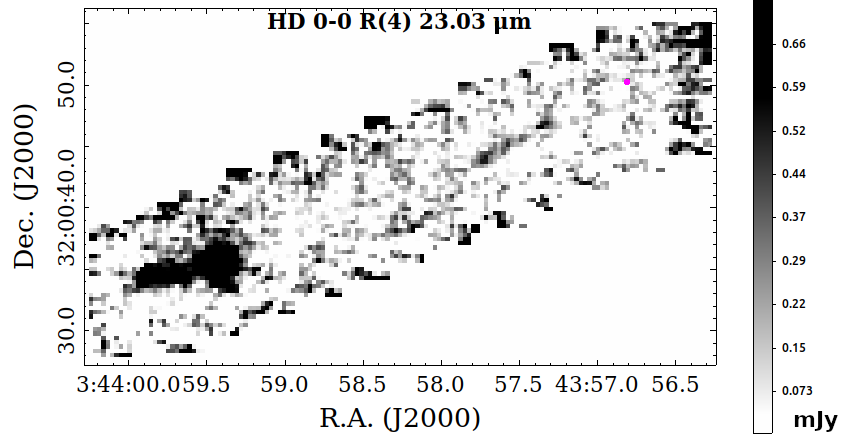}\\
     \includegraphics[width=0.47\textwidth]{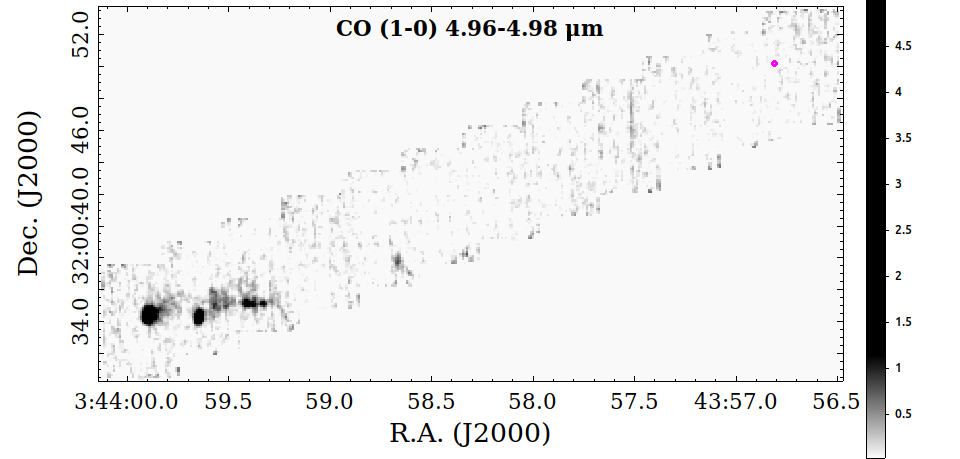}\\
     \includegraphics[width=0.47\textwidth]{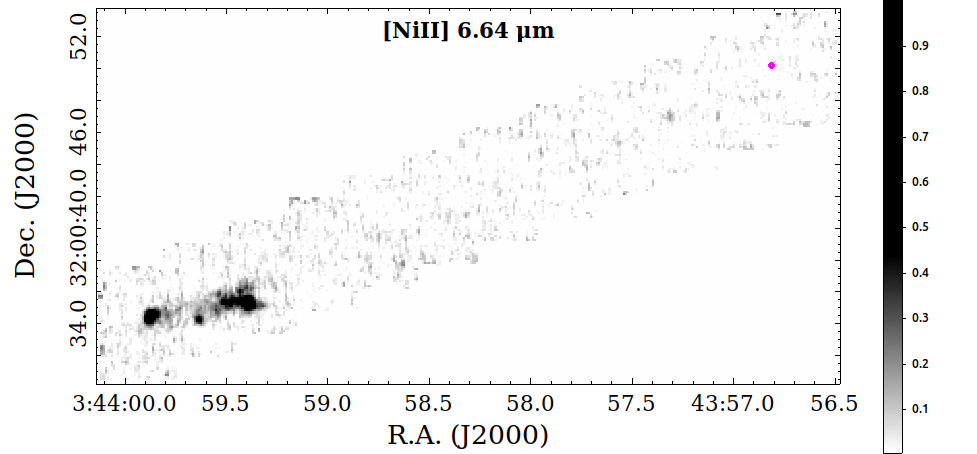}\\
     \includegraphics[width=0.47\textwidth]{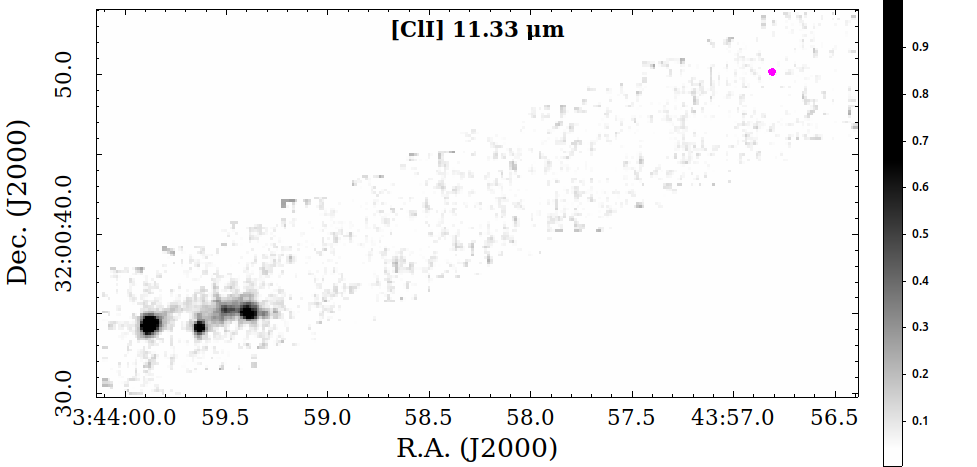}\\

    \caption{
   Additional line maps of some bright features detected along HH\,211. From top to bottom: HD\,0-0\,R(4) at 23.03\,$\mu$m, CO (1-0) lines from 4.96 to 4.98\,$\mu$m, \NiII at 6.64\,$\mu$m, \ClI at 11.33\,$\mu$m. The magenta circle shows the position of the ALMA mm continuum source. Integrated flux is in mJy\,pixel$^{-1}$. }
    \label{fig:all_maps}
\end{figure}

Figure\,\ref{fig:all_maps} presents the MIRI-MRS intensity integrated maps of the following transitions (from top to bottom): HD\,0-0\,R(4) at 23.03\,$\mu$m, CO (1-0) lines (from P29 to P31) from 4.96 to 4.98\,$\mu$m, \NiII at 6.64\,$\mu$m, \ClI at 11.33\,$\mu$m.

\section{$A_V$ estimate in the inner-jet region and Boltzmann plots}
\label{sec:Av_jet}

\begin{figure}
    \centering
    \includegraphics[width=0.49\textwidth]{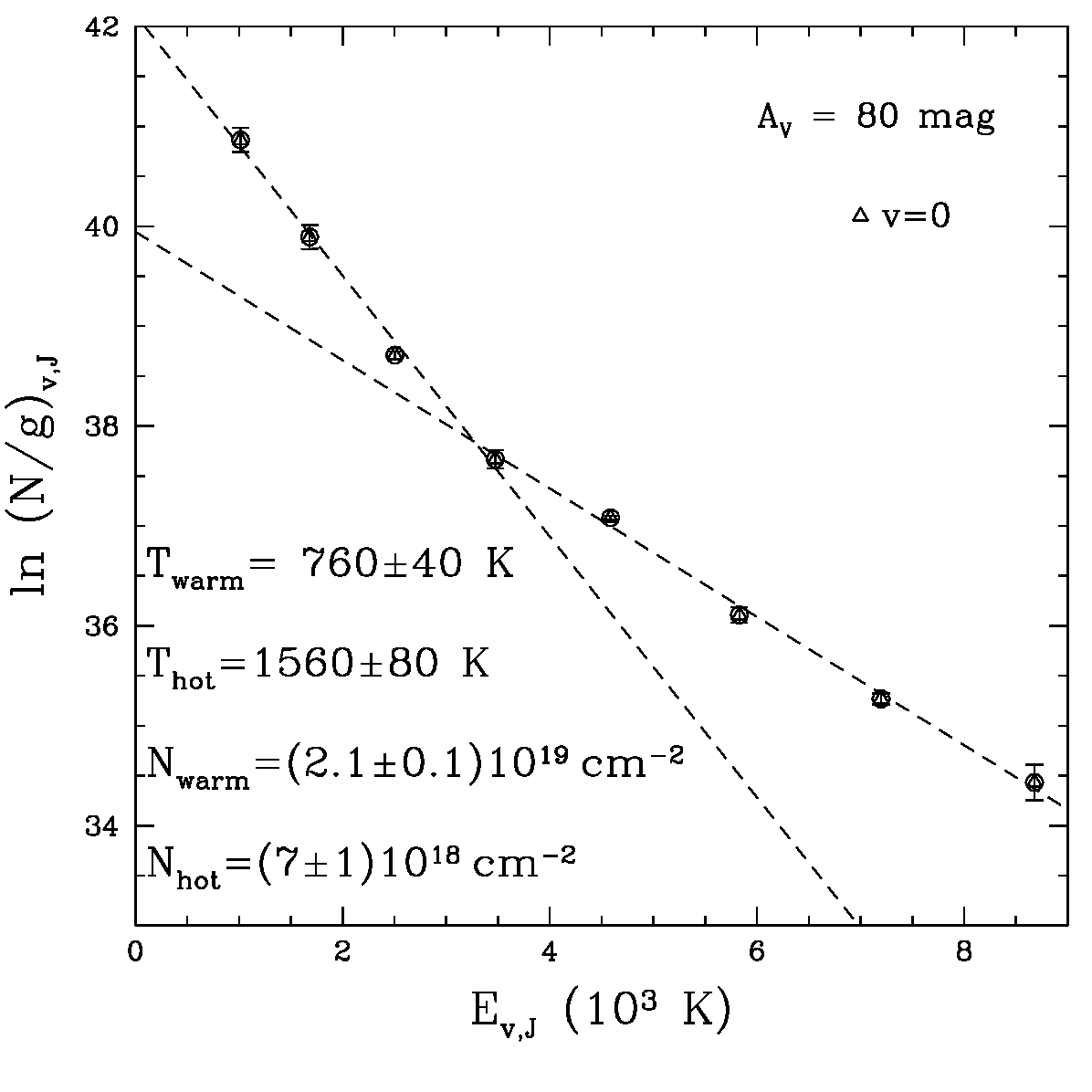}\\
    \caption{Ro-vibrational diagram towards the inner jet of HH\,211 (R.A.(J2000): $03^h43^m57.^s176$, Dec.(J2000): $ +32\degr00\arcmin48.\arcsec18$) at $\sim$4.$\arcsec$5 away from the  source. Two temperature components are fitted. \Av = 80\,mag  best fits the two gas components. }
    \label{fig:Av_jet}
\end{figure}

Figure\,\ref{fig:Av_jet} shows the H$_2$ ro-vibrational diagram for the blue-shifted inner jet extracted at R.A.(J2000): $03^h43^m57.^s176$, Dec.(J2000): $+32\degr00\arcmin48.\arcsec18$, namely about 4.$\arcsec$5 (towards the SE) away from the position of the mm ALMA source. Only pure rotation lines are detected and reported, as the the $v=1$ lines are too faint to be detected. Two temperature components (warm and hot) are visible and fitted. Values of $T({\rm H_2})$ and $N({\rm H_2})$ for both components are reported. Fits were obtained by varying \Av values, until the best fit (highest correlation coefficient) was derived. Inferred uncertainty in \Av is about 10\,mag.

\begin{figure}
    \centering
    \includegraphics[width=0.49\textwidth]{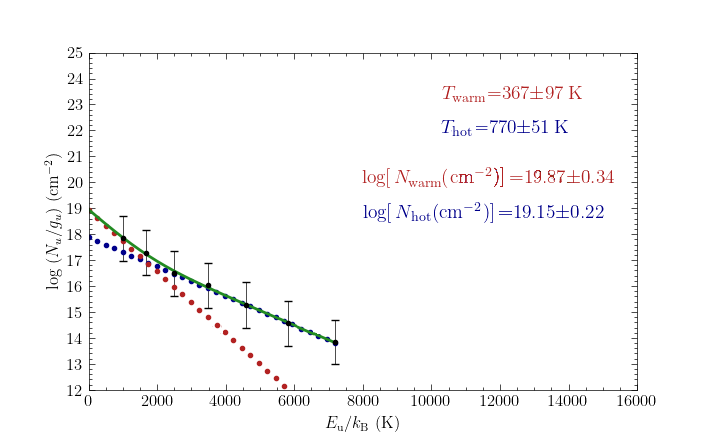}\\
    \includegraphics[width=0.49\textwidth]{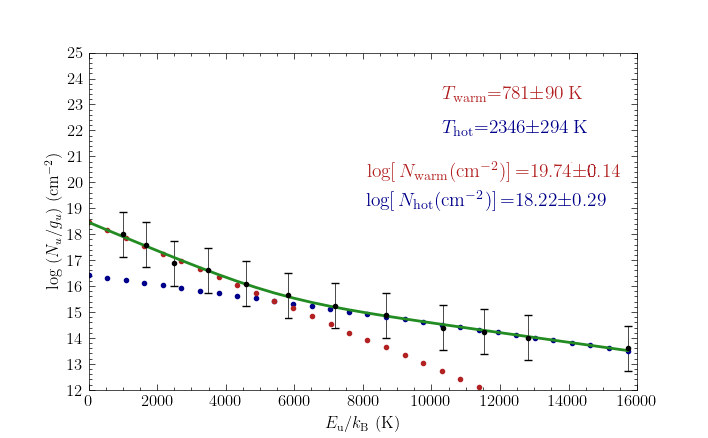}\\
    \caption{Examples of pixel-by-pixel ro-vibrational diagram towards HH\,211. Top panel: ro-vibrational diagram in the blue-shifted inner jet (R.A.(J2000): $03^h43^m57.^s469$, Dec.(J2000): $+32\degr00\arcmin47.\arcsec33$), where only $v=0$ lines (up to S(7)) are detected. In this case, temperature of both components is low and column density is high. 
    Bottom panel: ro-vibrational diagram towards the wing of BS\,3 (R.A.(J2000): $03^h43^m59.^s277$,  Dec.(J2000): $+32\degr00\arcmin35.\arcsec13$), where both $v=0$ and $v=1$ lines are detected. In this case, the temperature of both components is high and the column density is low.}
    \label{fig:RV_components}
\end{figure}

Figure\,\ref{fig:RV_components} show two examples of Boltzmann plots derived by our routine to compute the excitation maps. Spectra were extracted from a single spaxel in the blue-shifted inner jet (top panel) and in the wings of BS\,3 (bottom panel).

\section{Additional tables}
\label{sec:tables}

Table~\ref{tab:h2_lines} provides the list of identified H$_2$ transitions along with their theoretical wavelength (in $\mu$m), energy of the upper level (in K), and corresponding MIRI Channel/Grating.

Table~\ref{table:atomic_lines} provides a list of the forbidden atomic lines identified along the HH\,211 flow. The table also reports the ionisation potential of the species, transition ID, wavelength, excitation temperature of the upper level, the corresponding MIRI Channel/Grating and the outflow regions where the line was detected.

Table~\ref{tab:molecule_lines} lists the HD transitions detected as well as their wavelength, energy of the upper level, and corresponding MIRI Channel/Grating.

Table~\ref{tab:size} reports, for each identified knot and each species, its \mbox{FWHM} in arcseconds (measured orthogonally to the jet axis), the corresponding size (deconvolved for the \mbox{FWHM} of the corresponding $\mbox{PSF}$), and its radius in au.

As an example of the extracted spectra, Table~\ref{tab:line_fluxes_Knot2} lists identified lines, wavelengths, line fluxes and uncertainties in Knot\,2. Spectrum and identified lines are also shown in Fig.\,\ref{fig:knot2_spec}.

For each knot and bow-shock identified, Table~\ref{tab:FeII_kinematics} reports measured radial velocities and their uncertainties, along with the total velocities derived by assuming an average inclination angle of 11$\degr$ with respect  to the plane of the sky.

For each knot and bow-shock identified, Table~\ref{tab:H2_kinematics} reports the radial velocity of the H$_2$ 0-0\,S(1) and 0-0\,S(7) lines, the tangential velocity of the H$_2$ 1-0\,S(1) line as measured by \citet{Ray.ea.2023}, the total velocity of the hot and warm component, and the inclination angle of the feature with respect to the plane of the sky.

\begin{table*}
\centering
\caption{List of detected H$_2$ pure rotational lines in the jet and bow-shocks. }
\label{tab:h2_lines}
\begin{tabular}{lcccccccc}
\hline \hline
Line & $\lambda$ & $E_{\rm up}$& MIRI Channel/Grating \\
 & ($\mu$m) & (K) &   \\

&  &  & & \\
\hline
H$_2$\,(1-1) S(9) & 4.95409 & 15725 & ch1-SHORT \\
H$_2$\,(0-0) S(8) & 5.05312 & 8677  & ch1-SHORT  \\
H$_2$\,(1-1) S(8) & 5.33005 & 14220 & ch1-SHORT  \\
H$_2$\,(0-0) S(7) & 5.51116 & 7197  &  ch1-MEDIUM \\
H$_2$\,(1-1) S(7) & 5.81086 & 12817 & ch1-MEDIUM \\
H$_2$\,(0-0) S(6) & 6.10856 & 5830  &  ch1-MEDIUM \\
H$_2$\,(1-1) S(6) & 6.43835 & 11521 &  ch1-MEDIUM \\
H$_2$\,(0-0) S(5) & 6.90952 & 4586  &   ch1-LONG \\
H$_2$\,(1-1) S(5) & 7.28012 & 10341 &   ch1-LONG \\
H$_2$\,(0-0) S(4) & 8.02505 & 3474  & ch2-SHORT \\
H$_2$\,(1-1) S(4) & 8.45302 & 9286  & ch2-SHORT \\
H$_2$\,(0-0) S(3) & 9.66491 & 2504  &  ch2-MEDIUM \\
H$_2$\,(1-1) S(3) & 10.1778 & 8365  &  ch2-LONG \\
H$_2$\,(0-0) S(2) & 12.2786 & 1682  &  ch3-SHORT \\
H$_2$\,(1-1) S(2) & 12.9276 & 7584  &  ch3-SHORT \\
H$_2$\,(0-0) S(1) & 17.0348 & 1015  &  ch3-LONG \\
 \hline
\end{tabular}
\tablefoot{The H$_2$(1–1) S(1) line at 17.9323\,$\mu$m is blended with the bright \FeII transition at 17.92324\,$\mu$m and thus not detected.}\\
\end{table*}

\begin{table*}[ht]
\caption{List of detected atomic lines in the jet and bow-shocks.}
\begin{tabular}{lllcccc}
\hline
\hline
Ion & I.P.$^a$ & Line ID & $\lambda$ &  $E_{\rm up}$  & MIRI Channel/Grating & Region\\\relax
& (eV) & & ($\mu$m)  & (K)   & & line detected in \\
\hline
\FeI  & 0 &  a$^5$D$_3$-a$^5$D$_4$ & 24.04233 & 598.4$^*$ & ch4-MEDIUM & jet; BS\,1, 2, 3, and 4 $<$5$\sigma$ \\\relax
\FeII &  7.90 & a$^4$F$_{9/2}$-a$^6$D$_{9/2}$ & 5.3401693 & 2694.25&  ch1-SHORT & jet, BS\,1, 2, 3, 4\\
& &       a$^4$F$_{7/2}$-a$^6$D$_{5/2}$ & 5.6739070 & 3496.42&  ch1-SHORT & BS\,3$<$5$\sigma$\\
& &       a$^4$F$_{9/2}$-a$^6$D$_{7/2}$ & 6.721277 & 2694.25&   ch1-LONG & jet, BS\,1, 2, 3 $\leq$5$\sigma$\\
& &       a$^6$D$_{5/2}$-a$^6$D$_{9/2}$ & 14.977170 & 960.65 & ch3-MEDIUM & BS\,1, 2, 3 $\leq$5$\sigma$ \\
& &       a$^4$F$_{7/2}$-a$^4$F$_{9/2}$ & 17.936 & 3496.42&  ch3-LONG / ch4-SHORT & jet, BS\,1, 2, 3, 4\\
& &       a$^4$F$_{5/2}$-a$^4$F$_{7/2}$ & 24.5192 & 4083.22&  ch4-LONG & BS\,3\\
& &       a$^6$D$_{7/2}$-a$^6$D$_{9/2}$ & 25.988390 & 553.62$^*$ & ch4-LONG & jet, BS\,1, 2, 3, 4 \\\relax
\NiII & 7.64 & $^2$D$_{3/2}$-$^2$D$_{5/2}$ & 6.6360 & 2168.15   &  ch1-LONG & jet, BS\,1, 2, 3, 4\\
& &         $^4$F$_{7/2}$-$^4$F$_{9/2}$ & 10.6822 & 13423.83    &  ch2-LONG & BS\,1, 2, 3$<$5$\sigma$\\
& &       $^4$F$_{5/2}$-$^4$F$_{7/2}$ & 12.7288 & 14554.16  & ch3-SHORT  & BS\,1, 2, 3$<$5$\sigma$\\\relax
\ArII & 15.76 &  $^2$P$_{1/2}$-$^2$P$_{3/2}$ &  6.985274 & 2059.73$^*$ &  ch1-LONG & BS\,3$<$5$\sigma$\\\relax 
\CoII & 7.88 &a$^3$F$_{3}$-a$^3$F$_{4}$ & 10.522727 & 1367.30$^*$ & ch2-LONG & BS\,3$<$5$\sigma$\\\relax 
\ClI & 0 & $^2$P$_{1/2}$-$^2$P$_{3/2}$ & 11.333352 & 1269.51$^*$ & ch2-LONG & BS\,1, 2, 3, 4; jet$\leq$3$\sigma$\\\relax 
\ClII  & 23.81 & $^3$P$_{1}$-$^3$P$_{2}$ & 14.3678 & 1001.39$^*$ &  ch3-MEDIUM & BS\,3$<$5$\sigma$\\\relax 
\NeII & 21.56 & $^2$P$_{1/2}$-$^2$P$_{3/2}$ & 12.813548 & 1122.85$^*$ &  ch3-SHORT & BS\,3; BS\,1, 2, 4$<$5$\sigma$\\\relax 
\SI & 0 & $^3$P$_{1}$-$^3$P$_{2}$ & 25.2490 & 569.83$^*$ &  ch4-LONG & jet, BS\,1, 2, 3, 4\\\relax
[\ion{S}{iii}] & 23.24 & $^3$P$_{2}$-$^3$P$_{1}$ & 18.71303 & 1198.59 & ch4-SHORT & BS\,3; BS\,1 and BS\,2$<$5$\sigma$\\
 \hline
\end{tabular}

$^a$ Ionisation potential of the $X^{i-1}$ ion \\
$^*$ Fundamental transition to the ground state\\
\label{table:atomic_lines}
\end{table*}

\begin{table}
\centering
\caption{List of detected HD transitions. }
\label{tab:molecule_lines}
\begin{tabular}{lccc}
\hline \hline
Species & $\lambda$  & $E_{\rm up}$ & MIRI Channel/Grating \\
 & ($\mu$m) & (K)  &   \\
\hline
HD  0-0\,R(10)  &    11.57346	&    6657.4   &  ch2-LONG   \\
HD  0-0\,R(9)	&    12.47181	&    5503.7   &  ch3-SHORT  \\
HD  0-0\,R(8)	&    13.59265	&    4445.2   &  ch3-MEDIUM \\
HD  0-0\,R(7)	&    15.25104	&    3487.4   &  ch3-MEDIUM \\
HD  0-0\,R(6)	&    16.89381	&    2635.8   &  ch3-LONG   \\
HD  0-0\,R(5)	&    19.43100	&    1895.3   &  ch4-SHORT  \\
HD  0-0\,R(4)	&    23.03376	&    1270.7   &  ch4-MEDIUM \\
 \hline
\end{tabular}
\end{table}

\begin{table*}
\centering
\caption{Size of knots in the inner jet }
\label{tab:size}
\begin{tabular}{lccc}
\hline \hline
Feature & \mbox{FWHM}         & size    & radius \\
Name    & (\arcsec)      & (\arcsec)   & (au) \\
\hline\\
\SI $\mbox{PSF}$: 0.98\arcsec \\
\hline	              
Knot\,2 red  &	1.06   &  0.40$\pm$0.10	 & 64$\pm$16     \\
Knot\,1 red  &	1.03   &  0.32$\pm$0.10	 & 51$\pm$16    \\
Knot\,1      &	1.02   &  0.28$\pm$0.08	 & 45$\pm$13   \\
Knot\,2      &	1.04   &  0.35$\pm$0.08	 & 56$\pm$13     \\
Knot\,3      &	1.06   &  0.40$\pm$0.12	 & 64$\pm$20     \\
Knot\,4      &	1.11   &  0.52$\pm$0.10	 & 83$\pm$16     \\
\hline\\
\FeI $\mbox{PSF}$: 0.924\arcsec\\
\hline
Knot\,2 red  &  0.978   &  0.32$\pm$0.10	  & 51$\pm$16  	\\
Knot\,1 red  &  0.992   &  0.36$\pm$0.10 	  & 58$\pm$16    \\
Knot\,1      &  0.976   &  0.31$\pm$0.10	  & 50$\pm$16   \\
Knot\,2      &  0.977   &  0.32$\pm$0.10	  & 51$\pm$16     \\
Knot\,3      &  0.995	&  0.37$\pm$0.10	  & 59$\pm$16    	\\
Knot\,4      & 1.002    &  0.39$\pm$0.10	  & 62$\pm$16     \\
\hline\\
\FeII (17\,$\mu$m) $\mbox{PSF}$: 0.803\arcsec\\
\hline
Knot\,2 red  &  0.850   &  0.28$\pm$0.12     & 45$\pm$20     \\
Knot\,1 red  & $\cdots$ &  $\cdots$          & $\cdots$    \\
Knot\,1      &  0.900   &  0.40$\pm$0.12	  & 65$\pm$20   \\
Knot\,2      &  0.905   &  0.42$\pm$0.10	  & 67$\pm$16     \\
Knot\,3      &  0.913 	&  0.43$\pm$0.12     & 70$\pm$20     	\\
Knot\,4      &  1.03    &  0.64$\pm$0.10	  & 103$\pm$30     \\
\hline\\
\FeII (26\,$\mu$m) $\mbox{PSF}$: 0.98\arcsec\\
\hline
Knot\,2 red  & 1.02 & 0.29$\pm$0.08 	  & 47$\pm$12 	\\
Knot\,1 red  & 1.02 & 0.29$\pm$0.08 	  & 47$\pm$12    \\
Knot\,1      & 1.02 & 0.29$\pm$0.1	      & 47$\pm$16 \\
Knot\,2      & 1.04 & 0.35$\pm$0.08	      & 56$\pm$12 \\
Knot\,3      & 1.07 & 0.42$\pm$0.12 	  & 69$\pm$20    \\
Knot\,4      & 1.14 &  0.58$\pm$0.12      & 93$\pm$20    \\
\hline\\
H$_2$ 0-0\,S(1) $\mbox{PSF}$: 0.73\arcsec\\
\hline
Knot\,2 red  & 1.350 & 1.14$\pm$0.08 	  & 182$\pm$12 	\\
Knot\,1 red  & 0.98  & 0.65$\pm$0.18      & 104$\pm$29    \\
Knot\,1      & 1.00  & 0.7$\pm$0.2      & 112$\pm$32 \\
Knot\,2      & 1.289 & 1.06$\pm$0.06	  & 170$\pm$10 \\
Knot\,3      & 1.310 & 1.10$\pm$0.08 	  & 177$\pm$12    \\
Knot\,4      & 1.345 &  1.13$\pm$0.06     & 181$\pm$10    \\
\hline\\
H$_2$ 0-0\,S(7) $\mbox{PSF}$: 0.3\arcsec\\
\hline
Knot\,2 red  & 0.45 &  0.35$\pm$0.12 	  &  56$\pm$20 	\\
Knot\,1 red  & 0.48 &  0.37$\pm$0.12       & 60$\pm$20    \\
Knot\,1      & 0.73 &  0.67$\pm$0.1	       & 107$\pm$16 \\
Knot\,2      & 0.78 &  0.72$\pm$0.08	  & 115$\pm$12 \\
Knot\,3      & 0.76 &  0.70$\pm$0.08 	  & 112$\pm$12    \\
Knot\,4      & 0.86 &  0.80$\pm$0.06       & 129$\pm$10    \\
\hline
\end{tabular}
\end{table*}

\begin{table}
\centering
\caption{Detected lines and fluxes in Knot\,2. }
\label{tab:line_fluxes_Knot2}
\begin{tabular}{lcc}
\hline \hline
Line & $\lambda$ & $F\pm\Delta~F$ \\
 & ($\mu$m) & (10$^{-15}$erg\,cm$^{-2}$\,s$^{-1}$)    \\
&  &   \\
\hline
H$_2$\,(1-1) S(9) & 4.95  &  0.6$\pm$0.1  \\
H$_2$\,(0-0) S(8) & 5.05  &  5.48$\pm$0.08   \\
H$_2$\,(1-1) S(8) & 5.33  &  0.4$\pm$0.1  \\
H$_2$\,(0-0) S(7) & 5.51  &  25.3$\pm$0.1   \\
H$_2$\,(1-1) S(7) & 5.81  &  0.6$\pm$0.1   \\
H$_2$\,(0-0) S(6) & 6.11  &  11.70$\pm$0.08   \\
H$_2$\,(0-0) S(5) & 6.91  &  48.10$\pm$0.08   \\
H$_2$\,(1-1) S(5) & 7.28  &  0.3$\pm$0.1   \\
H$_2$\,(0-0) S(4) & 8.02  &  20.00$\pm$0.04  \\
H$_2$\,(0-0) S(3) & 9.66  &  28.90$\pm$0.03  \\
H$_2$\,(0-0) S(2) & 12.28 &  8.14$\pm$0.03   \\
H$_2$\,(0-0) S(1) & 17.03 &  7.44$\pm$0.02  \\
\hline\\[-7pt]
HD\,0-0 R(6) & 16.89 &  0.10$\pm$0.02 \\
HD\,0-0 R(5) & 19.43 &  0.19$\pm$0.03 \\
HD\,0-0 R(4) & 23.03 &  0.29$\pm$0.04 \\
\hline\\[-7pt]
CO$_2$\,v=2 & 14.96--14.99 & 0.16$\pm$0.05 \\
\hline\\[-7pt]
\FeII a$^4$F$_{9/2}$-a$^6$D$_{9/2}$ & 5.34 &  0.41$\pm$0.08  \\
\FeII a$^6$F$_{5/2}$-a$^6($F$_{9/2}$ & 14.98 &  0.09$\pm$0.03  \\
\FeII a$^4$F$_{7/2}$-a$^4$F$_{9/2}$ & 17.92 &  0.13$\pm$0.02  \\
\FeII a$^4$F$_{5/2}$-a$^4$F$_{7/2}$ & 25.99 &  5.38$\pm$0.07  \\
 \hline\\[-7pt]
\FeI  a$^5$D$_3$-a$^5$D$_4$  & 24.04 & 2.66$\pm$0.05  \\
\hline\\[-7pt] 
 \SI  $^3$P$_{1}$-$^3$P$_{2}$ & 25.25 & 13.50$\pm$0.07  \\
 \hline\\[-7pt]
\NiII $^2$D$_{3/2}$-$^2$D$_{5/2}$ & 6.63 & 0.39$\pm$0.09  \\
\hline
\end{tabular}
\end{table}

\begin{table}
\centering
\caption{\FeII radial velocity values along the HH\,211 flow}
\label{tab:FeII_kinematics}
\begin{tabular}{lcc}
\hline \hline
Feature & ${\rm v}_{\rm r}$ & ${\rm v}_{\rm tot}$ \\
 & \multicolumn{2}{c}{(km\,s$^{-1}$)}  \\
\hline
Knot\,2 red &  25$\pm$5 &  130$\pm$25       \\
Knot\,1 red &  25$\pm$5  & 130$\pm$25      \\
Knot\,1     &   -25$\pm$5   &  130$\pm$25    \\
Knot\,2     &    -25$\pm$5   &  130$\pm$25 \\
Knot\,3     &    -25$\pm$5   &  130$\pm$25 \\
Knot\,4     &    -30$\pm$5   &  160$\pm$25\\
Knot\,5     &    -20$\pm$5   &  105$\pm$25\\ 
BS\,4       &     -20$\pm$5   &  105$\pm$5 \\ 
BS\,3       &     5$\pm$5    &  25$\pm$25\\ 
BS\,2       &     -15$\pm$5   &  80$\pm$25 \\
BS\,1       &     -20$\pm$5   &  105$\pm$25  \\
 \hline
\end{tabular}
\end{table}

\begin{table*}
\centering
\caption{H$_2$ velocities, inclination along the HH\,211 flow }
\label{tab:H2_kinematics}
\begin{tabular}{lcccccc}
\hline \hline
Feature & ${\rm v}_{\rm r} (0-0\,S(1))$ & ${\rm v}_{\rm r} (0-0\,S(7))$ & ${\rm v}_{\rm tg} (1-0\,S(1))^a$ & ${\rm v}_{\rm tot}(\rm H)^b$ & ${\rm v}_{\rm tot}(\rm W)^c$ & $i^d$ \\
 & (km\,s$^{-1}$) & (km\,s$^{-1}$) & (km\,s$^{-1}$)  & (km\,s$^{-1}$) &  (km\,s$^{-1}$) & ($\degr$)  \\
\hline
Knot\,2R &    12$\pm$6  &    17$\pm$5  &  $\cdots$  &  89$\pm$26$^e$  & 63$\pm$31$^e$  & $\cdots$	\\
Knot\,1R &    10$\pm$6  &    18$\pm$7  &   $\cdots$ &  94$\pm$34$^e$  & 52$\pm$31$^e$  & $\cdots$		\\
Knot\,1 &    -11$\pm$6  &   -20$\pm$6  &  $\cdots$  &  105$\pm$31$^e$ & 58$\pm$31$^e$  & $\cdots$		\\
Knot\,2 &    -12$\pm$5  &   -17$\pm$5  &  79$\pm$9  & 81$\pm$10   & 58$\pm$24  & 12$\pm$1	\\
Knot\,3 &    -13$\pm$5  &   -14$\pm$5  &  81$\pm$10   & 82$\pm$11 & 76$\pm$29  & 10$\pm$2    \\
Knot\,4 &    -14$\pm$5  &   -14$\pm$5  &  77$\pm$10   & 78$\pm$11 & 78$\pm$29  & 10$\pm$2    \\
Knot\,5 &    -5$\pm$5  &    -17$\pm$5  &  83$\pm$9   & 85$\pm$10  & 25$\pm$24  & 12$\pm$1    \\ 
BS\,4   &     -9$\pm$5 &    -17$\pm$5 &  115$\pm$9   & 116$\pm$10 & 62$\pm$34  & 8$\pm$1    \\ 
BS\,3   &     -8$\pm$5 &    -12$\pm$5 &  34$\pm$13   & 36$\pm$14  & 24$\pm$15  & 19$\pm$2    \\ 
BS\,2   &     -8$\pm$5 &    -11$\pm$5 &  82$\pm$8   & 83$\pm$9   & 60$\pm$38   & 8$\pm$2    \\
BS\,1   &     -5$\pm$5  &    -8$\pm$5  &  83$\pm$8   & 83$\pm$9  & 52$\pm$52   & 6$\pm$4    \\
\hline
\end{tabular}
\tablefoot{$^a$Tangential velocity of the 1-0\,S(1) line measured by \citet{Ray.ea.2023}.
$^b$Total velocity of the hot H$_2$ component derived from ${\rm v}_{\rm r} (0-0\,S(7))$ and ${\rm v}_{\rm tg} (1-0\,S(1))$. $^c$Total velocity of the warm H$_2$ component derived from ${\rm v}_{\rm r} (0-0\,S(1))$ and $i$. $^d$ Inclination angle to the plane of the sky as derived from ${\rm v}_{\rm r} (0-0\,S(7))$ and ${\rm v}_{\rm tg} (1-0\,S(1))$. $^e$Total velocity computed assuming a jet average inclination angle of 11$\degr$ with respect to the plane of the sky.}\\
\end{table*}

\section{Additional MIRI-MRS velocity maps}
\label{sec:add_vmaps}

\begin{figure}
        \includegraphics[width=0.47\textwidth]{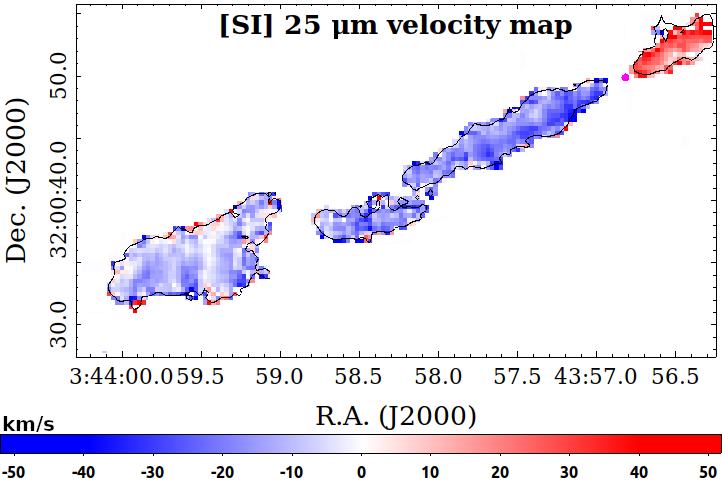} \vspace{0.4cm}
        \\
        \includegraphics[width=0.47\textwidth]{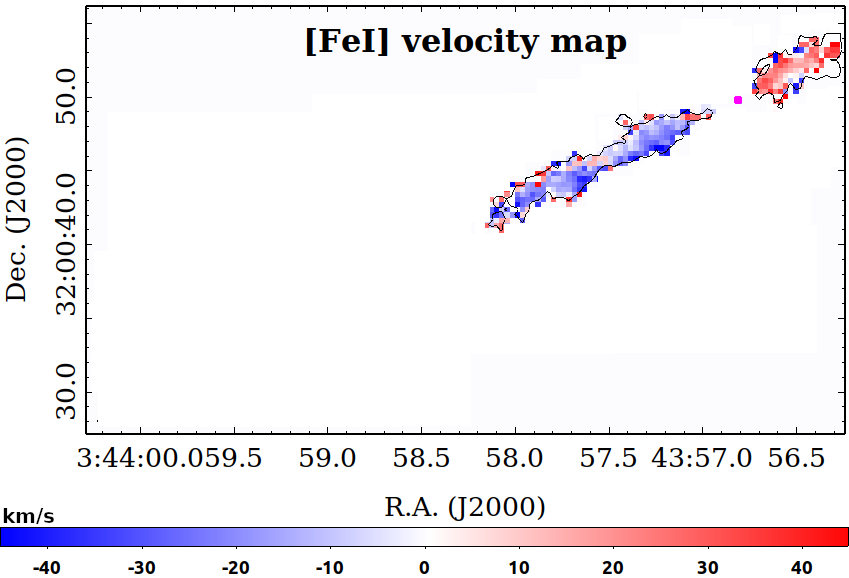} \vspace{0.4cm}
        \\
         \includegraphics[width=0.47\textwidth]{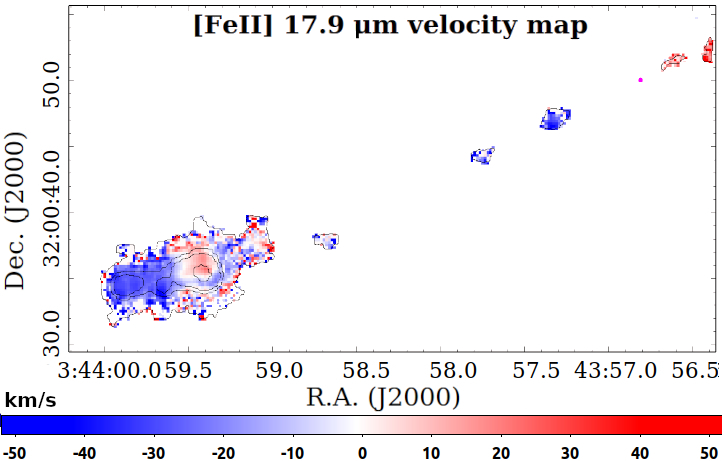}
    \caption{\SI at 25\,$\mu$m (top panel) and \FeI at 24\,$\mu$m (middle panel) and \FeII at 17.9\,$\mu$m (bottom panel) velocity maps. Black contours in the top and bottom panels show the integrated continuum-subtracted line intensity at 5$\sigma$ (0.8\,mJy\,pixel$^{-1}$). Bottom panel black contours are at 5, 50, 100, and 500\,$\sigma$ (namely, 0.5, 5, 10, and 50\,mJy\,pixel$^{-1}$
    Only velocities within a 5$\sigma$ threshold are plotted. The magenta circle shows the position of the ALMA mm continuum source.}
    \label{fig:atomic_vmap}
\end{figure}

\begin{figure}
        \includegraphics[width=0.48\textwidth]{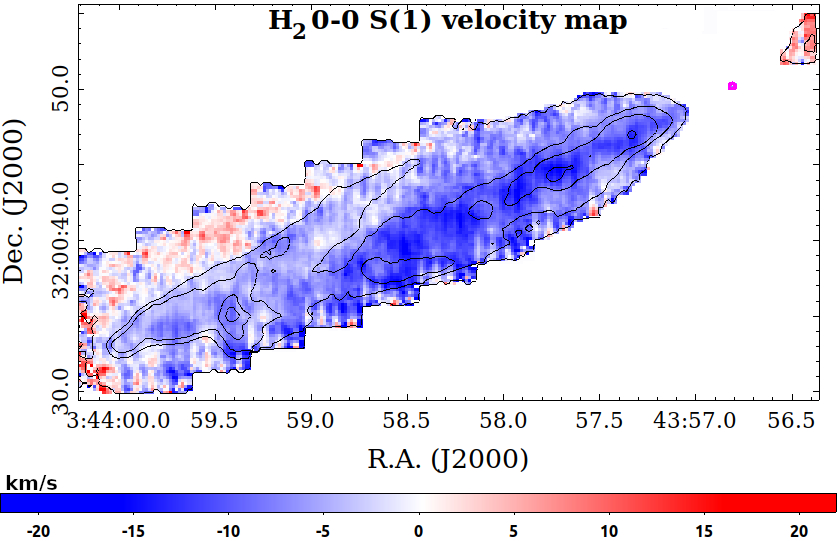}\vspace{0.4cm}
        \\
        \includegraphics[width=0.48\textwidth]{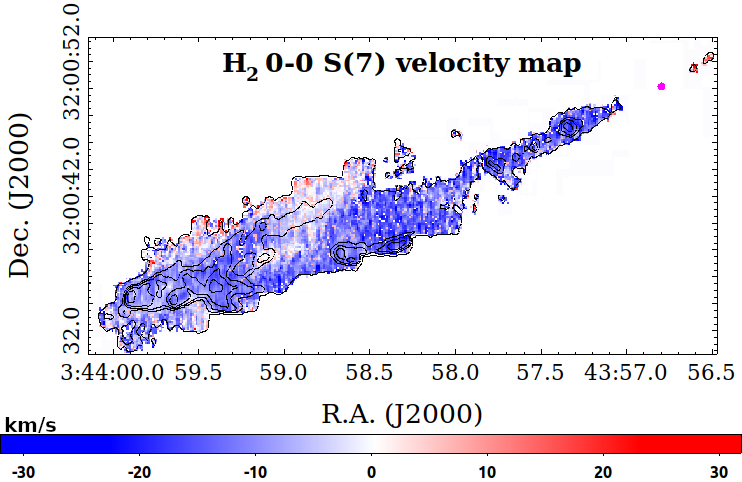}
    \caption{Top panel: H$_2$ velocity map of the 0-0\,S(1) at 17$\mu$m. Black contours show the integrated continuum-subtracted line intensity at 5, 10, 50, 100, and 200\,$\sigma$ (namely, 0.5, 1, 5, 10, and 20\,mJy\,pixel$^{-1}$). Bottom panel: H$_2$ velocity map of the 0-0\,S(7) at 5\,$\mu$m. Black contours show the integrated continuum-subtracted line intensity at 5, 50, 100, 200, and 500\,$\sigma$ (namely, 0.1, 1, 2, 4, and 10\,mJy\,pixel$^{-1}$). Only velocities within a 5$\sigma$ threshold are plotted in both panels. The magenta circle shows the position of the ALMA mm continuum source.    }
    \label{fig:H2_vmap}
\end{figure}

In this section, additional velocity maps of the brightest detected atomic species (\SI at 25\,$\mu$m and \FeI at 24\,$\mu$m; top and bottom panel of Figure\,\ref{fig:atomic_vmap}, respectively) and H$_2$ transitions (0-0\,S(1) and 0-0\,S(7); top and bottom panel of Figure\,\ref{fig:H2_vmap}, respectively) are presented. 

\end{document}